\documentclass[superscriptaddress,amsmath,amssymb,aps,pra,floatfix,twocolumn]{revtex4-2}

\usepackage{graphicx}
\usepackage{dcolumn}
\usepackage{bm}
\usepackage[utf8]{inputenc}
\usepackage{physics}
\usepackage{mathtools}
\usepackage[justification=RaggedRight, singlelinecheck=false, caption=false]{subfig}
\usepackage[colorlinks=true,linkcolor=blue]{hyperref}
\usepackage[dvipsnames]{xcolor}

\begin{document}

\title{Averaging method and coherence applied to Rabi oscillations in a two-level system}

\author{L. Chalkopiadis}
\affiliation{National and Kapodistrian University of Athens, Department of Physics, Panepistimiopolis, 15784 Zografos, Athens, Greece}
\author{C. Simserides}
\email{csimseri@phys.uoa.gr}
\affiliation{National and Kapodistrian University of Athens, Department of Physics, Panepistimiopolis, 15784 Zografos, Athens, Greece}

\date{\today}

\begin{abstract}
We study Rabi oscillations in a two-level system within the semiclassical approximation as an archetype test field of the Averaging Method (AM). The population transfer between the two levels is approached within the first and the second order AM. We systematically compare AM predictions with the rotating wave approximation (RWA) and with the complete numerical solution  utilizing standard algorithms (NRWA). We study both the resonance ($\Delta = 0$) and out-of-resonance ($\Delta \ne 0$) cases, where $\Delta = \omega-\Omega$, and $\hbar \Omega = E_2-E_1$ is the two-level energetic separation, while $\omega$ is the  (cyclic) frequency of the electromagnetic field. We introduce three types of dimensionless factors $\epsilon$, i.e., $\Omega_{\textrm{R}}/\Delta$, 
$\Omega_{\textrm{R}}/\Sigma$, and
$\Omega_{\textrm{R}}/\omega$, where
$\Omega_{\textrm{R}}$ is the Rabi (cyclic) frequency and  $\Sigma = \omega + \Omega$ and explore the range of $\epsilon$ where the AM results are equivalent to NRWA.
Finally, by allowing for a phase difference in the initial electron wave functions, we explore the prospects coherence can offer. We  illustrate that even with equal initial probabilities at the two levels, but with phase difference, strong oscillations can be generated and manipulated.
\end{abstract}

\maketitle


\section{Introduction}   
\label{sec:Introduction} 
The Averaging Method (AM) in nonlinear dynamical systems belongs to asymptotic methods~\cite{Sanders:2007}. The simplest form of averaging is {\it periodic} averaging, which deals with solving a perturbation problem of the standard form 
\begin{equation}\label{standardform}
\dot{\mathbf{x}} = \epsilon f(\mathbf{x},t) + 
\epsilon^2 g(\mathbf{x},t) + \dots
\end{equation}
We write the periodic (with period $T$) function $f$ as
\begin{equation}\label{f}
f(\mathbf{x},t) = \overline{f}(\mathbf{x}) +\widetilde{f}(\mathbf{x},t).
\end{equation}
$\overline{f}(\mathbf{x})$ is an {\it idiotypic} temporal average of $f$ in the regime $[0,T]$: We average over $t$, but assuming $\mathbf{x}(t)$ constant, hypothesizing that $\mathbf{x}(t)$ is a slowly varying function, i.e.,
\begin{equation}\label{mesitimi}
\overline{f}(\mathbf{x})= \frac{1}{T} \int_{0}^{T} f(\mathbf{x},t) dt.
\end{equation}
Similarly with $f(\mathbf{x},t)$, we treat $g(\mathbf{x},t)$, $\dots$.
Below we use similar separation of functions $f, g, h \dots$ to  $\overline{f}, \overline{g}, \overline{h}, \dots$ and 
$\widetilde{f}, \widetilde{g}, \widetilde{h}, \dots$.
Characteristic problems solved with AM can be found in Refs.~\cite{Sanders:2007, Verhulst:1975}. In recent years, among other applications, the  AM has been used in Robotics~\cite{ScienceRobotics1}, Engineering~\cite{Ogundele:2021, Mani:2020, WilsonEng:2020, Rega:2020, Du:2020}, Biology~\cite{Wilson:2020} and Physics~\cite{WilsonPhys:2020, Fajman:2021, Fajman:2020, Podvigina:2020, Gokler:2021}.

Oscillations of electron probabilities between (usually two) energy levels due to the presence of an oscillating perturbing electric field are usually termed \textit{Rabi oscillations}, although originally Rabi studied magnetic moment in the presence of a magnetic field~\cite{Rabi:1937}. Rabi oscillations can be treated fully  quantum-mechanically~\cite{Loudon:2000,Simserides:2016}, but here we use the semiclassical approach and the two-level system (2LS) as a archetype: Our aim is merely to examine the first and mainly the second order AM and compare it with standard Rotating Wave Approximation (RWA) as well as with a full numerical treatment without any compromise, NRWA.

We obtained the idea of using the 2LS in the semiclassical approach as a benchmark for the AM by coming across Ref.~\cite{Batista:2015}. However, we develop AM in a much different way:  
We introduce three types of
small quantities $\epsilon$, i.e., 
$\frac{\Omega_{\textrm{R}}}{\Delta}$,
$\frac{\Omega_{\textrm{R}}}{\Sigma}$ and
$\frac{\Omega_{\textrm{R}}}{\omega}$ 
for non-resonance ($\Delta \ne 0$), 
but only one type of $\epsilon$, i.e.,
$\frac{\Omega_{\textrm{R}}}{\omega}$,
for resonance ($\Delta = 0$).
$\Delta = \omega -\Omega$,  $\Sigma = \omega + \Omega$.
$\hbar \Omega = E_2-E_1$ is the two-level energetic separation, while $\omega$ is the (cyclic) frequency of the electromagnetic (EM) field. $\Omega_{\textrm{R}} := \mathcal{P} \mathcal{E}_0 / \hbar$ is the Rabi (cyclic) frequency, where $\mathcal{P}$ is the non-diagonal matrix element of the dipole moment along the electric field direction and $\mathcal{E}_0$ is the electric field amplitude. 

We use first as well as second order AM, cf. Sec.~\ref{sec:AM}. We systematically compare the AM results with NRWA numerical results as well as with RWA. We explore the range of $\epsilon$ parameters for AM to work successfully, i.e., so that the second order AM results are equivalent with the NRWA numerical results. We also we explore the prospects coherence can offer.

The article is organized as follows: In Sec.~\ref{sec:RWA-NRWA} we define RWA and NRWA within the semiclassical approach. In Sec.~\ref{sec:AM} we introduce the AM. In Sec.~\ref{sec:NRWA-RWA-AM} we compare our results for AM, RWA and NRWA. Finaly, in Sec.~
\ref{sec:Conclusion} we state our conclusions.

\vspace{-0.5cm}

\section{Rabi oscillations with or without Rotating Wave Approximation} 
\label{sec:RWA-NRWA} 
Rabi oscillations of electron probabilities $|C_1(t)|^2$ and $|C_2(t)|^2$, as functions of time $t$, in a 2LS interacting with an electromagnetic field, are described by~\cite{Simserides:2016}
\begin{equation}\label{2LS}
\left.
\begin{aligned}
&\dot{C}_1(t) = C_2(t) \dfrac{i\Omega_{\textrm{R}} }{2} \,
\bigg[ e^{i(\omega - \Omega)t}  +  e^{-i(\omega + \Omega)t} \bigg]\\
&\dot{C}_2(t) = C_1(t) \dfrac{i \Omega_{\textrm{R}} }{2} \,
\bigg[ e^{ i (\omega + \Omega)t}  +  e^{-i (\omega - \Omega)t} \bigg]
\end{aligned}
\right..
\end{equation}
Using $\Delta = \omega -\Omega$ and $\Sigma = \omega + \Omega$, we obtain
\begin{align}\label{2LS_M}
\hspace{-0.5cm}
\begin{bmatrix}
\dot{C_1}(t)\\
\dot{C_2}(t)
\end{bmatrix} \! = \!
\frac{i \Omega_{\textrm{R}} }{2}
\begin{bmatrix}
0 & e^{+i\Delta t}+e^{-i\Sigma t} \\
e^{-i\Delta t}+e^{+i\Sigma t} & 0
\end{bmatrix}
\begin{bmatrix}
C_1(t) \\
C_2(t)
\end{bmatrix}
\end{align}
or 
\begin{align}\label{2LS_Mx}
\dot{\mathbf{x}}(t) = i \mathbf{A}(t) \; \mathbf{x}(t).
\end{align}
To appreciate the complexity of Eq.~\eqref{2LS_Mx}, we notice that trying the eigenvector-eigenvalue method, i.e., solutions of the form $\mathbf{x}(t)=\mathbf{v} e^{i\lambda t}$, we obtain $\mathbf{A}(t) \mathbf{v} = \lambda \mathbf{v}$, i.e., a different  eigenvector-eigenvalue problem for each $t$.

Equation~\eqref{2LS_M} describes a separation into counter-rotating terms containing $\Delta$ or $\Sigma$. Taking into account only terms containing $\Delta$ is termed Rotating Wave Approximation (RWA). We shall call the full problem containing both the $\Delta$ and the $\Sigma$ terms NRWA (no RWA). To solve Eq.~\eqref{2LS_M} numerically, within NRWA, we utilize matlab, using trapezoid and Runge-Kutta (4,5) algorithms. The probability to find the electron at the lower (higher) level is 
$P_1(t)=|C_1(t)|^2$ ($P_2(t)=|C_2(t)|^2$).

RWA stems from the assumption that if $\omega$ is close to $\Omega$, the $\Delta$ terms are slow and the $\Sigma$ terms are fast. Hence, in any remarkable time scale, the fast terms are somehow expected to have negligible effect. RWA is the claim that we can ignore the fast terms, i.e.,
\begin{align}\label{2LS_M_RWA}
\begin{bmatrix}
\dot{C_1}(t)\\
\dot{C_2}(t)
\end{bmatrix}=\frac{i \Omega_{\textrm{R}} }{2}
\begin{bmatrix}
0 & e^{+i\Delta t} \\
e^{-i\Delta t} & 0
\end{bmatrix}
\begin{bmatrix}
C_1(t) \\
C_2(t)
\end{bmatrix}.
\end{align}
The quantitative difference between NRWA and RWA is expored in the work, rather systematically. The analytical solution within RWA is known~\cite{Simserides:2016}. 

For example, with initial conditions $C_1(0)=1$, $C_2(0)=0$, we obtain
\begin{equation}\label{C1oftsquaredRWA}
P_{1,\textrm{RWA}}(t)=
1 - \dfrac{\Omega_{\textrm{R}}^2}{\Omega_{\textrm{R}}^2 + \Delta^2} \sin^2(\lambda t),
\end{equation}
\begin{equation} \label{C2oftsquaredRWA}
P_{2,\textrm{RWA}}(t)=
\dfrac{\Omega_{\textrm{R}}^2}{\Omega_{\textrm{R}}^2 + \Delta^2} \; \sin^2(\lambda t),
\end{equation}
$2\lambda = \sqrt{\Omega_{\textrm{R}}^2 + \Delta^2}$. Hence, 
the period of the oscillations, 
\begin{equation}\label{TRWA}
T_{\textrm{RWA}} = \frac{2\pi}{\sqrt{\Omega_{\textrm{R}}^2+\Delta^2}}, 
\end{equation}
and the maximum transfer percentage, 
\begin{equation}\label{ARWA}
\mathcal{A}_{\textrm{RWA}} = \frac{\Omega_{\textrm{R}}^2}{\Omega_{\textrm{R}}^2+\Delta^2}.
\end{equation}
The period at resonance ($\Delta = 0$), $T_{\textrm{RWA,0}}= 2\pi/\Omega_{\textrm{R}}$, offers a convenient time scale, which will be exploited later, in the presentation of our results. However, we notice that when $\Omega_{\textrm{R}}$ is significant, even at resonance,
the frequency of the oscillations predicted by RWA, $f_{\textrm{RWA}}=1/T_{\textrm{RWA}}$, 
does not coincide with the main frequency of NRWA, which,  additionally, has richer frequency content.

If we assume the initial condition $C_1(0) = \frac{1}{\sqrt{2}}e^{i \theta}$ and $C_2(0) = \frac{1}{\sqrt{2}}e^{i \phi}$, i.e., we place the electron, at time zero with equal probability at both levels, $|C_1(0)|^2 = |C_2(0)|^2 = \frac{1}{2}$, but we allow phase to vary, we obtain
\begin{align} \label{P1phasi}
P_1(t) = \frac{1}{2} 
& - \frac{\Omega_R \Delta }{2(\Omega_R^2 + \Delta^2)} \cos(\theta - \phi)(1 - \cos(2\lambda t)) \nonumber \\
& + \frac{\Omega_R}{2 \sqrt{\Omega_R^2 + \Delta^2}} \sin(2\lambda t) \sin(\theta - \phi) 
\end{align}
\begin{align} \label{P2phasi}
P_2(t) = \frac{1}{2} 
& + \frac{\Omega_R \Delta }{2(\Omega_R^2 + \Delta^2)} \cos(\theta - \phi)(1-\cos(2\lambda t)) \nonumber \\
& - \frac{\Omega_R}{2 \sqrt{\Omega_R^2 + \Delta^2}} \sin(2\lambda t) \sin(\theta - \phi) 
\end{align}
$2\lambda = \sqrt{\Omega_{\textrm{R}}^2 + \Delta^2}$.
In case of resonance ($\Delta = 0 \Rightarrow \omega = \Omega$), we have
\begin{align} \label{phasiDeltaZero1}
P_1(t) = \frac{1}{2} + \frac{1}{2} \sin(\Omega_R t) \sin(\theta- \phi)
\end{align}
\begin{align} \label{phasiDeltaZero2}
P_2(t) = \frac{1}{2} - \frac{1}{2} \sin(\Omega_R t) \sin(\theta- \phi) 
\end{align}
Hence, if $\theta - \phi = 2\kappa \pi$ with $\kappa = 0, 1, 2,...$, then in case of resonance, Rabi oscillations disappear, i.e., the probabilities are $P_1(t) = P_2(t ) = \frac{1}{2}$,  constantly.

Finally, we notice that if $\Omega_{\textrm{R}} >> \Delta$ and $\Omega_{\textrm{R}} >> \Sigma$, so that in Eq.~\eqref{2LS} each exponential term  within the square brackets can be approximated by $1$, then, 
\begin{equation}\label{C1oftsquaredext}
P_{1*}(t)=\cos^2(\Omega_{\textrm{R}}t),
\end{equation}
\begin{equation} \label{C2oftsquaredext}
P_{2*}(t)=\sin^2(\Omega_{\textrm{R}}t),
\end{equation}
Hence, the period of the oscillations, 
\begin{equation}\label{Text}
T_{*} = \frac{\pi}{\Omega_{\textrm{R}}}, 
\end{equation}
and the maximum transfer percentage, 
\begin{equation}\label{Aext}
\mathcal{A}_{*} =1.
\end{equation}

\section{Averaging Method} 
\label{sec:AM} 
For non-resonance (Subsec.~\ref{subsec:NR}), we employ three types of small quantities $\epsilon$, i.e., 
$\frac{\Omega_{\textrm{R}}}{\Delta}$,
$\frac{\Omega_{\textrm{R}}}{\Sigma}$ and
$\frac{\Omega_{\textrm{R}}}{\omega}$.
Unavoidably, when $\Delta$ becomes smaller,
at some point, $\frac{\Omega_{\textrm{R}}}{\Delta}$ gets so large that non-resonant AM is not successful anymore.
Hence, resonance must be treated via a different path, using only one type of $\epsilon$, i.e.,
$\frac{\Omega_{\textrm{R}}}{\omega}$ (Subsec.~\ref{subsec:R}).
This will become evident also in the Results, Sec.~
\ref{sec:NRWA-RWA-AM}. 

\subsection{Non-Resonance} 
\label{subsec:NR} 
\subsubsection{First order AM} 
\label{subsubsec:1stAM} 
In case of non resonance ($\Delta \neq 0 \Rightarrow \omega \neq \Omega$), we can write Eq.~\ref{2LS_M} in the AM form of Eq.~\ref{standardform},
\begin{align}\label{NRCstandard}
\dot{\mathbf{x}}(t) =\underbrace{\frac{i \Omega_{\textrm{R}} }{2}
	\begin{bmatrix}
	0 & e^{+i\Delta t}+e^{-i\Sigma t} \\
	e^{-i\Delta t}+e^{+i\Sigma t} & 0
	\end{bmatrix} \mathbf{x}(t)}_{\epsilon f(\mathbf{x},t)},
\end{align}
i.e., we call $\epsilon f(\mathbf{x},t)$ the RHS of Eq.~\ref{2LS_M} and 
\begin{equation}\label{giszero}
g(\mathbf{x},t)=0.
\end{equation}
Equation~\eqref{NRCstandard} includes two periods, $T_{\Delta} = \frac{2\pi}{\Delta}$ and $T_{\Sigma} = \frac{2\pi}{\Sigma}$. If $\frac{T_{\Delta}}{T_{\Sigma}}$ is a rational number, the system is {\it periodic} with common period, $T$, the least common multiple of $T_{\Delta}$ and $T_{\Sigma}$. [In numerical calculations, since $T_{\Delta}$ and $T_{\Sigma}$ are represented as floats $\frac{T_{\Delta}}{T_{\Sigma}}$ is always a rational number.] 

In our case, combining \eqref{mesitimi} and  Eqs.~\eqref{NRCstandard}, we obtain 
\begin{align}\label{NRCstandardp}
\epsilon \overline{f}(\mathbf{x}) = 
\frac{i \Omega_{\textrm{R}} }{2}
\begin{bmatrix}
0&0\\
0&0
\end{bmatrix} 
\textbf{x}(t) =
\begin{bmatrix}
0&0\\
0&0
\end{bmatrix}. 
\end{align}
Combining Eqs.~\eqref{f},~\eqref{NRCstandard} and ~\eqref{NRCstandardp}, we obtain
\begin{align}\label{epsilontildef}
\epsilon \widetilde{f}(\mathbf{x},t) =\frac{i \Omega_{\textrm{R}} }{2}
\begin{bmatrix}
0 & e^{+i\Delta t}+e^{-i\Sigma t} \\
e^{-i\Delta t}+e^{+i\Sigma t} & 0
\end{bmatrix} \mathbf{x}(t).
\end{align}

In first order AM, we define
\begin{equation}\label{transformationfirst}
\mathbf{x}(t) = \mathbf{y}(t) + 
\epsilon \; \mathbf{w}[\mathbf{y}(t),t], 
\end{equation}
where $\mathbf{w}$, a function of $\mathbf{y}(t)$ and $t$, 
is defined so that
\begin{equation}\label{w}
\widetilde{f}(\mathbf{y},t)= 
\frac{\partial \mathbf{w}}{\partial t}.
\end{equation}
The reason behind this definition becomes clear when we perform detailed calculations, cf. Eq.\eqref{Rearranging}. 
After the detailed calculations shown in Appendix~\ref{Appendix1stAM}, we obtain
\begin{align}\label{AMfirstorder}
& \dot{\mathbf{y}} = \epsilon \; \overline{f}(\mathbf{y}) + \\ 
& \epsilon^2 \; \left(\frac{\partial \overline{f}(\mathbf{y})}{\partial \mathbf{y}} \mathbf{w} + 
\frac{\partial \widetilde{f}(\mathbf{y},t)}{\partial \mathbf{y}} \mathbf{w} + g(\mathbf{y},t) - \frac{\partial \mathbf{w}}{\partial \mathbf{y}} \overline{f}(\mathbf{y})\right) + \mathcal{O}(\epsilon^3) \nonumber
\end{align}
In Eq~\eqref{AMfirstorder}, if we ignore terms of order $\epsilon^2$ and above, we obtain the first order AM of Eq.~\eqref{standardform}, i.e., 
\begin{equation}
\dot{\mathbf{y}}= \epsilon \; \overline{f}(\mathbf{y}).
\end{equation}
In our case, $\overline{f}(\mathbf{y}) = 0$, therefore,
\begin{equation}
\dot{\mathbf{y}}=0
\end{equation}
Therefore, $\mathbf{y}$ is a constant, i.e.,
\begin{equation}\label{ysolution}
\mathbf{y}=\begin{bmatrix}
y_{10} \\ y_{20}
\end{bmatrix}.
\end{equation}
We determine $y_{10}, y_{20}$ by applying the initial conditions.

Finally, from Eqs.~\eqref{epsilontildef}, \eqref{w}, \eqref{taylor2}, we obtain
\begin{equation}\label{w2}
\epsilon \mathbf{w}(\mathbf{y},t) = 
\frac{i \Omega_{\textrm{R}}}{2} 
\begin{bmatrix}
0 & \frac{e^{+i\Delta t}}{i\Delta}-\frac{e^{-i\Sigma t}}{i\Sigma} \\
-\frac{e^{-i\Delta t}}{i\Delta}+\frac{e^{+i\Sigma t}}{i\Sigma} & 0
\end{bmatrix} \mathbf{y}(t)
\end{equation}

\subsubsection{Second order AM} 
\label{subsubsec:2ndAM} 
In second order AM, we start from Eq.~\eqref{AMfirstorder},
and call $\mathbf{h}(\mathbf{y},t)$ the $\mathcal{O}(\epsilon^2)$ function, i.e.,
\begin{align} \label{AMsecondorderstart}
& \dot{\mathbf{y}} = \epsilon \overline{f}(\mathbf{y}) + \\ 
& \epsilon^2 \underbrace{\left(\frac{\partial \overline{f}(\mathbf{y})}{\partial \mathbf{y}} \mathbf{w} + 
	\frac{\partial \widetilde{f}(\mathbf{y},t)}{\partial \mathbf{y}} \mathbf{w} + g(\mathbf{y},t) - \frac{\partial \mathbf{w}}{\partial \mathbf{y}} \overline{f}(\mathbf{y})\right)}_{\mathbf{h}(\mathbf{y},t)} + \mathcal{O}(\epsilon^3). \nonumber
\end{align}
Then, as with Eq.~\eqref{f}, we define
\begin{equation}\label{h}
\mathbf{h}(\mathbf{y},t) = \overline{h}(\mathbf{y}) + \widetilde{h}(\mathbf{y},t),
\end{equation}
and similarly to Eq.\eqref{transformationfirst}, we define
\begin{equation}\label{transformationsecond}
\mathbf{y}(t)=\mathbf{z}(t) + \epsilon^2 \; \mathbf{u}[\mathbf{z}(t),t],
\end{equation}
where $\mathbf{u}$ is a function of $\mathbf{z}(t)$ and $t$,  defined so that
\begin{equation}\label{u}
\widetilde{h}(\mathbf{z},t) = \frac{\partial \mathbf{u}}{\partial t}.
\end{equation}
Function $\mathbf{u}$ in Eq.~\eqref{u}, in second order AM, has the same role as function $\mathbf{w}$ in Eq.~\eqref{w}, in first order AM.
After detailed calculations shown in Appendix~\ref{Appendix2ndAM}, we obtain
\begin{equation}\label{AMsecondorder}
\dot{\mathbf{z}} = \epsilon \; \overline{f}(\mathbf{z}) + \epsilon^2 \; \overline{h}(\mathbf{z}) + \mathcal{O}(\epsilon^3)
\end{equation}
In Eq.~\eqref{AMsecondorder}, if we ignore terms of order  $\epsilon^3$ and above, we obtain the second order AM of
Eq.~\eqref{standardform}, i.e.,
\begin{equation}\label{secondorder}
\dot{\mathbf{z}}= \epsilon \; \overline{f}(\mathbf{z}) + \epsilon^2 \; \overline{h}(\mathbf{z}),
\end{equation}
where
\begin{align}\label{h1}
& \mathbf{h}(\mathbf{z},t) = \frac{\partial \overline{f}(\mathbf{z}) }{\partial \mathbf{z}} \mathbf{w}(\mathbf{z},t) + \\ 
& \frac{\partial \widetilde{f}(\mathbf{z},t) }{\partial \mathbf{z}} \mathbf{w}(\mathbf{z},t) + g(\mathbf{z},t) - \frac{\partial \mathbf{w}(\mathbf{z},t)}{\partial \mathbf{z}} \overline{f}(\mathbf{z}).
\nonumber
\end{align}
In our case, $\overline{f}(\mathbf{z}) = 0$ and $g(\mathbf{z},t) = 0$, cf. Eqs.~\eqref{NRCstandardp} and \eqref{giszero}, respectively. Hence, 
\begin{equation}\label{h2}
\mathbf{h}(\mathbf{z}(t),t) = \frac{\partial \widetilde{f}(\mathbf{z},t)}{\partial \mathbf{z}} \mathbf{w}(\mathbf{z},t).
\end{equation}
Therefore, via Eqs.~\eqref{NRCstandardp} and \eqref{h2}, 
Eq.~\eqref{secondorder} reads
\begin{equation}\label{dotz}
\dot{\mathbf{z}}=\epsilon^2 \; \overline{\frac{\partial \widetilde{f}(\mathbf{z},t)}{\partial \mathbf{z}} \mathbf{w}(\mathbf{z},t)}.
\end{equation}

Combining Eqs.~\eqref{epsilontildef},~\eqref{w2}~and~\eqref{h2}, \begin{align} \label{h3}
&\epsilon^2 \mathbf{h}(\mathbf{z}(t),t)= \\
& i \left(\frac{\Omega_R}{2}\right)^2 \left(\frac{1}{\Sigma} 
\begin{bmatrix}
1 & 0 \\
0 & -1
\end{bmatrix} 
+ \frac{1}{\Delta}
\begin{bmatrix}
-1 & 0 \\
0 & 1
\end{bmatrix}\right)\mathbf{z}(t) \nonumber + \\
& i \left(\frac{\Omega_R}{2}\right)^2
\frac{1}{\Delta} \begin{bmatrix}
-e^{-i(\Delta + \Sigma)t} & 0 \\
0 & e^{i(\Delta + \Sigma) t}
\end{bmatrix}\mathbf{z}(t) \nonumber + \\
& i \left(\frac{\Omega_R}{2}\right)^2 
\frac{1}{\Sigma} 
\begin{bmatrix}
e^{i(\Delta + \Sigma)t} & 0 \\
0 & -e^{-i(\Delta + \Sigma) t}
\end{bmatrix}\mathbf{z}(t) \nonumber
\end{align}
where 
\begin{align} \label{h_average2}
&\epsilon^2 \overline{h}(\mathbf{z}) = \epsilon^2 \overline{\bigg[\frac{\partial \widetilde{f}(\mathbf{z},t)}{\partial \mathbf{z}} \mathbf{w}(\mathbf{z},t)\bigg]} = \\
& i (\frac{\Omega_{\textrm{R}}}{2})^2 \bigg(\frac{1}{\Sigma} \begin{bmatrix}
1 & 0 \\
0 & -1 
\end{bmatrix} + \frac{1}{\Delta} \begin{bmatrix}
-1 & 0 \\
0 & 1 
\end{bmatrix}\bigg)\mathbf{z}(t). \nonumber
\end{align}
From Eqs~\eqref{h}, \eqref{h3}, \eqref{h_average2} and $\Delta + \Sigma = 2\omega$ we have
\begin{align}\label{tildeh}
& \epsilon^2 \tilde{h}(\mathbf{z},t) = \\
& i (\frac{\Omega_R}{2})^2 \left(\frac{1}{\Delta} \begin{bmatrix}
- e^{-i2\omega t} & 0 \\
0 & e^{i2\omega t}
\end{bmatrix} + \frac{1}{\Sigma} \begin{bmatrix}
e^{i2\omega t} & 0 \\
0 & -e^{-i2\omega t}
\end{bmatrix} 
\right) \mathbf{z}(t) \nonumber
\end{align}
Therefore, Eq.~\eqref{dotz} becomes
\begin{equation}\label{zdotz}
\dot{\mathbf{z}} = i A \begin{bmatrix}
-1 & 0 \\
0 & 1
\end{bmatrix}\mathbf{z}(t), 
\end{equation}
\begin{equation} \label{equationA}
A = (\frac{\Omega_{\textrm{R}} }{2})^2(\frac{2\Omega}{\omega^2 - \Omega^2}).
\end{equation}
The solution of Eq.~\eqref{zdotz} is
\begin{equation}
\mathbf{z}(t)=\begin{bmatrix}
z_1(t) \\ z_2(t) \end{bmatrix} = \begin{bmatrix}
z_{10} e^{-iAt} \\ z_{20} e^{iAt}
\end{bmatrix}
\end{equation}
Also, $\mathbf{u}$ is calculated from the Eqs~\eqref{u}, \eqref{tildeh}
\begin{align}
& \epsilon^2 \mathbf{u}(\mathbf{z},t) = \\
& i(\frac{\Omega_{\textrm{R}}}{2})^2 \begin{bmatrix}
\frac{e^{-i2\omega t}}{i 2 \omega \Delta} + \frac{e^{i2\omega t}}{i 2 \omega \Sigma}& 0 \\
0 & \frac{e^{i2\omega t}}{i 2 \omega \Delta} + \frac{e^{-i2\omega t}}{i 2 \omega \Sigma}
\end{bmatrix} \mathbf{z}(t) \nonumber
\end{align}

Finally, we epitomize our AM results. For \textit{first order}:
\begin{equation}
\mathbf{x} = \mathbf{y} + \epsilon \; \mathbf{w}(\mathbf{y},t),
\end{equation} 
\begin{equation}
\mathbf{y} = 
\begin{bmatrix}
y_{10} \\ y_{20}
\end{bmatrix},
\end{equation}
\begin{equation}
\epsilon \mathbf{w}(\mathbf{y},t) = i \frac{\Omega_{\textrm{R}}}{2} 
\begin{bmatrix}
0 & \frac{e^{+i\Delta t}}{i\Delta}-\frac{e^{-i\Sigma t}}{i\Sigma} \\
-\frac{e^{-i\Delta t}}{i\Delta}+\frac{e^{+i\Sigma t}}{i\Sigma} & 0
\end{bmatrix} \mathbf{y}.
\end{equation}
For \textit{second order}:
\begin{equation}
\mathbf{x}(t) = \mathbf{z}(t) + \epsilon \; \mathbf{w}(\mathbf{z},t) + \epsilon^2 \; \mathbf{u}(\mathbf{z},t), 
\end{equation}
\begin{equation}
\mathbf{z}(t) = 
\begin{bmatrix}
z_1 \\ z_2 \end{bmatrix} = 
\begin{bmatrix}
z_{10} e^{-iAt} \\ z_{20} e^{iAt}
\end{bmatrix},
\end{equation}
\begin{equation}
\epsilon \; \mathbf{w}(\mathbf{z},t) = 
i \frac{\Omega_{\textrm{R}}}{2} 
\begin{bmatrix}
0 & \frac{e^{+i\Delta t}}{i\Delta}-\frac{e^{-i\Sigma t}}{i\Sigma} \\
-\frac{e^{-i\Delta t}}{i\Delta}+\frac{e^{+i\Sigma t}}{i\Sigma} & 0
\end{bmatrix} \mathbf{z},
\end{equation}
\begin{equation}
\epsilon^2 \mathbf{u}(\mathbf{z},t) = i \left(\frac{\Omega_{\textrm{R}}}{2}\right)^2 
\begin{bmatrix}
\frac{e^{-i2\omega t}}{i 2 \omega \Delta} + \frac{e^{i2\omega t}}{i 2 \omega \Sigma}& 0 \\
0 & \frac{e^{i2\omega t}}{i 2 \omega \Delta} + \frac{e^{-i2\omega t}}{i 2 \omega \Sigma}
\end{bmatrix} \mathbf{z}.
\end{equation}
Three types of $\epsilon$ occur in our first and second order equations: $\frac{\Omega_{\textrm{R}}}{\Delta}$, $\frac{\Omega_{\textrm{R}}}{\Sigma}$ and
$\frac{\Omega_{\textrm{R}}}{\omega}$.
The constants $y_{10}$, $y_{20}$, $z_{10}$, $z_{20}$ are  calculated from the initial conditions.

\subsection{Resonance} 
\label{subsec:R} 
Similarly, we treat the resonant case ($\Delta = 0 => \omega = \Omega$). Details can be found in Ref.~\cite{Chalkopiadis:2021}.
Below we summarize our AM results. For \textit{first order}:
\begin{equation}
\mathbf{x} = \mathbf{y} + \epsilon \; \mathbf{w}(\mathbf{y},t),
\end{equation} 
\begin{equation}
y(t) = 
\begin{bmatrix}
y_1 \\ y_2
\end{bmatrix}= 
\begin{bmatrix}
A_{11} \cos(\frac{\Omega_{\textrm{R}}}{2} t) + 
B_{11} \sin(\frac{\Omega_{\textrm{R}}}{2} t) \\  
A_{21} \cos(\frac{\Omega_{\textrm{R}}}{2} t) + 
B_{21} \sin(\frac{\Omega_{\textrm{R}}}{2} t) 
\end{bmatrix},
\end{equation}
\begin{equation}
\epsilon \; \mathbf{w}(\mathbf{y},t) = \frac{\Omega_{\textrm{R}}}{4\omega} \begin{bmatrix}
0 & -e^{-i2\omega t} \\ e^{i2\omega t} & 0
\end{bmatrix} \mathbf{y}.
\end{equation}
For \textit{second order}:
\begin{equation}
\mathbf{x}(t) = \mathbf{z}(t) + \epsilon \; \mathbf{w}(\mathbf{z},t) + \epsilon^2 \; \mathbf{u}(\mathbf{z},t),
\end{equation}
\begin{equation}
\mathbf{z}(t) = 
\begin{bmatrix}
z_1 \\ z_2
\end{bmatrix} =
\begin{bmatrix}
A_{12} \cos(Bt) + B_{12} \sin(Bt) \\  
A_{22} \cos(Bt) + B_{22} \sin(Bt) 
\end{bmatrix},
\end{equation}
where
\begin{equation}
B = \sqrt{\left[\left(\frac{\Omega_{\textrm{R}}}{2}\right)^2\frac{1}{2\omega}\right]^2 + \left(\frac{\Omega_{\textrm{R}}}{2}\right)^2}
\end{equation}
\begin{equation}
\epsilon \; \mathbf{w}(\mathbf{z},t) = \frac{\Omega_{\textrm{R}}}{4\omega} \begin{bmatrix}
0 & -e^{-i2\omega t} \\ e^{i2\omega t} & 0
\end{bmatrix} \; \mathbf{z},
\end{equation}
\begin{equation}
\epsilon^2 \; \mathbf{u}(\mathbf{z},t) = 
i \left(\frac{\Omega_{\textrm{R}}}{2}\right)^2 \frac{1}{2\omega^2} 
\sin{2 \omega t} 
\begin{bmatrix}
1 & 0 \\ 0 & -1
\end{bmatrix} \; \mathbf{z}.
\end{equation}
In the resonant case $\epsilon$ is introduced in the first and second order equations with just one form, $\frac{\Omega_{\textrm{R}}}{\omega}$. The constants $A_{11}$, $A_{21}$, $B_{11}$, $B_{21}$, $A_{12}$, $A_{22}$, $B_{12}$, $B_{22}$ are calculated from the initial conditions.

\begin{figure*}[t!]
	\centering
	\vspace{-0.5cm}
	\includegraphics[width=0.45\textwidth]{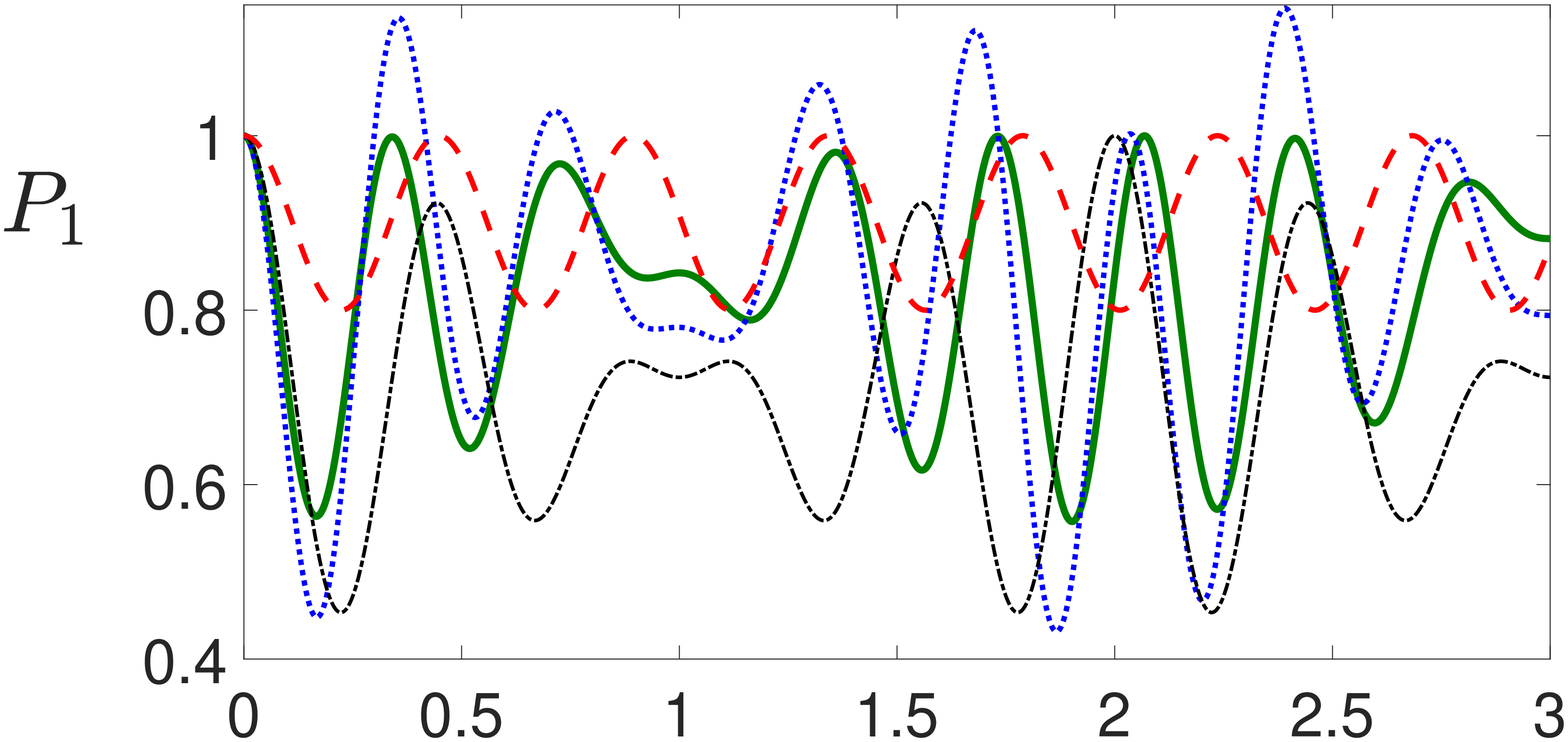}
	\includegraphics[width=0.45\textwidth]{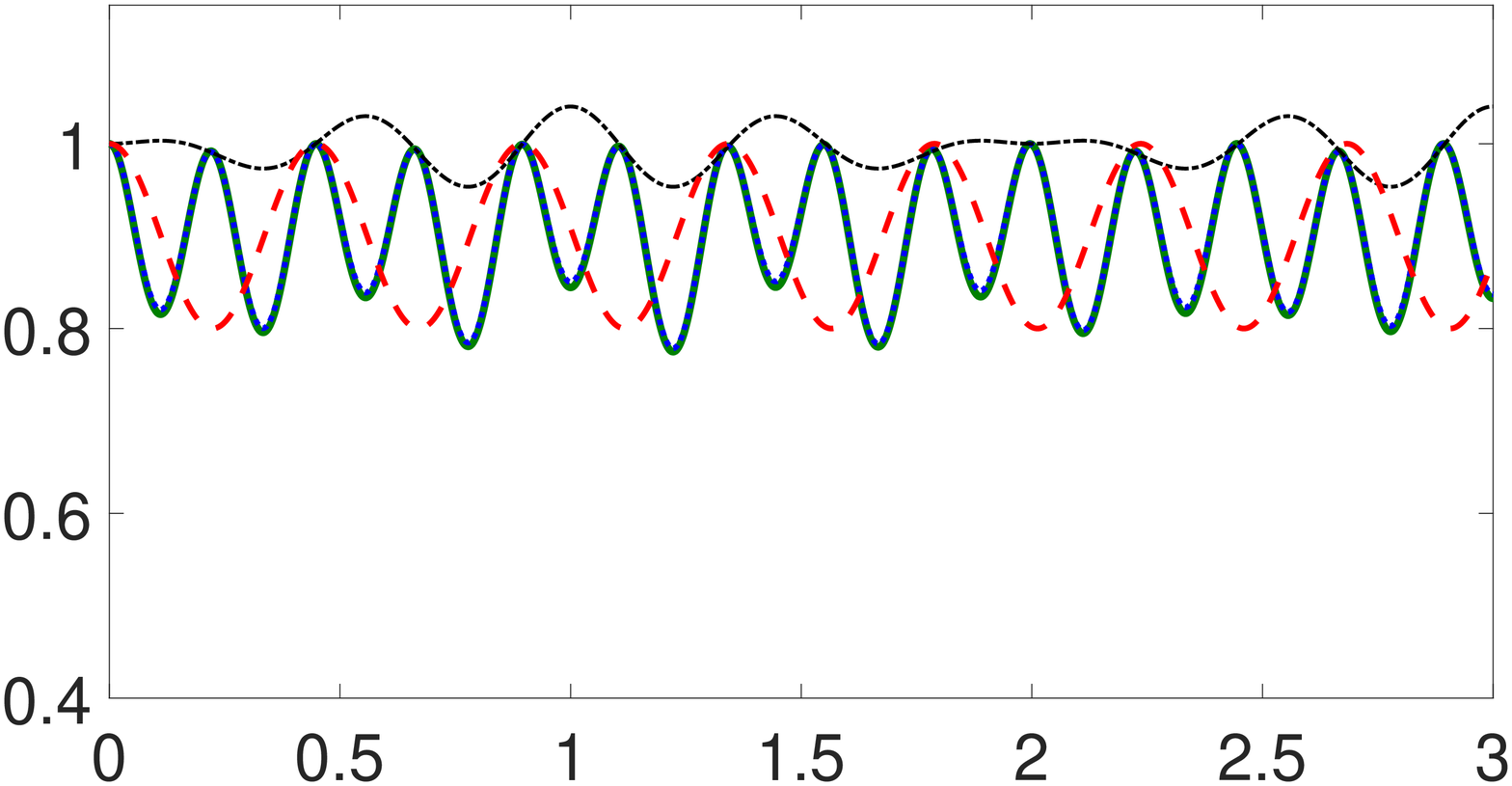}\\
	\vspace{-0.6cm}
	\includegraphics[width=0.45\textwidth]{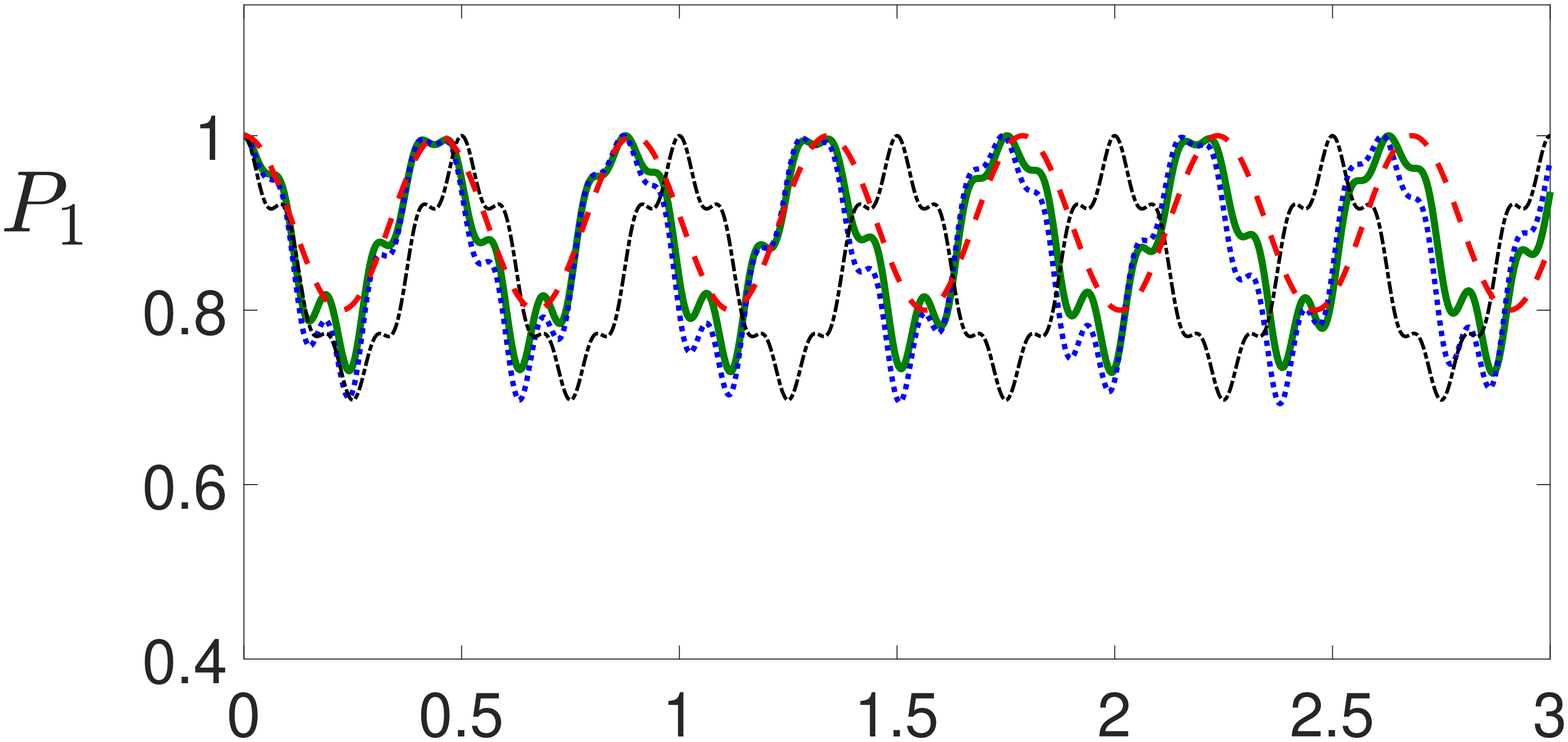}
	\includegraphics[width=0.45\textwidth]{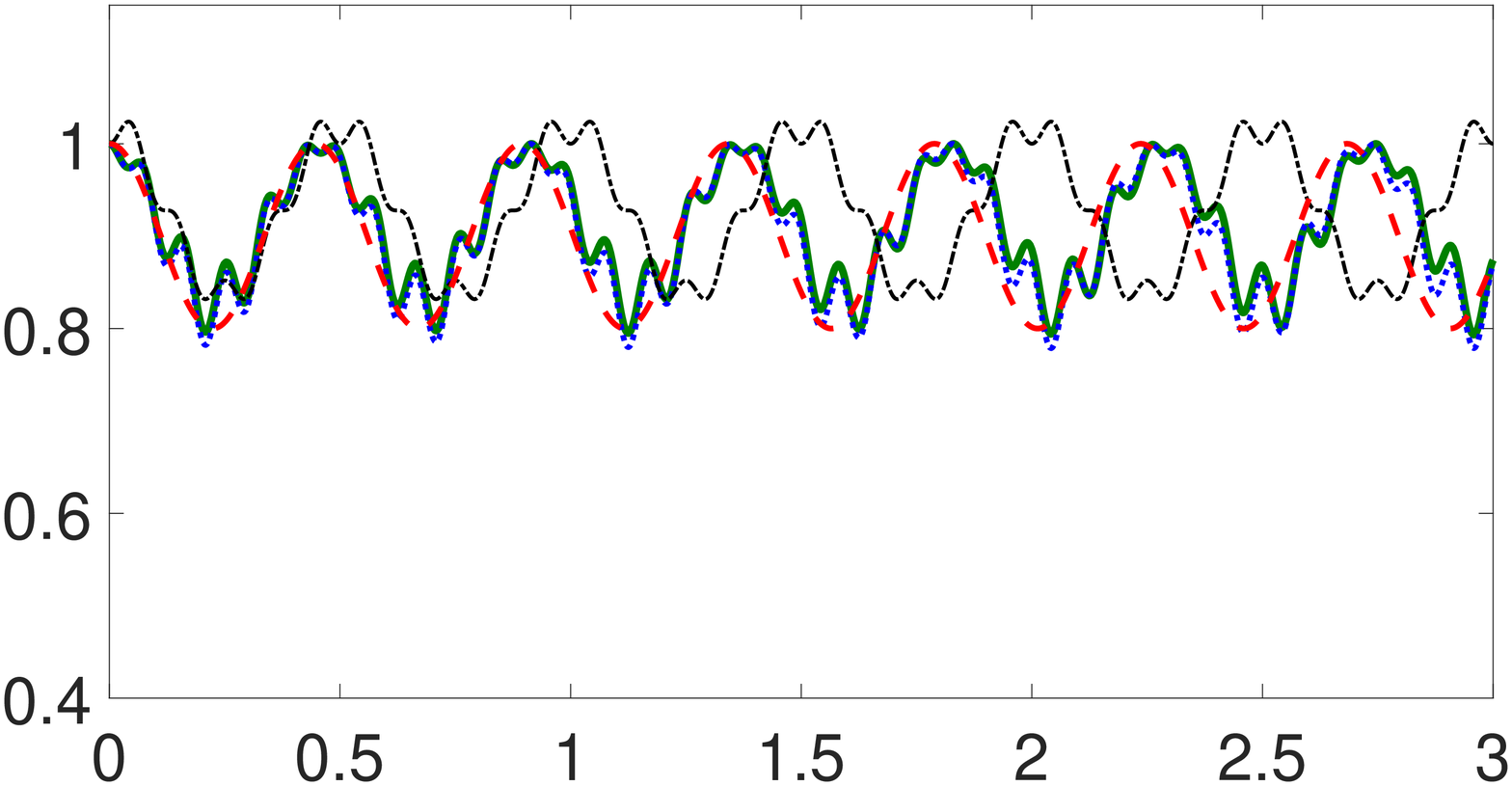}\\
	\vspace{-0.6cm}
	\includegraphics[width=0.45\textwidth]{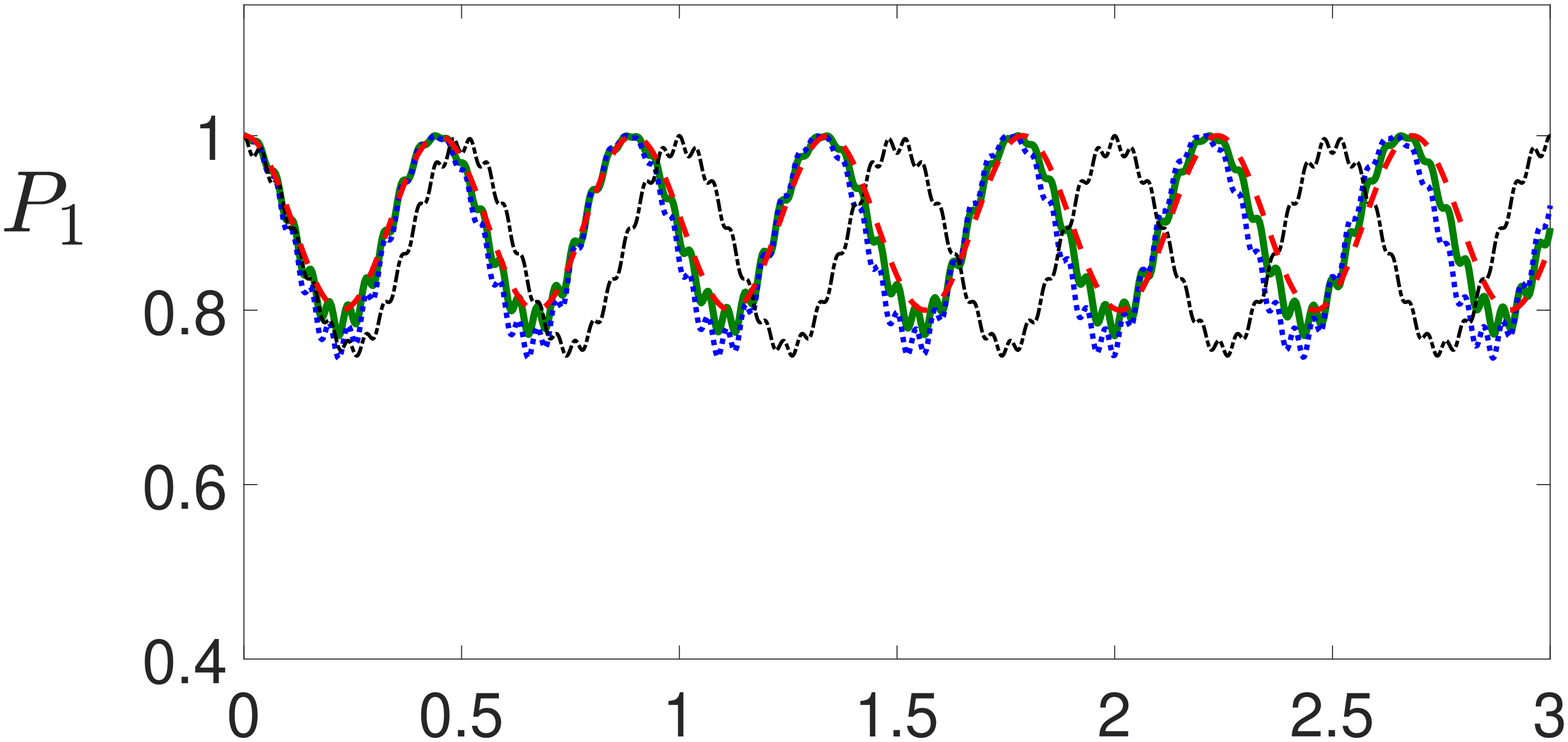}
	\includegraphics[width=0.45\textwidth]{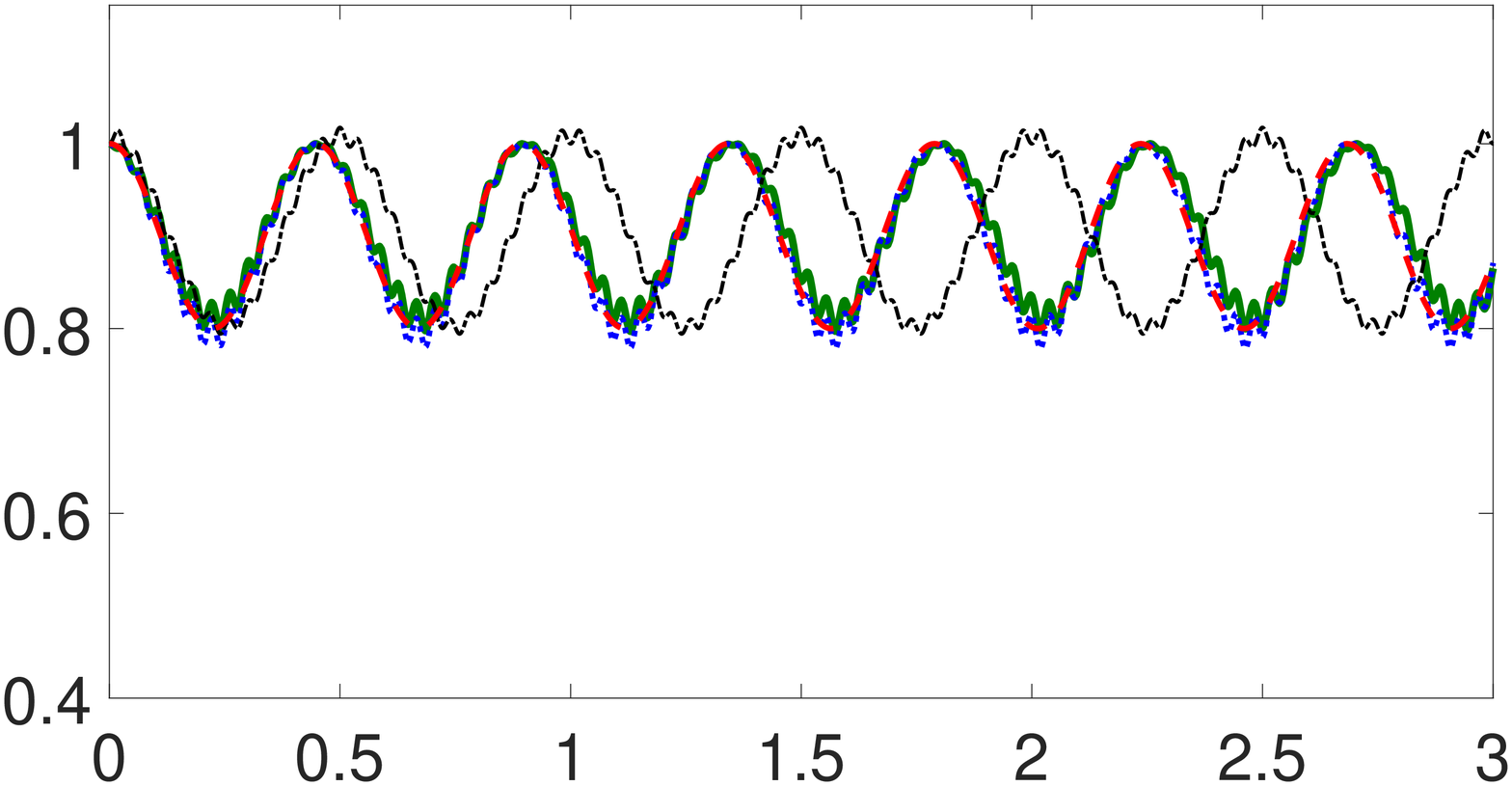}\\
	\vspace{-0.6cm}
	\includegraphics[width=0.45\textwidth]{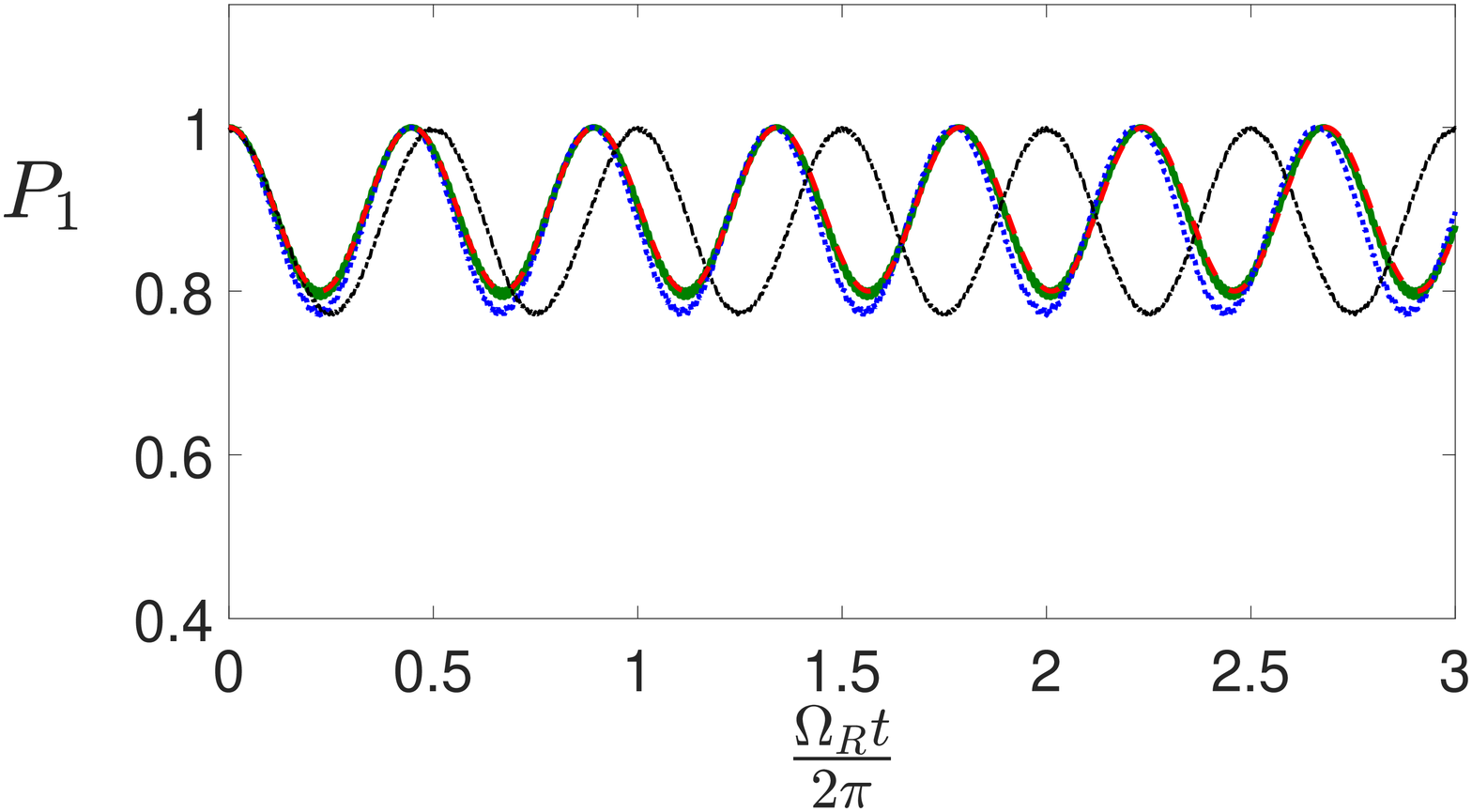}
	\includegraphics[width=0.45\textwidth]{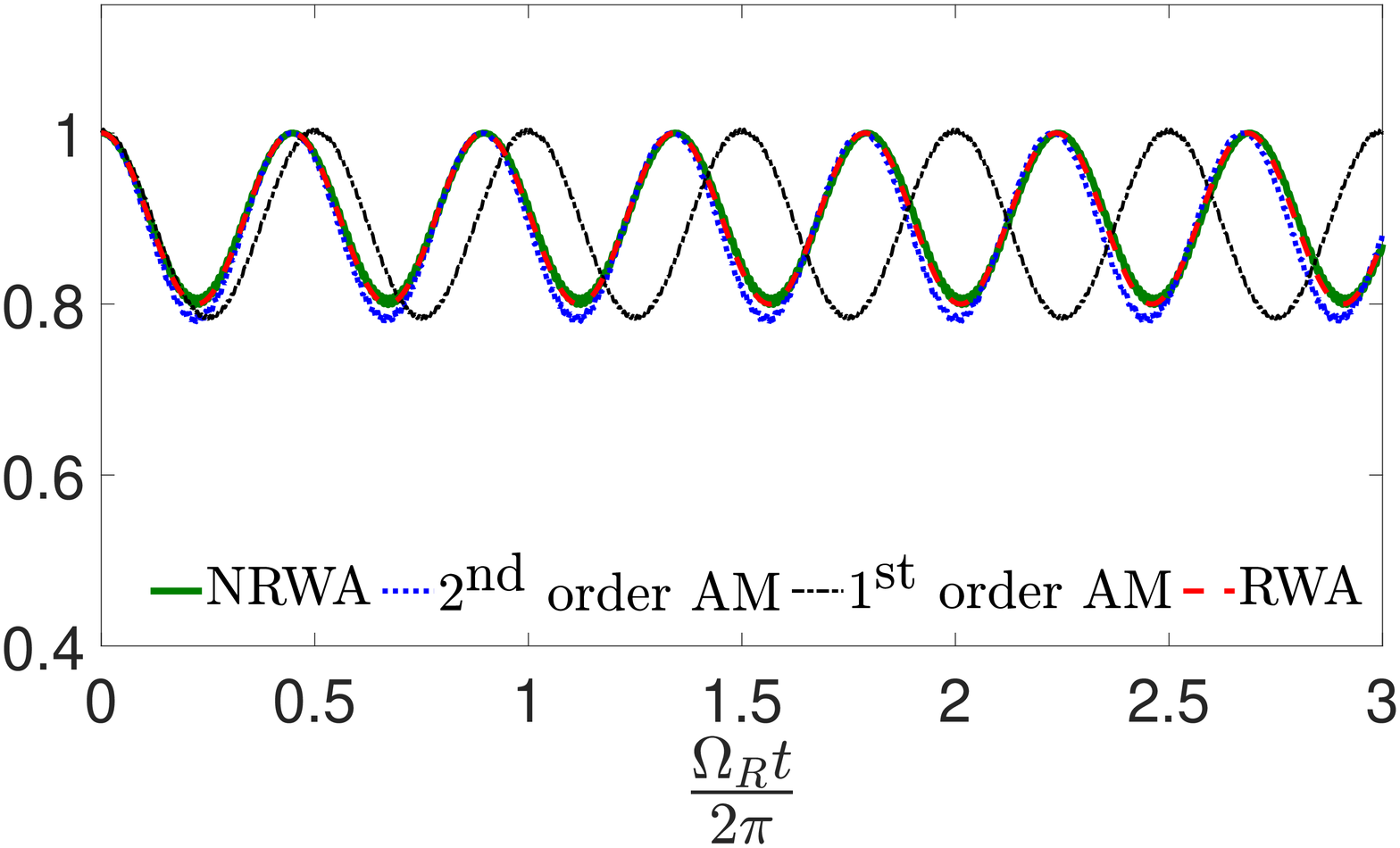}
	\vspace{-0.3cm}
	\caption{$\Delta \ne 0$. $P_1$ vs. $\frac{\Omega_{\textrm{R}} t}{2\pi}$ for $\epsilon_1 = \frac{\Omega_{\textrm{R}}}{\Delta} = -0.5$ (left column) and 
		$\epsilon_1 = \frac{\Omega_{\textrm{R}}}{\Delta} = 0.5$ (right column), varying 
		$\epsilon_2 = \frac{\Omega_{\textrm{R}}}{\Sigma}$. 
		(a), (b) $\epsilon_2 = 0.4$. 
		(c), (d) $\epsilon_2 = 0.1$. 
		(e), (f) $\epsilon_2 = 0.04$. 
		(g), (h) $\epsilon_2 = 0.01$.
		Lines correspond to 
		NRWA (continuous {\color{ForestGreen} ---}), 
		RWA (dashed {\color{red} $--$}), 
		second order AM (dotted {\color{blue} $\cdots$}), 
		first order AM (dash-dotted $\cdot -$).}
	\label{fig:epsilonSigma}
\end{figure*}

\section{Results with NRWA, RWA, first and second order AM} 
\label{sec:NRWA-RWA-AM}                                     
We compare our results of NRWA, RWA, first and second order AM. In the figures, in the horizontal axes we use the dimensionless quantity $\frac{\Omega_{\textrm{R}} t}{2\pi}$, i.e, time $t$ divided by $T_{\textrm{RWA,0}}$ and in the vertical axes we present the probability at the lower level, $P_1$. 
For non-resonance we employ three types of small quantities $\epsilon$, i.e., 
$\frac{\Omega_{\textrm{R}}}{\Delta}$,
$\frac{\Omega_{\textrm{R}}}{\Sigma}$ and
$\frac{\Omega_{\textrm{R}}}{\omega}$. 
Unavoidably, when $\Delta$ becomes smaller,
at some point, $\frac{\Omega_{\textrm{R}}}{\Delta}$ gets so large that non-resonant AM is not successful anymore and 
resonance must be treated via a different path, 
using just one type of $\epsilon$, i.e.,
$\frac{\Omega_{\textrm{R}}}{\omega}$.
First order AM is frequently away from the numerical solution. 
We include it in the figures below just for comparison.

\begin{figure*}[t!]
	\centering
	\vspace{-0.6cm}
	\includegraphics[width=0.45\textwidth]{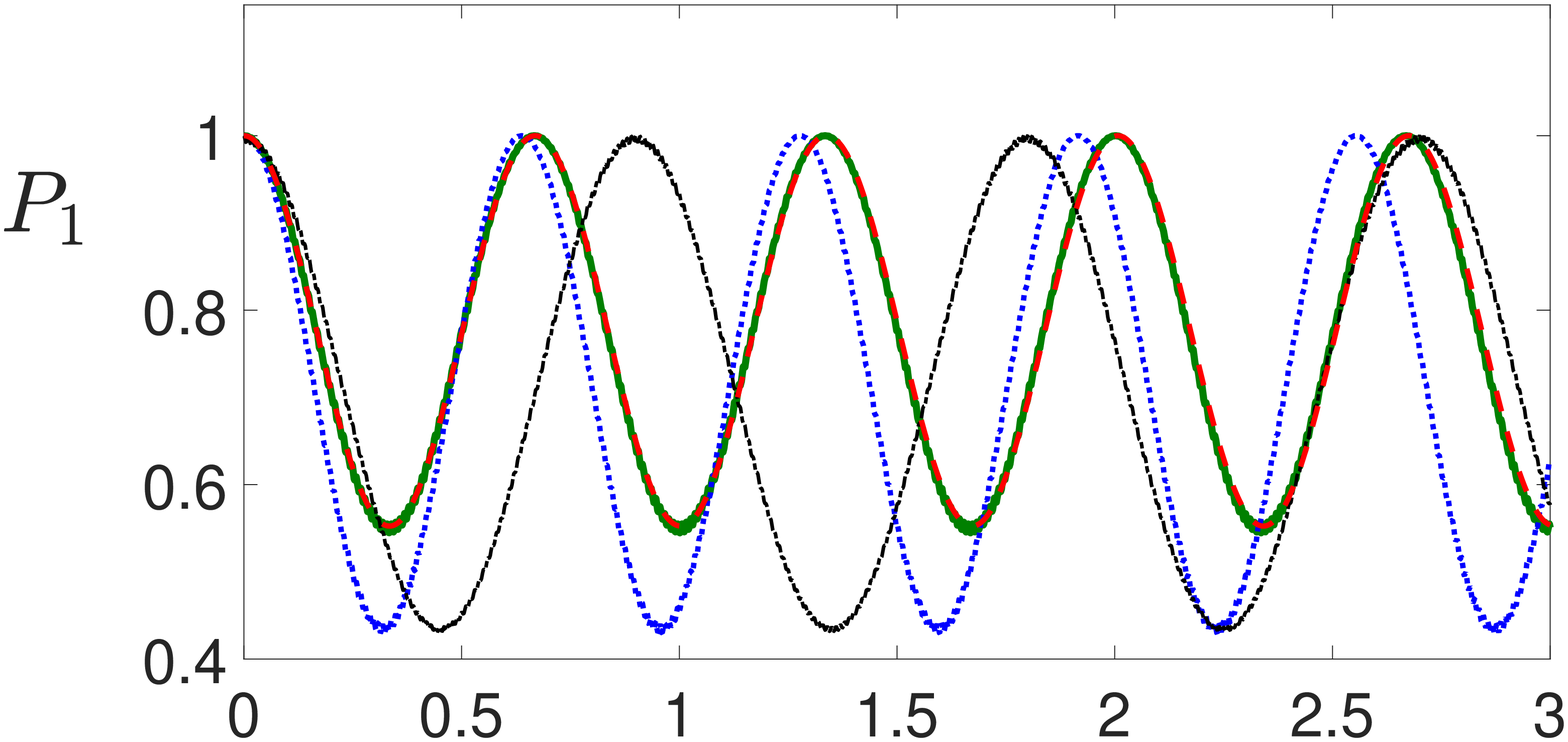}
	\includegraphics[width=0.45\textwidth]{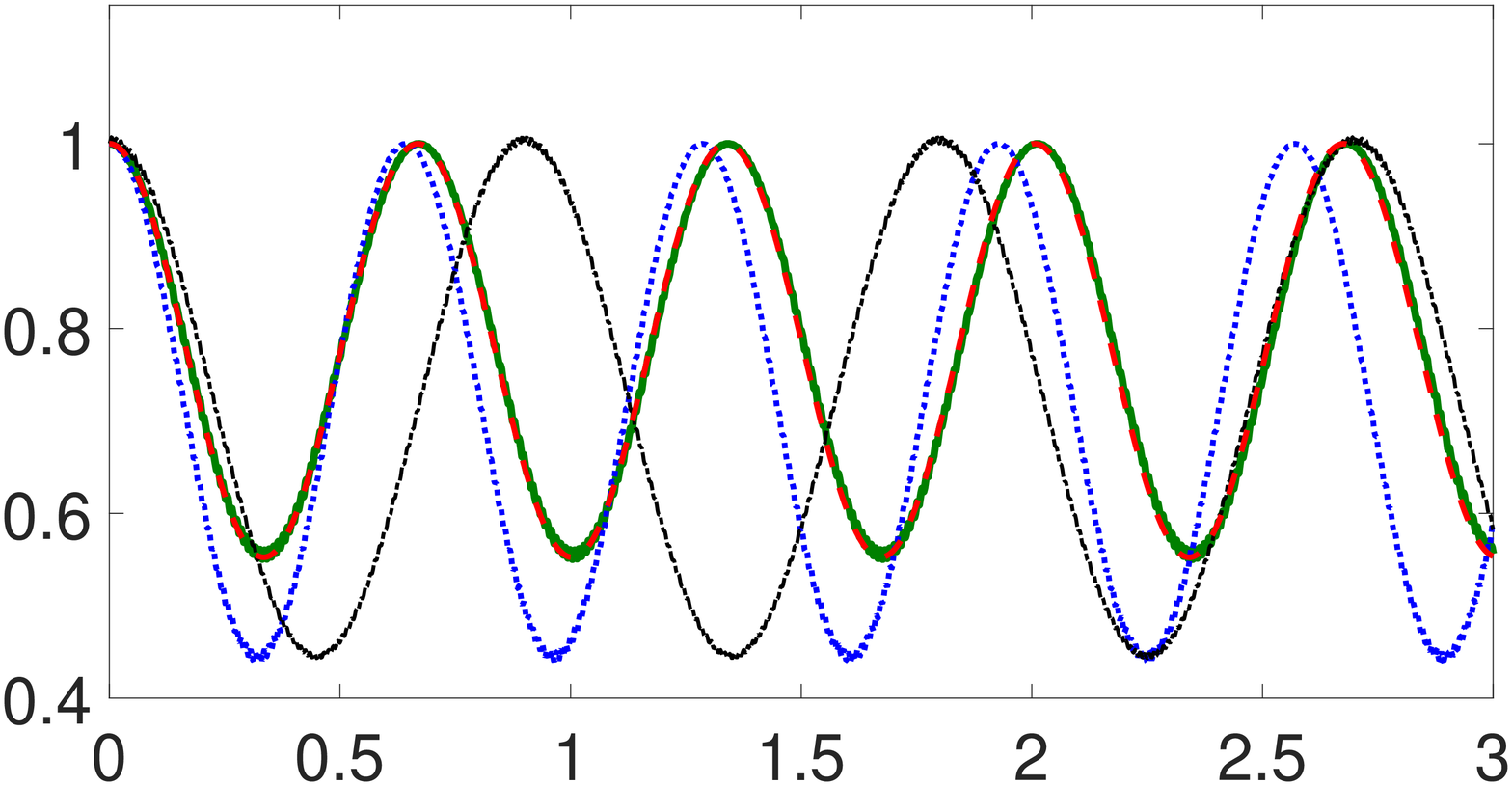}\\
	\vspace{-0.6cm}
	\includegraphics[width=0.45\textwidth]{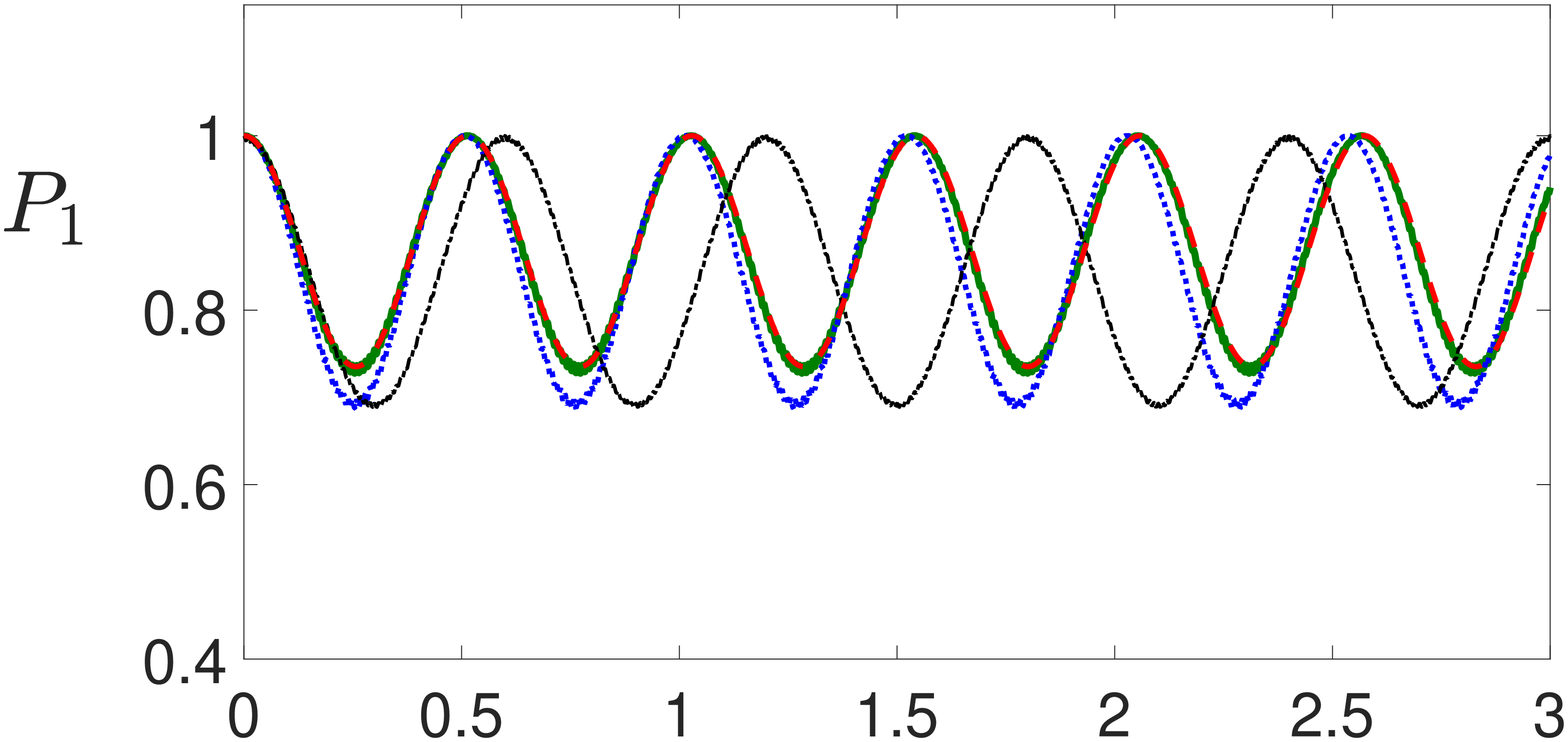}
	\includegraphics[width=0.45\textwidth]{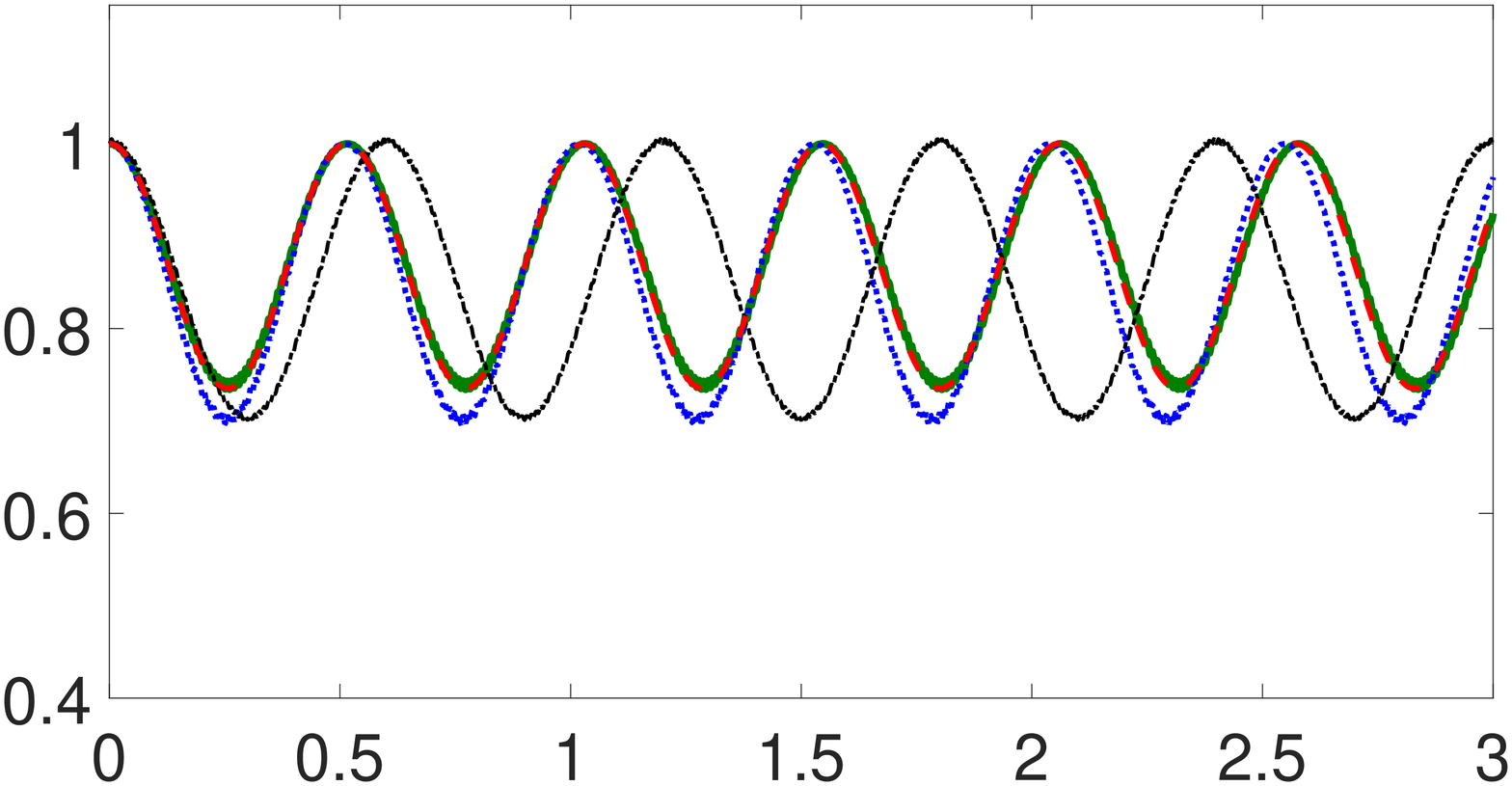}\\
	\vspace{-0.6cm}
	\includegraphics[width=0.45\textwidth]{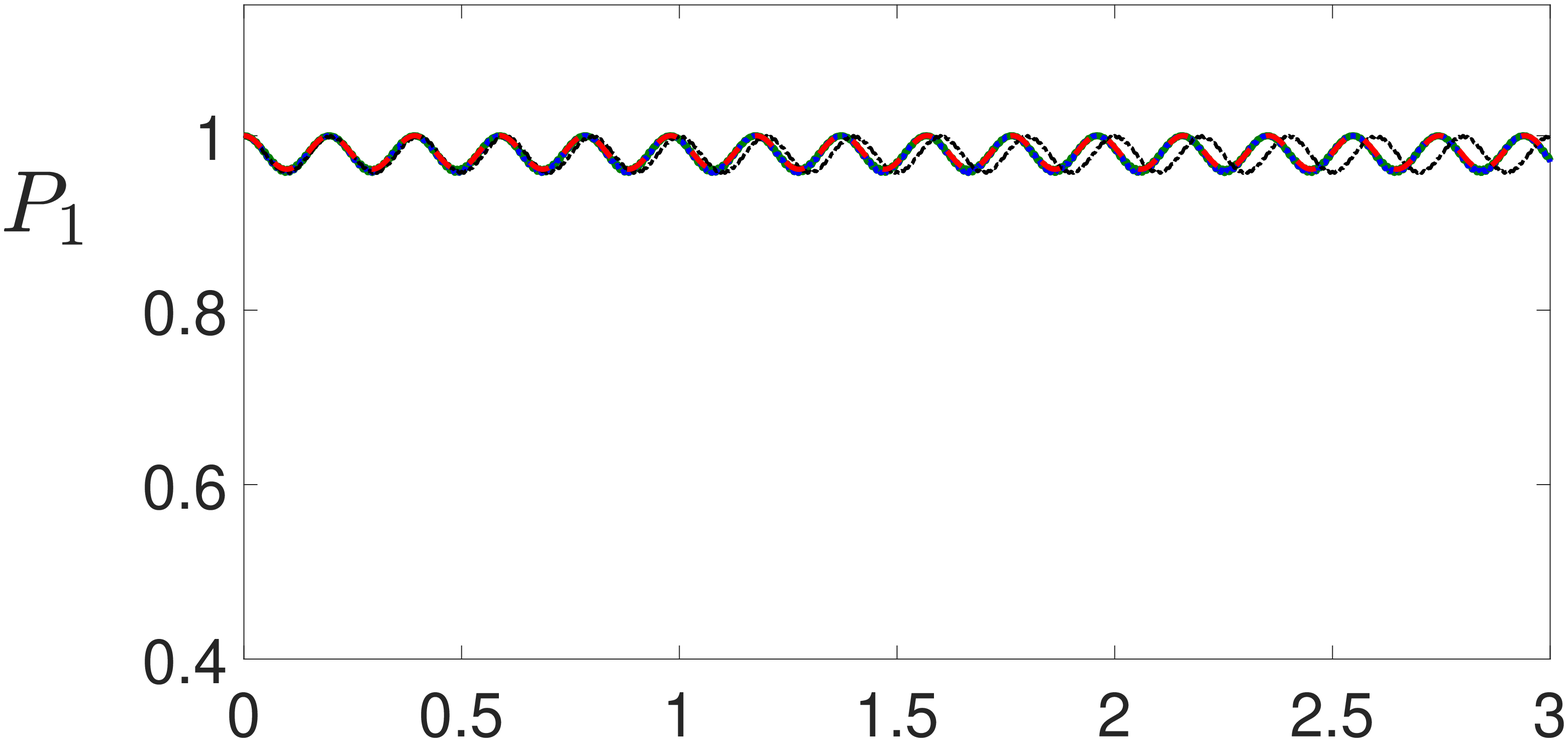}
	\includegraphics[width=0.45\textwidth]{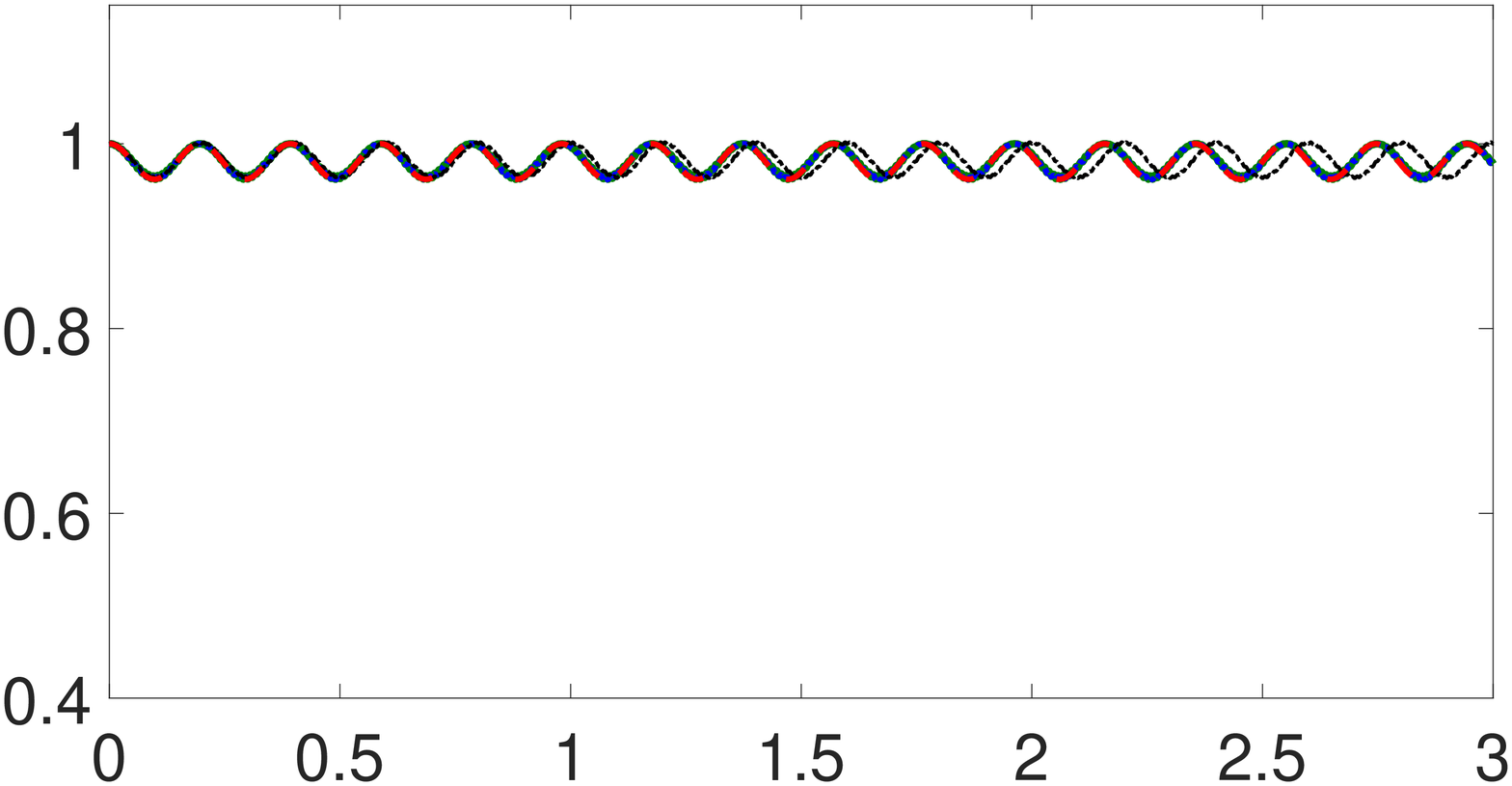}\\
	\vspace{-0.6cm}
	\includegraphics[width=0.45\textwidth]{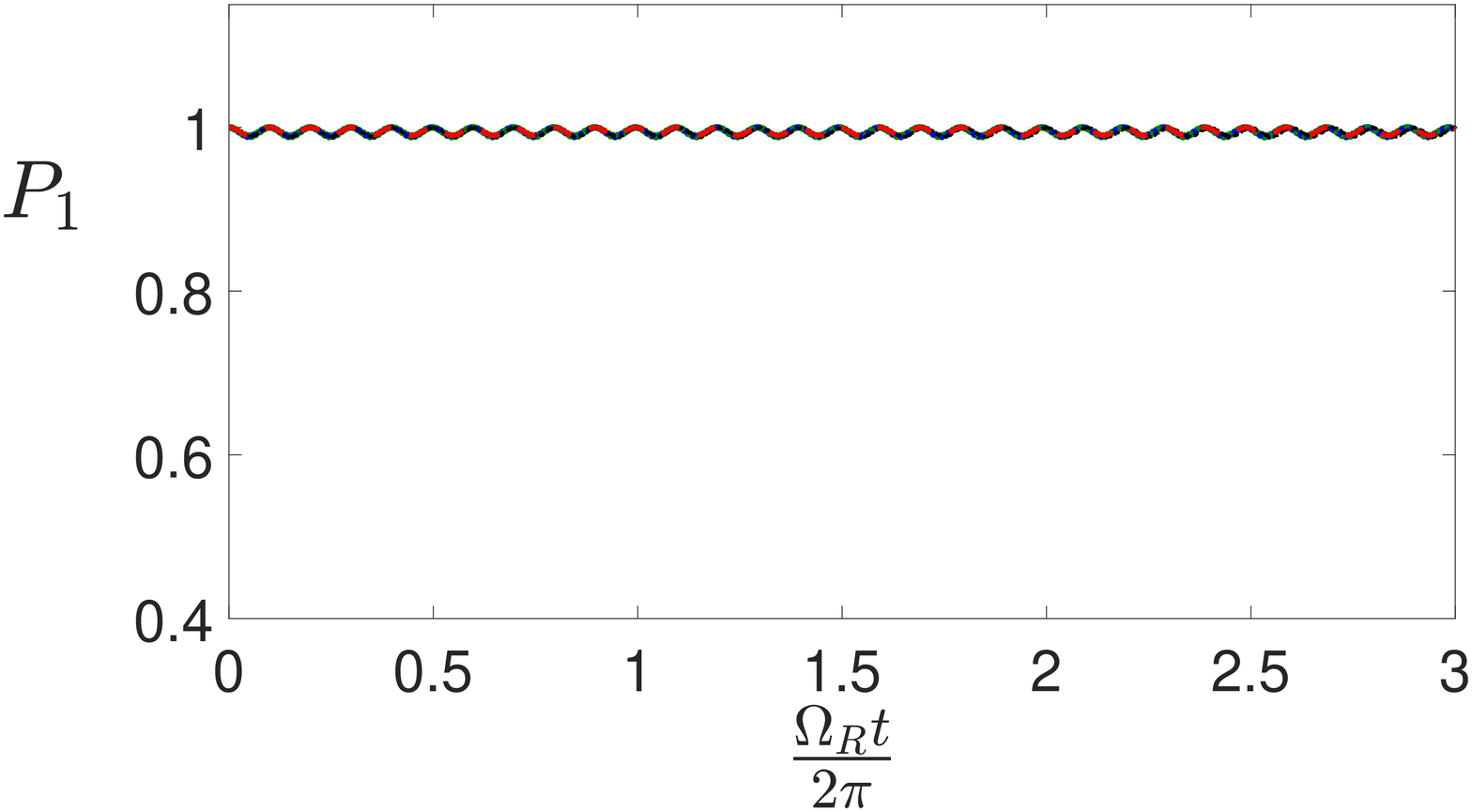}
	\includegraphics[width=0.45\textwidth]{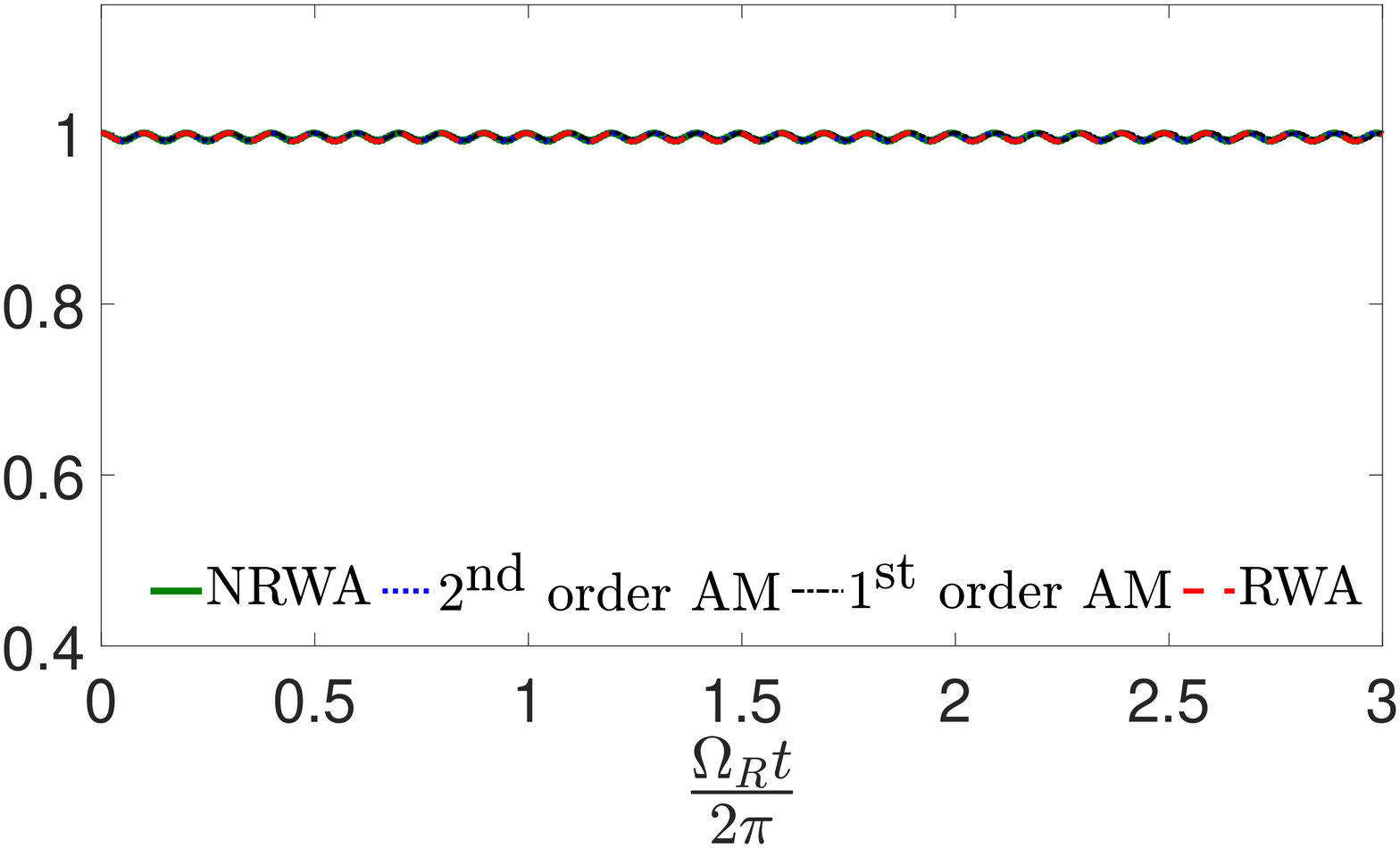}
	\vspace{-0.3cm}
	\caption{$\Delta \ne 0$. $P_1$ vs. $\frac{\Omega_{\textrm{R}} t}{2\pi}$, for $\epsilon_2 = \frac{\Omega_{\textrm{R}}}{\Sigma} = 0.01$, varying $\epsilon_1 = \frac{\Omega_{\textrm{R}}}{\Delta}$. (a) $\epsilon_1 = -0.9$. (b) $\epsilon_1 = 0.9$. (c) $\epsilon_1 = -0.6$. (d) $\epsilon_1 = 0.6$. (e) $\epsilon_1 = -0.2$. (f) $\epsilon_1 = 0.2$. (g) $\epsilon_1 = -0.1$. (h) $\epsilon_1 = 0.1$. Lines correspond to 
		NRWA (continuous {\color{ForestGreen} ---}), 
		RWA (dashed {\color{red} $--$}), 
		second order AM (dotted {\color{blue} $\cdots$}), 
		first order AM (dash-dotted $\cdot -$).}
	\label{fig:epsilonDelta}
\end{figure*}

\begin{figure*}[t!]
	\centering
	\vspace{-0.2cm}
	\includegraphics[width=0.45\textwidth]{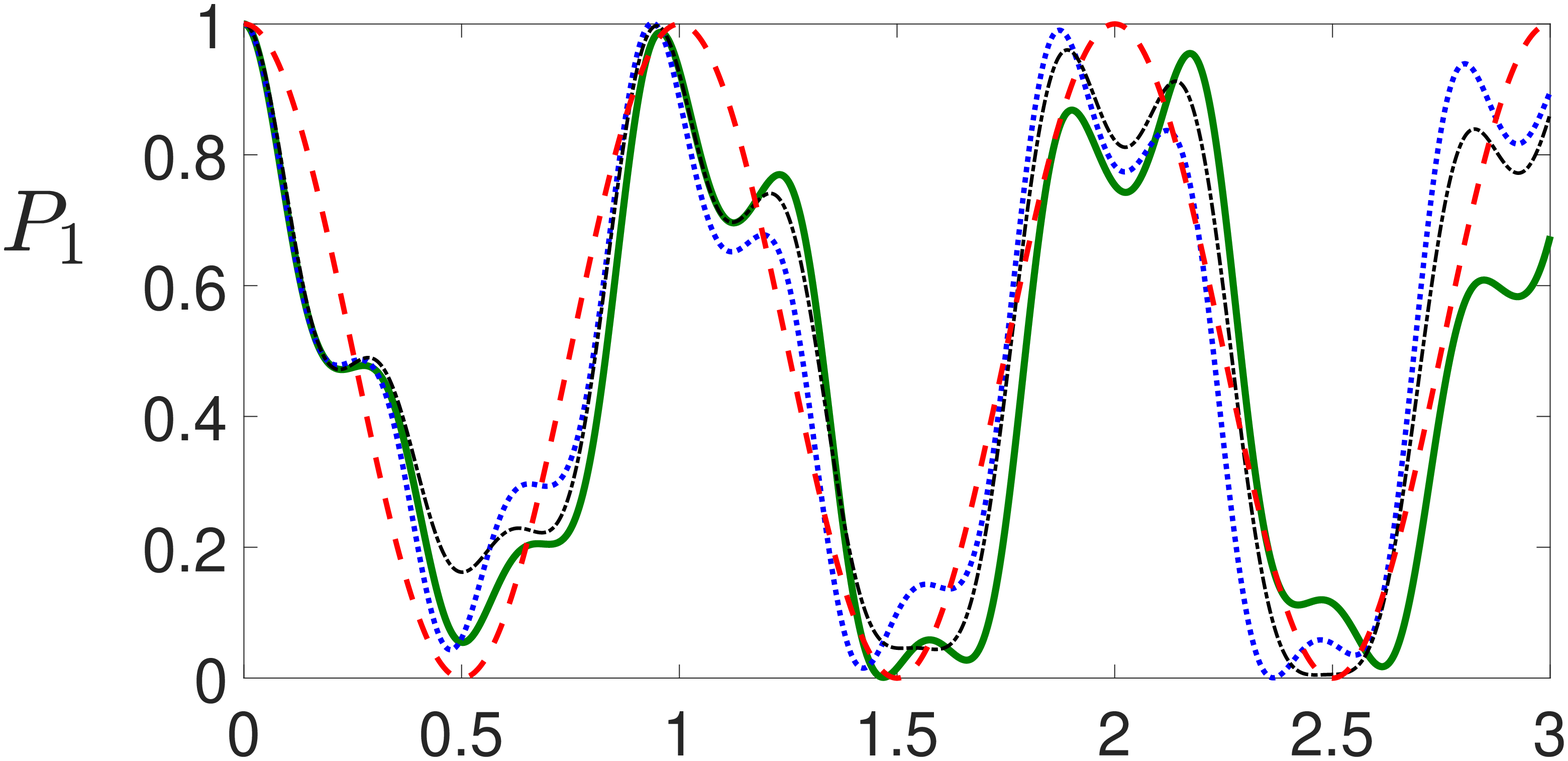}
	\includegraphics[width=0.45\textwidth]{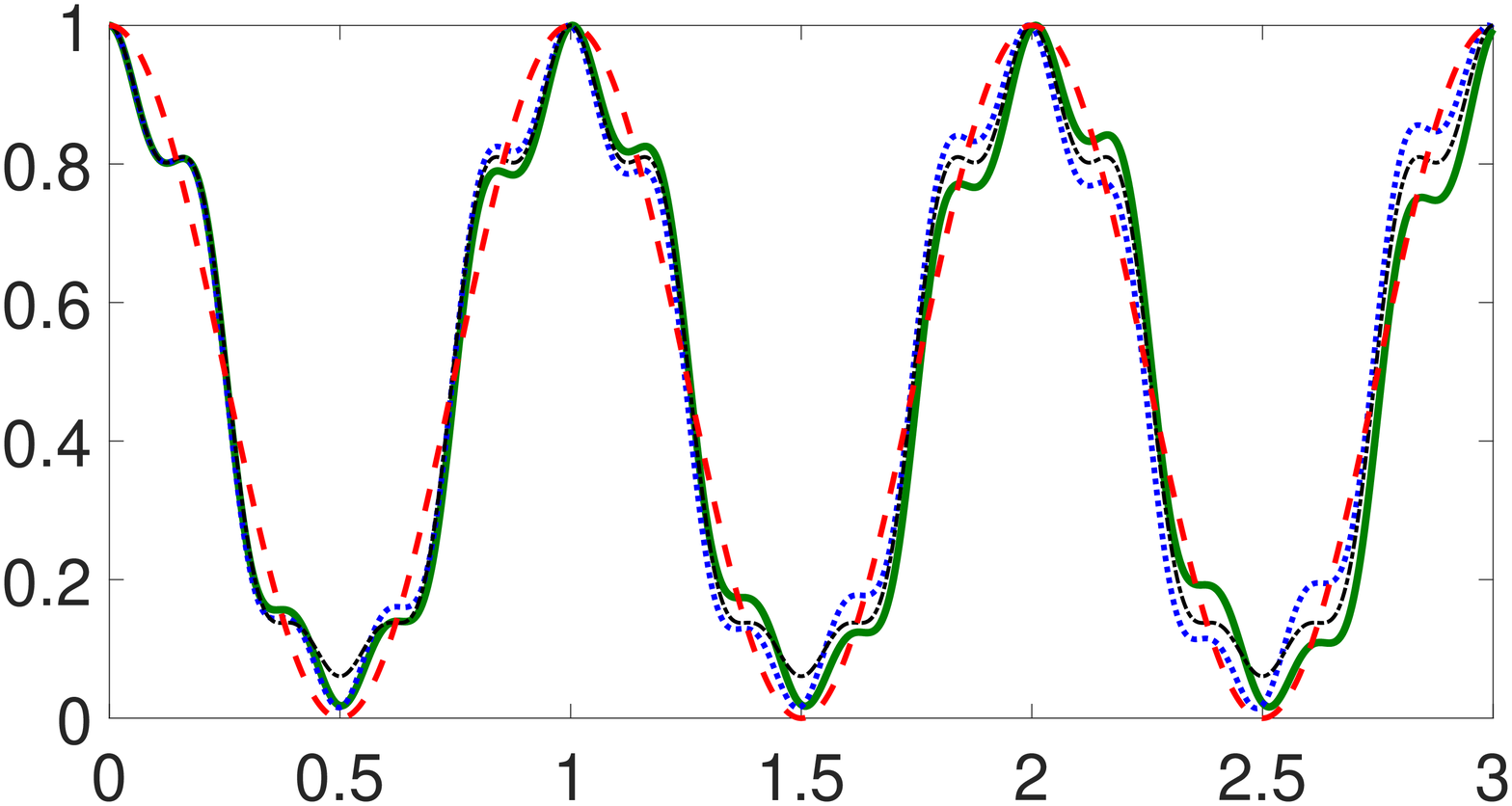}\\
	\vspace{-0.6cm}
	\includegraphics[width=0.45\textwidth]{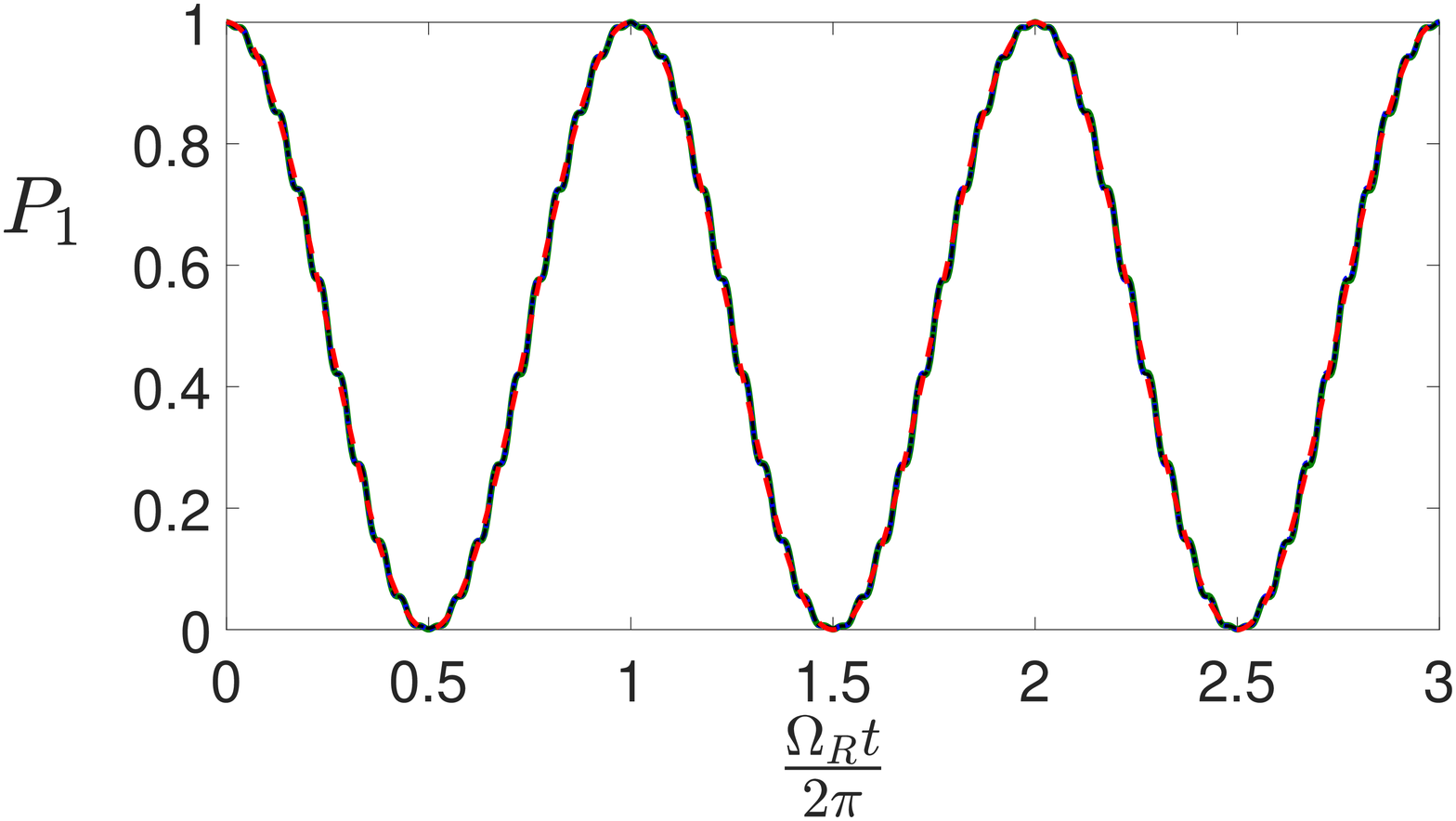}
	\includegraphics[width=0.45\textwidth]{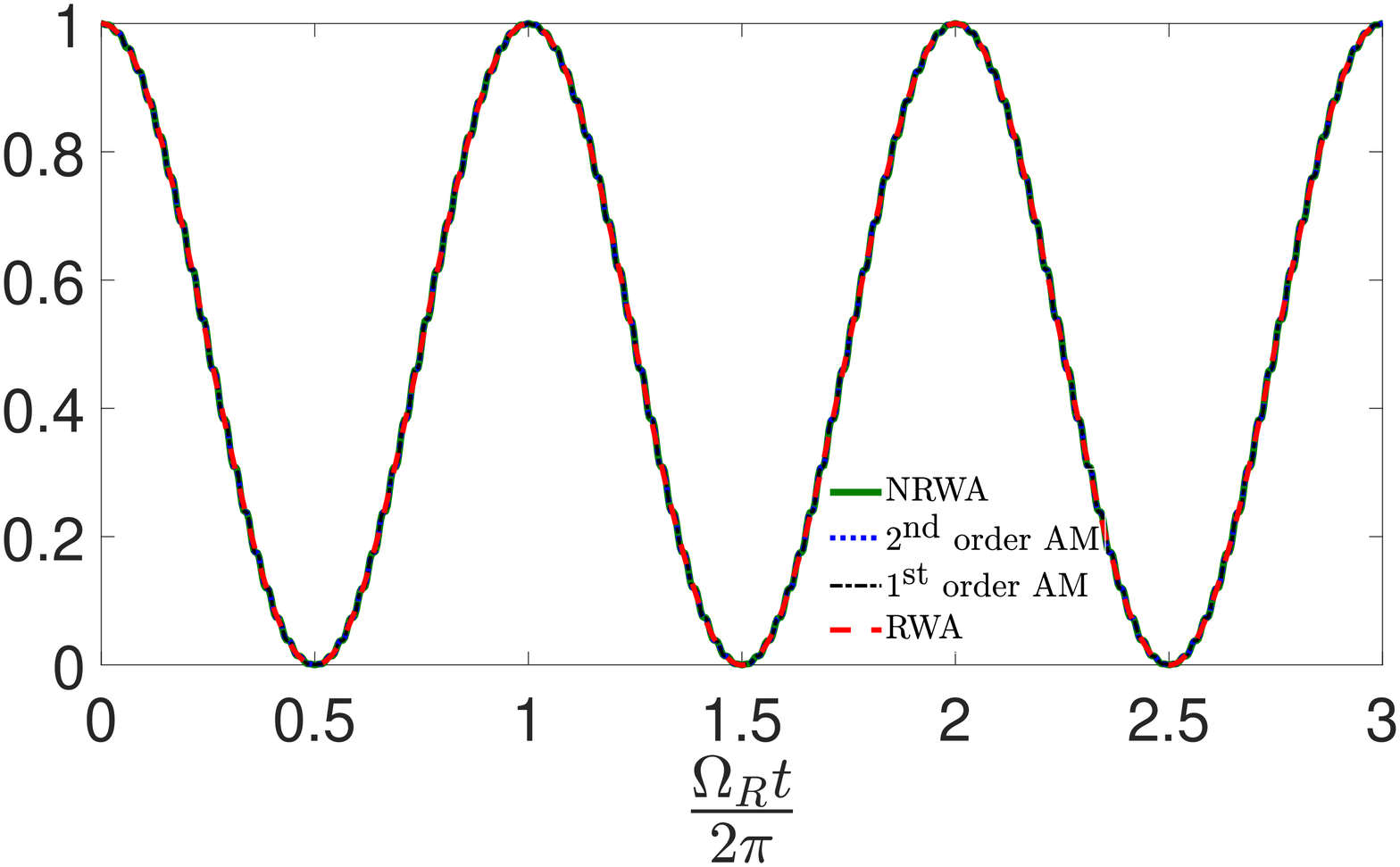}
	\vspace{-0.3cm}
	\caption{$\Delta = 0$. $P_1$ vs. $\frac{\Omega_{\textrm{R}} t}{2\pi}$, varying $\epsilon=\frac{\Omega_{\textrm{R}}}{\omega}$. 
		(a) $\epsilon=0.9$. (b) $\epsilon=0.5$. (c) $\epsilon=0.1$. (d) $\epsilon=0.05$.
		Lines correspond to 
		NRWA (continuous {\color{ForestGreen} ---}), 
		RWA (dashed {\color{red} $--$}), 
		second order AM (dotted {\color{blue} $\cdots$}), 
		first order AM (dash-dotted $\cdot -$).}
	\label{fig:epsilonResonance}
\end{figure*}

\subsection{Initial Condition (1,0)} 
\label{subsec:10} 
We assume initial condition $C_1(0) = 1$, $C_2(0) = 0$, placing the electron, at $ t = 0$, at the lower level.  Subsubsec.~\ref{subsubsec:AM-RWA-NS:non-resonance10} 
(\ref{subsubsec:AM-RWA-NS:resonance10}) refers to non-resonance (resonance).

\subsubsection{Non-resonance} 
\label{subsubsec:AM-RWA-NS:non-resonance10} 
In Fig.~\ref{fig:epsilonSigma} we modify $\epsilon_2 = \frac{\Omega_{\textrm{R}}}{\Sigma}$ keeping  
$\epsilon_1 = \frac{\Omega_{\textrm{R}}}{\Delta} = -0.5$ 
($\epsilon_1 = 0.5$) on the left (right) column. 
For $\epsilon_1 > 0$, as $\epsilon_2$ gets smaller, RWA is identified with NRWA. Second order AM is very close to NRWA in all cases. For $\epsilon_1 < 0$, as $\epsilon_2$ gets smaller, AM and RWA are identified with NRWA. The different behavior of AM for negative and positive $\epsilon_1$ stems from $\epsilon_3 = \frac{\Omega_{\textrm{R}}}{\omega}$ being different: 
for $\epsilon_1 > 0$, $\epsilon_3$ is smaller 
than for $\epsilon_1 < 0$.
In Fig.~\ref{fig:epsilonDelta} we modify $\epsilon_1 = \frac{\Omega_{\textrm{R}}}{\Delta}$ and keep $\epsilon_2 = \frac{\Omega_{\textrm{R}}}{\Sigma} = 0.01$. 
On left (right) column $\epsilon_1 < 0 $ ($\epsilon_1 > 0$). 
RWA is identified with NRWA, but not with second order AM. As $\epsilon_1$ gets smaller, AM gradually approaches NRWA. Oscillations diminish as $\epsilon_1$ becomes smaller. Oscillations at the same row but in different columns are a little different due to the different value of $\epsilon_3$.
In Figs.~\ref{fig:epsilonSigma},~\ref{fig:epsilonDelta}, the two panels of the same line seem similar, because $\epsilon_3 = \frac{\Omega_{\textrm{R}}}{\omega}$ are almost identical, except for the first line in Fig.~\ref{fig:epsilonSigma}. For example, in Fig.~\ref{fig:epsilonSigma}, the two panels of the last line have $\epsilon_1 = -0.5$, $\epsilon_2 = 0.01$, $\epsilon_3 = \frac{1}{49}$ and $\epsilon_1 = 0.5$, $\epsilon_2 = 0.01$, $\epsilon_3 = \frac{1}{51}$, respectively, while the two panels of the first line have $\epsilon_1 = -0.5$, $\epsilon_2 = 0.4$, $\epsilon_3 = 4$ and $\epsilon_1 = 0.5$, $\epsilon_2 = 0.4$, $\epsilon_3 = \frac{4}{9}$, respectively. Hence, second order AM is identified with NRWA when $\epsilon_1$, $\epsilon_2$ and $\epsilon_3$ are sufficiently small.

\subsubsection{Resonance} 
\label{subsubsec:AM-RWA-NS:resonance10} 
In Fig.~\ref{fig:epsilonResonance} we illustrate $P_1$ vs. $\frac{\Omega_{\textrm{R}} t}{2\pi}$, modifying $\epsilon=\frac{\Omega_{\textrm{R}}}{\omega}$. As $\epsilon$ gets smaller, AM is identified with NRWA.

\begin{figure*}[t!]
	\centering
	\vspace{-0.5cm}
	\includegraphics[width=0.45\textwidth]{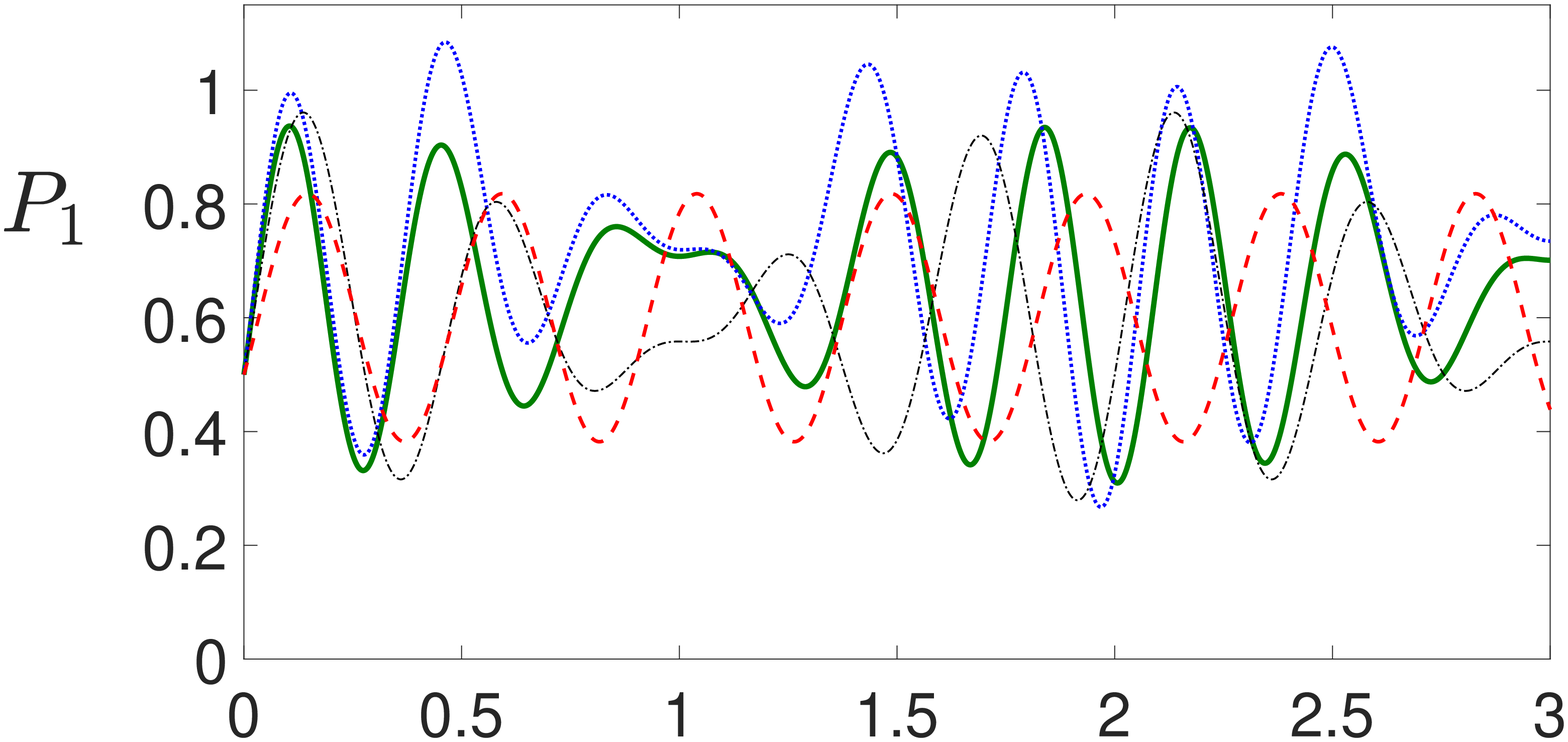}
	\includegraphics[width=0.45\textwidth]{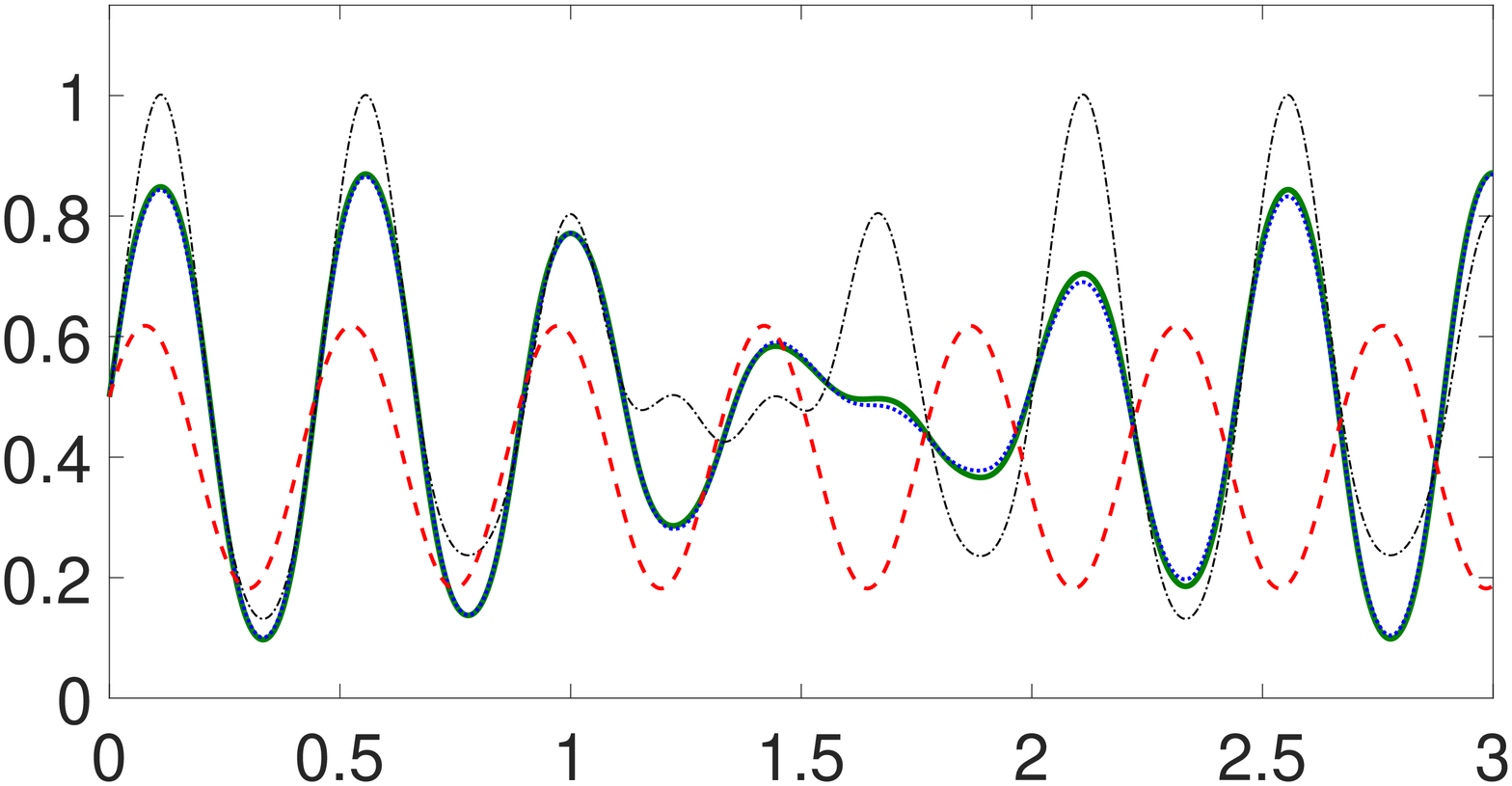}\\
	\vspace{-0.5cm}
	\includegraphics[width=0.45\textwidth]{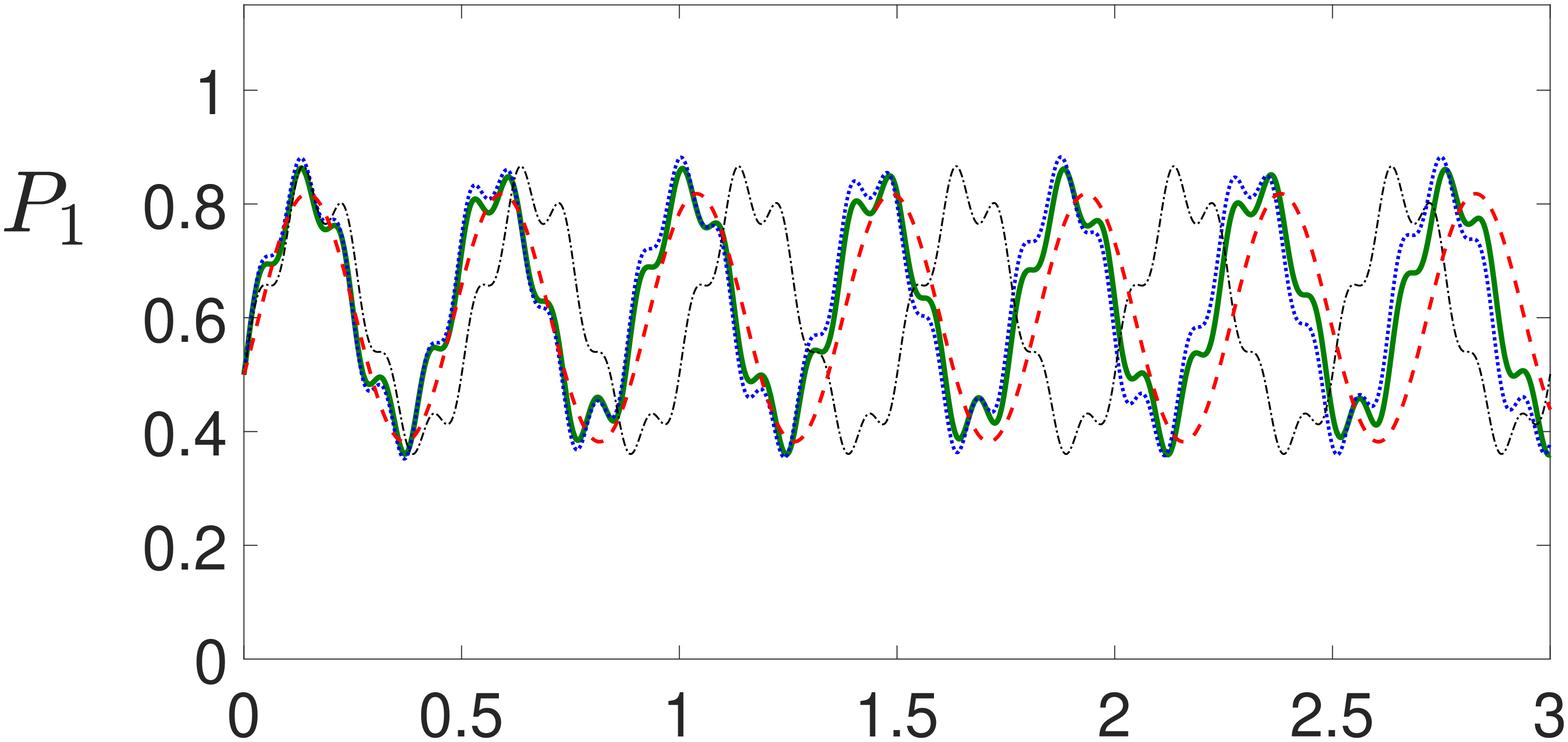}
	\includegraphics[width=0.45\textwidth]{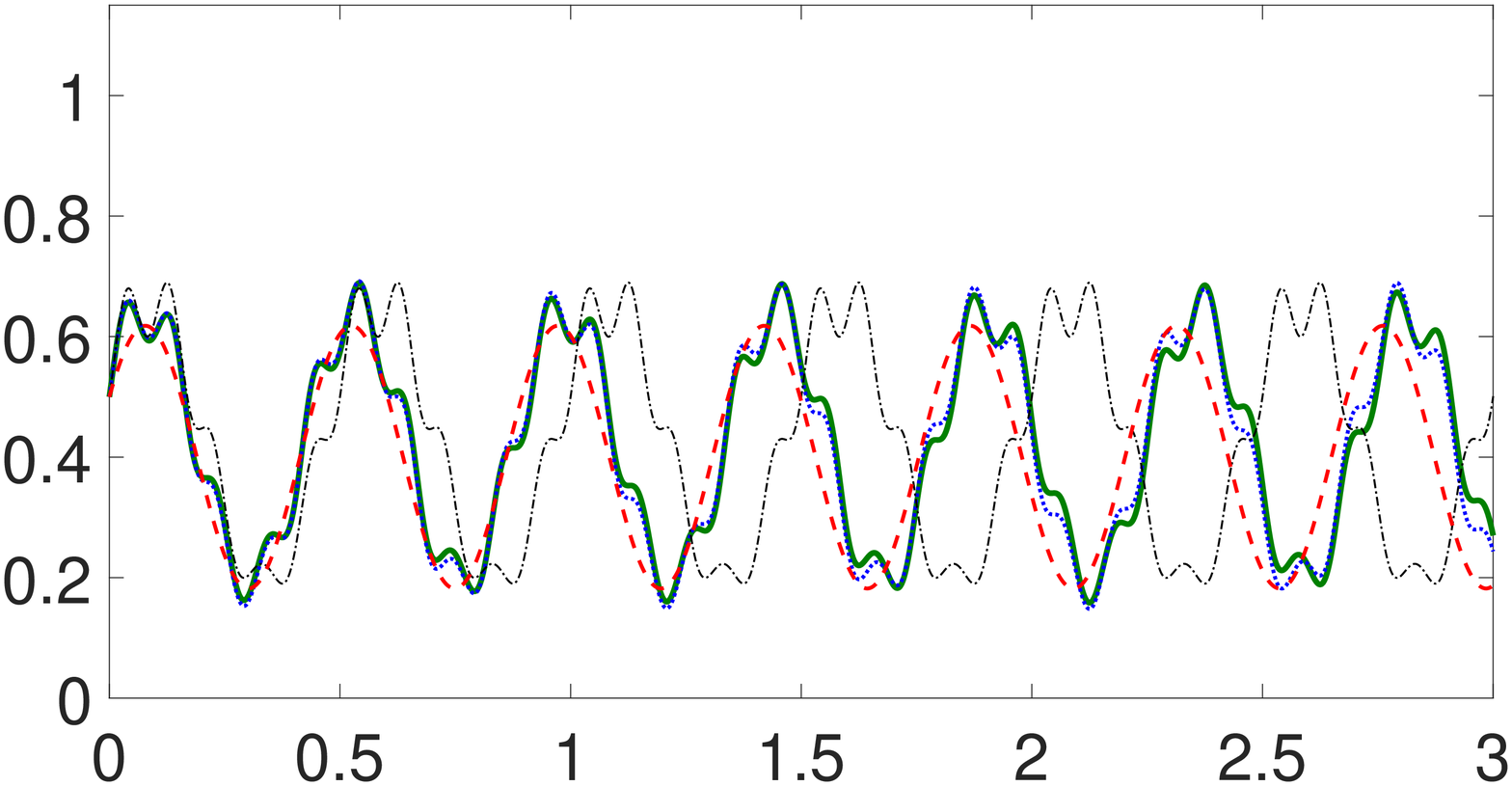}\\
	\vspace{-0.5cm}
	\includegraphics[width=0.45\textwidth]{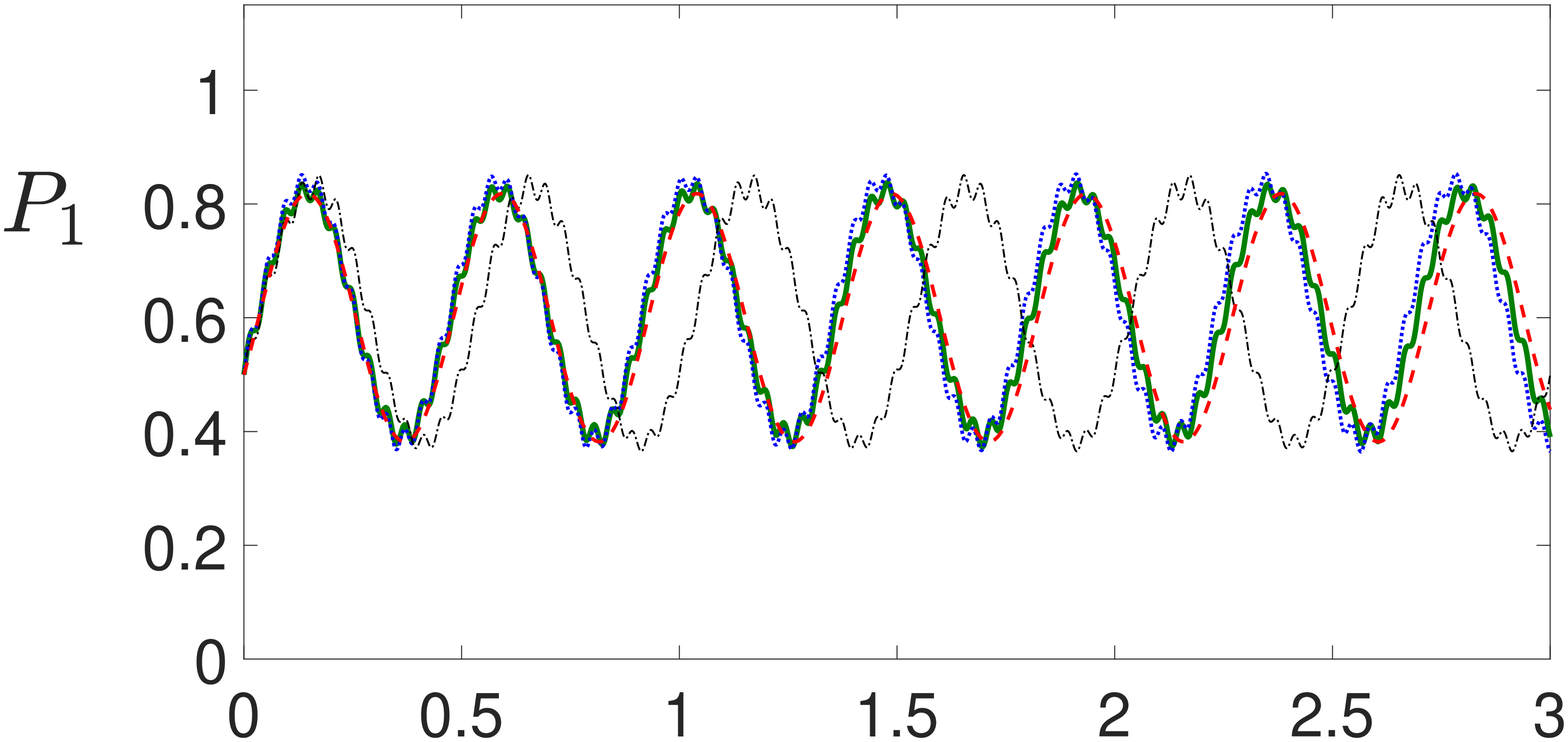}
	\includegraphics[width=0.45\textwidth]{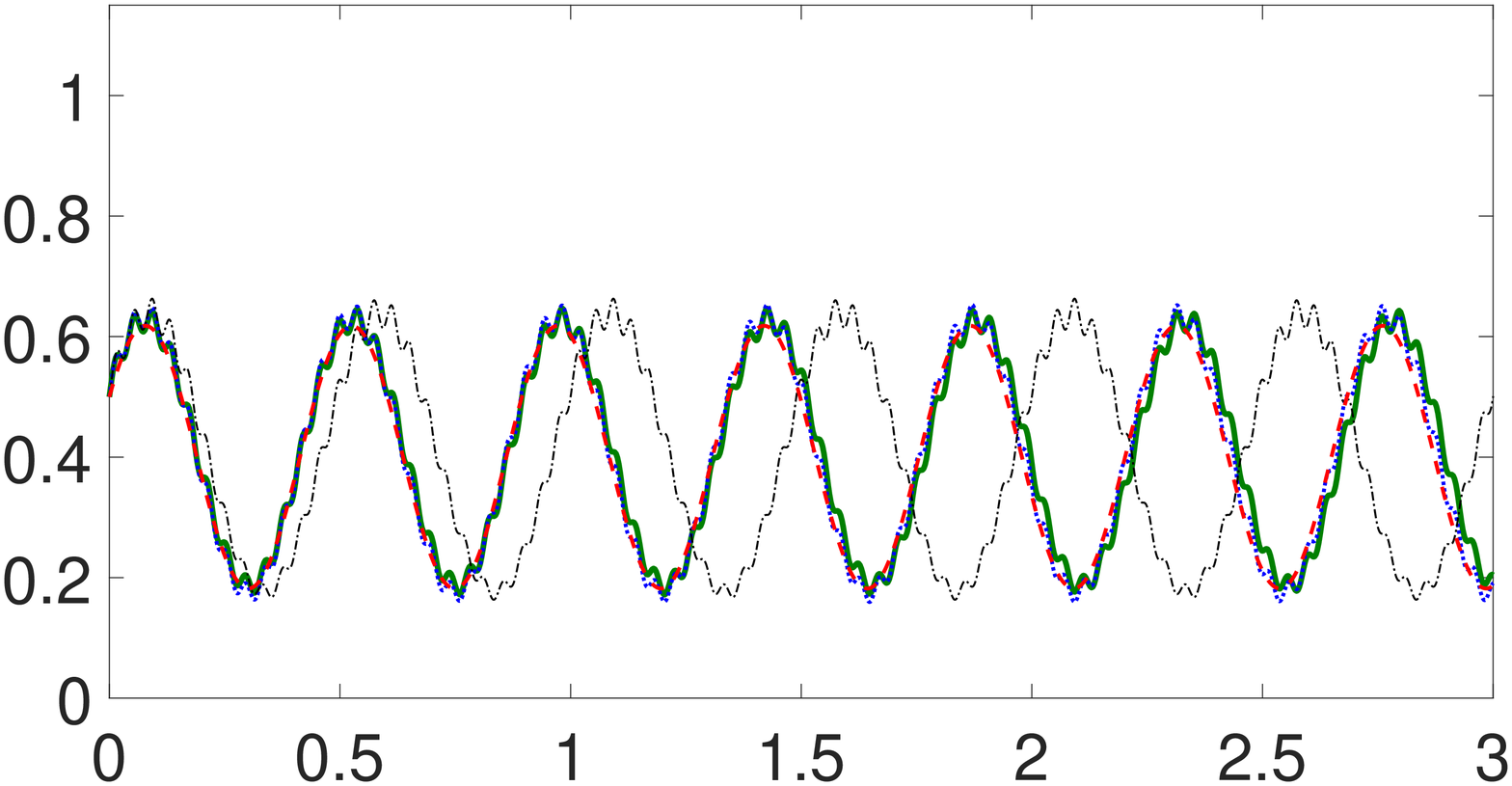}\\
	\vspace{-0.5cm}
	\includegraphics[width=0.45\textwidth]{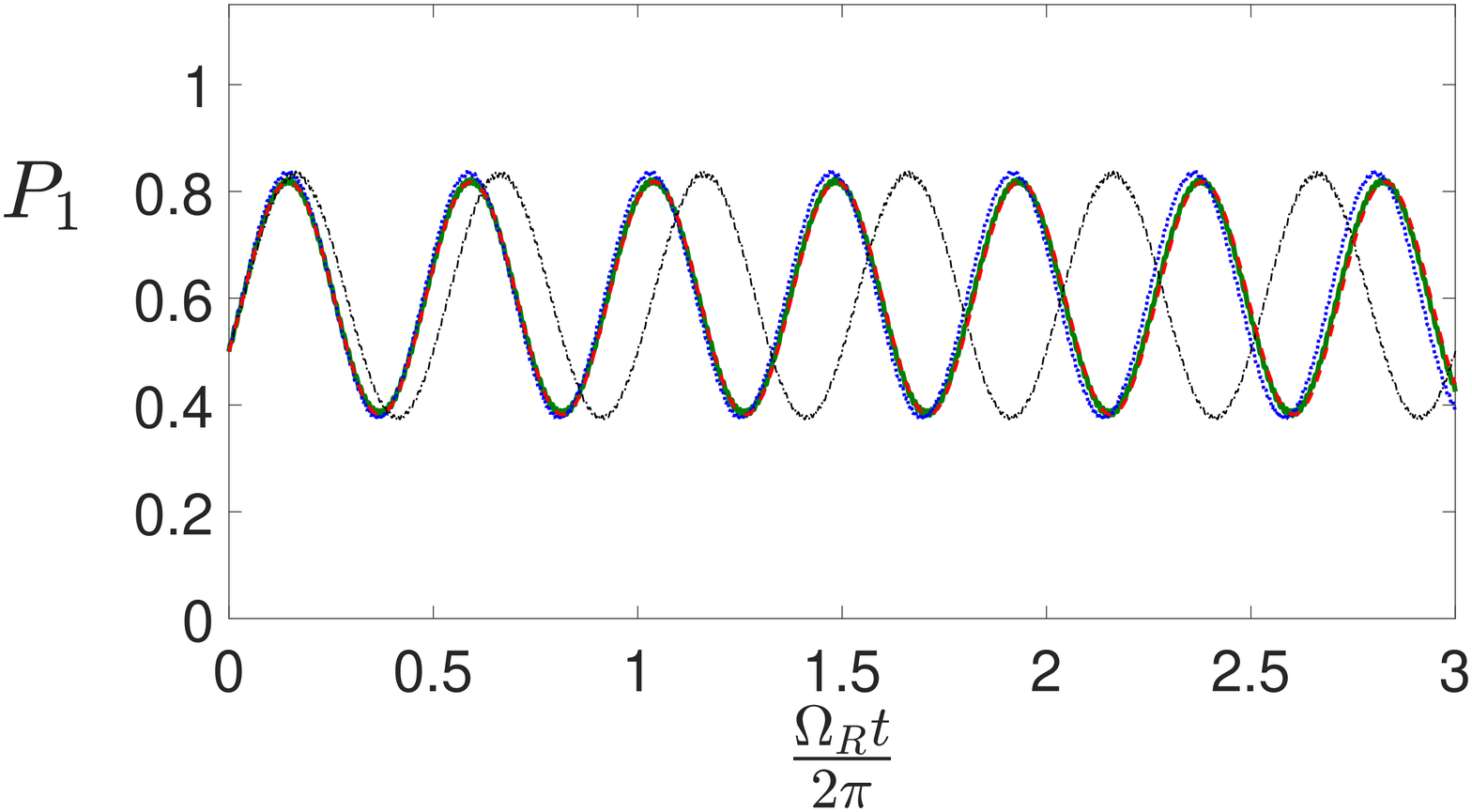}
	\includegraphics[width=0.45\textwidth]{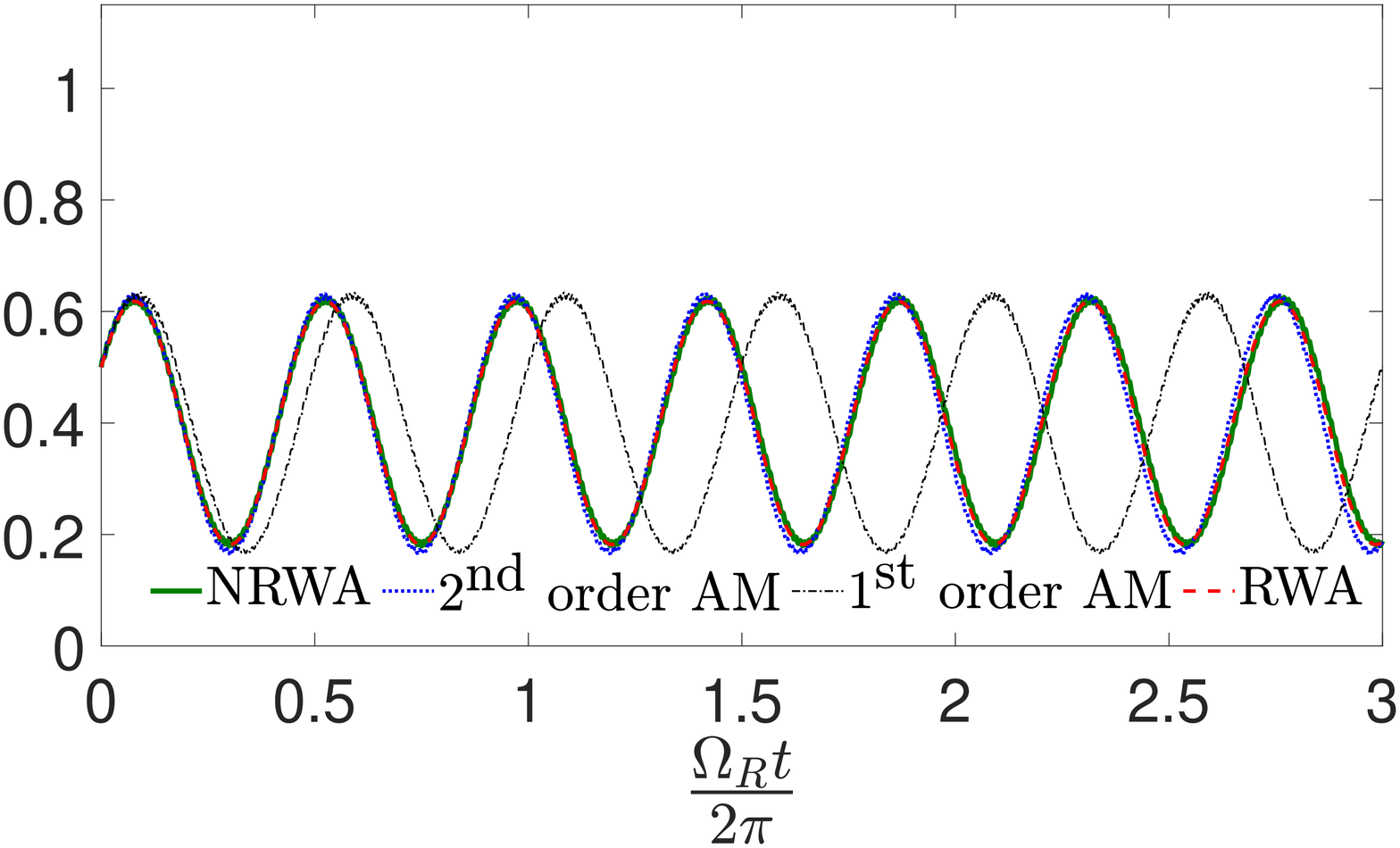}
	\vspace{-0.3cm}
	\caption{$\Delta \ne 0$. $P_1$ vs. $\frac{\Omega_{\textrm{R}} t}{2\pi}$ for $\epsilon_1 = \frac{\Omega_{\textrm{R}}}{\Delta} = -0.5$ (left column) and 
		$\epsilon_1 = \frac{\Omega_{\textrm{R}}}{\Delta} = 0.5$ (right column), varying 
		$\epsilon_2 = \frac{\Omega_{\textrm{R}}}{\Sigma}$ 
		with $\theta-\phi =\frac{\pi}{3}$.
		(a), (b) $\epsilon_2 = 0.4$. 
		(c), (d) $\epsilon_2 = 0.1$. 
		(e), (f) $\epsilon_2 = 0.04$. 
		(g), (h) $\epsilon_2 = 0.01$.
		Lines correspond to
		NRWA (continuous {\color{ForestGreen} ---}), 
		RWA (dashed {\color{red} $--$}), 
		second order AM (dotted {\color{blue} $\cdots$}), 
		first order AM (dash-dotted $\cdot -$).}
	\label{fig:epsilonSigmaphasi}
\end{figure*}

\subsection{Initial condition (1/2,1/2)} 
\label{subsec:hh} 
Here we explore the prospects coherence can offer, by defining a phase difference in the initial electron wave functions. Specifically, we assume initial condition $C_1(0) = \frac{1}{\sqrt{2}}e^{i \theta}$, $C_2(0) = \frac{1}{\sqrt{2}}e^{i \phi}$, i.e., at time zero, equal probability at both levels, $|C_1(0)|^2 = |C_2(0)|^2 = \frac{1}{2}$, but with a phase difference $\theta-\phi$. Subsubsec.~\ref{subsubsec:AM-RWA-NS:non-resonancehh} (\ref{subsubsec:AM-RWA-NS:resonancehh}) is devoted to  non-resonance (resonance).

\vspace{-0.5cm}

\subsubsection{Non-resonance} 
\label{subsubsec:AM-RWA-NS:non-resonancehh} 
In Fig.~\ref{fig:epsilonSigmaphasi} we vary $\epsilon_2 = \frac{\Omega_{\textrm{R}}}{\Sigma}$, keeping $\epsilon_1 = \frac{\Omega_{\textrm{R}}}{\Delta} = -0.5$ 
($\epsilon_1 = 0.5$) on the left (right) column, 
with $\theta - \phi = \frac{\pi}{3}$. Although the initial probabilities at the two levels are equal, phase difference of the initial wave functions leads to strong oscillations, a clear coherent phenomenon. Decreasing $\epsilon_2$, second order AM approaches NRWA.

\begin{figure*}[t!]
	\centering
	\vspace{-0.5cm}
	\includegraphics[width=0.45\textwidth]{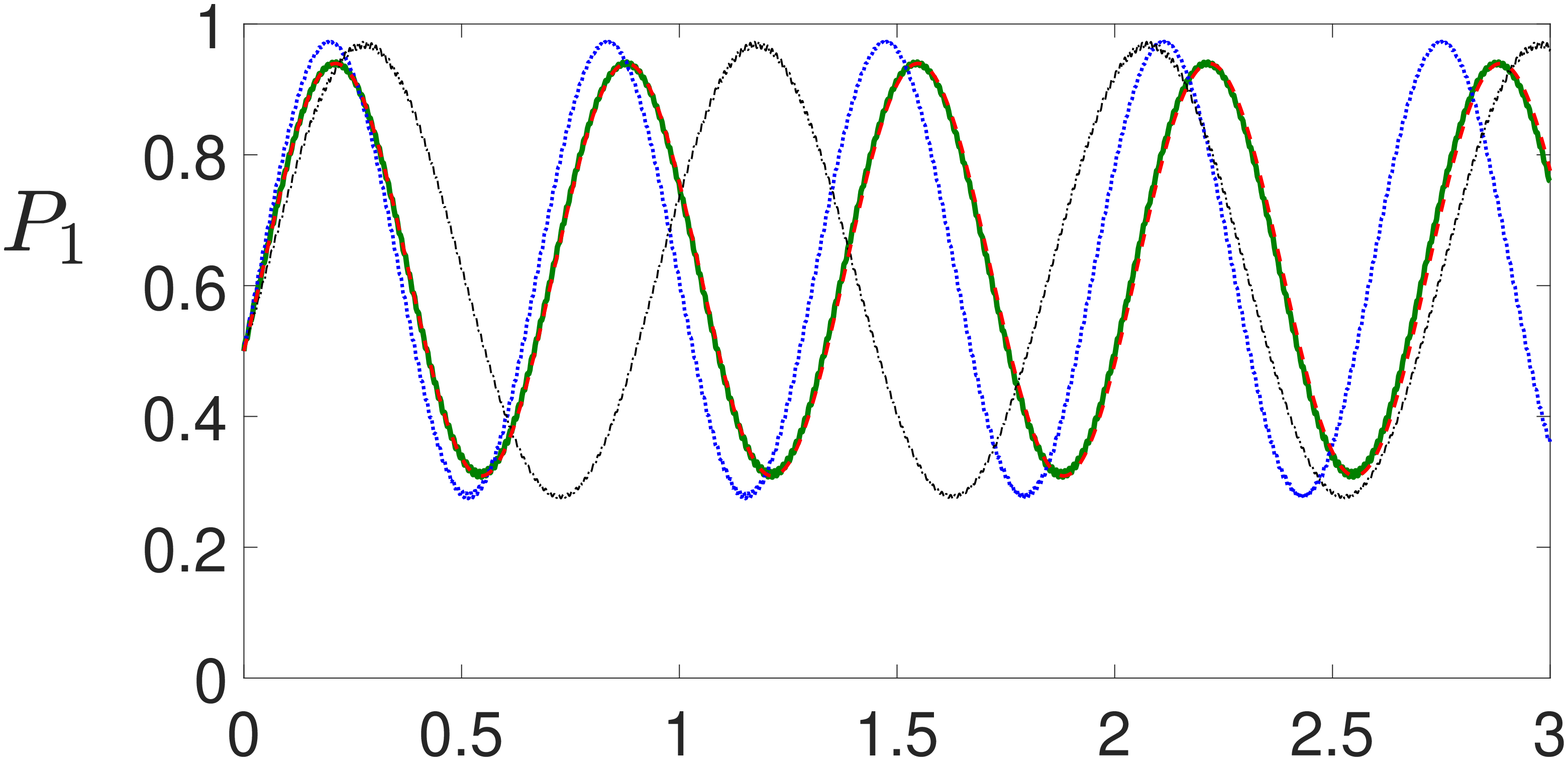}
	\includegraphics[width=0.45\textwidth]{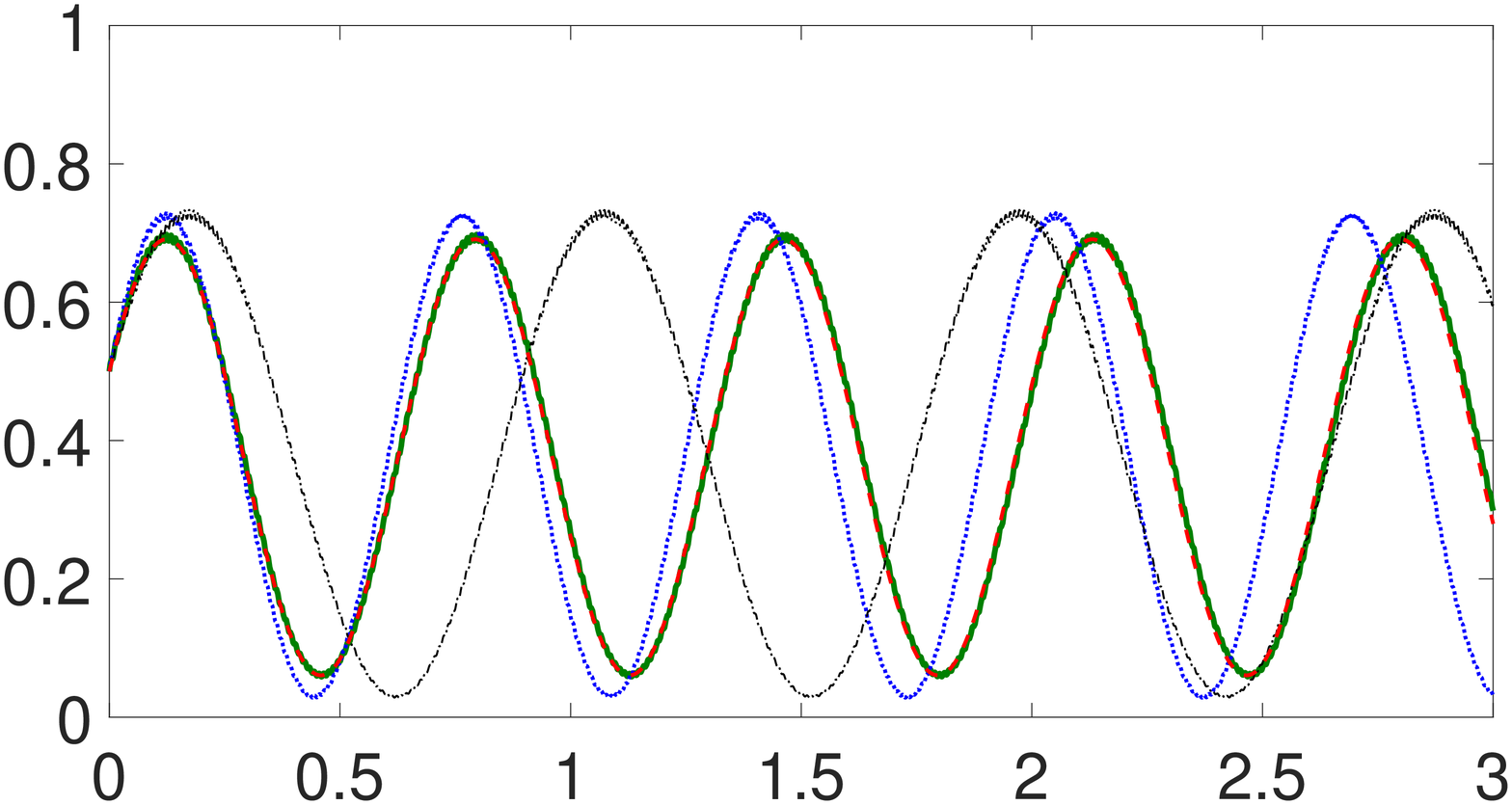}\\
	\vspace{-0.5cm}
	\includegraphics[width=0.45\textwidth]{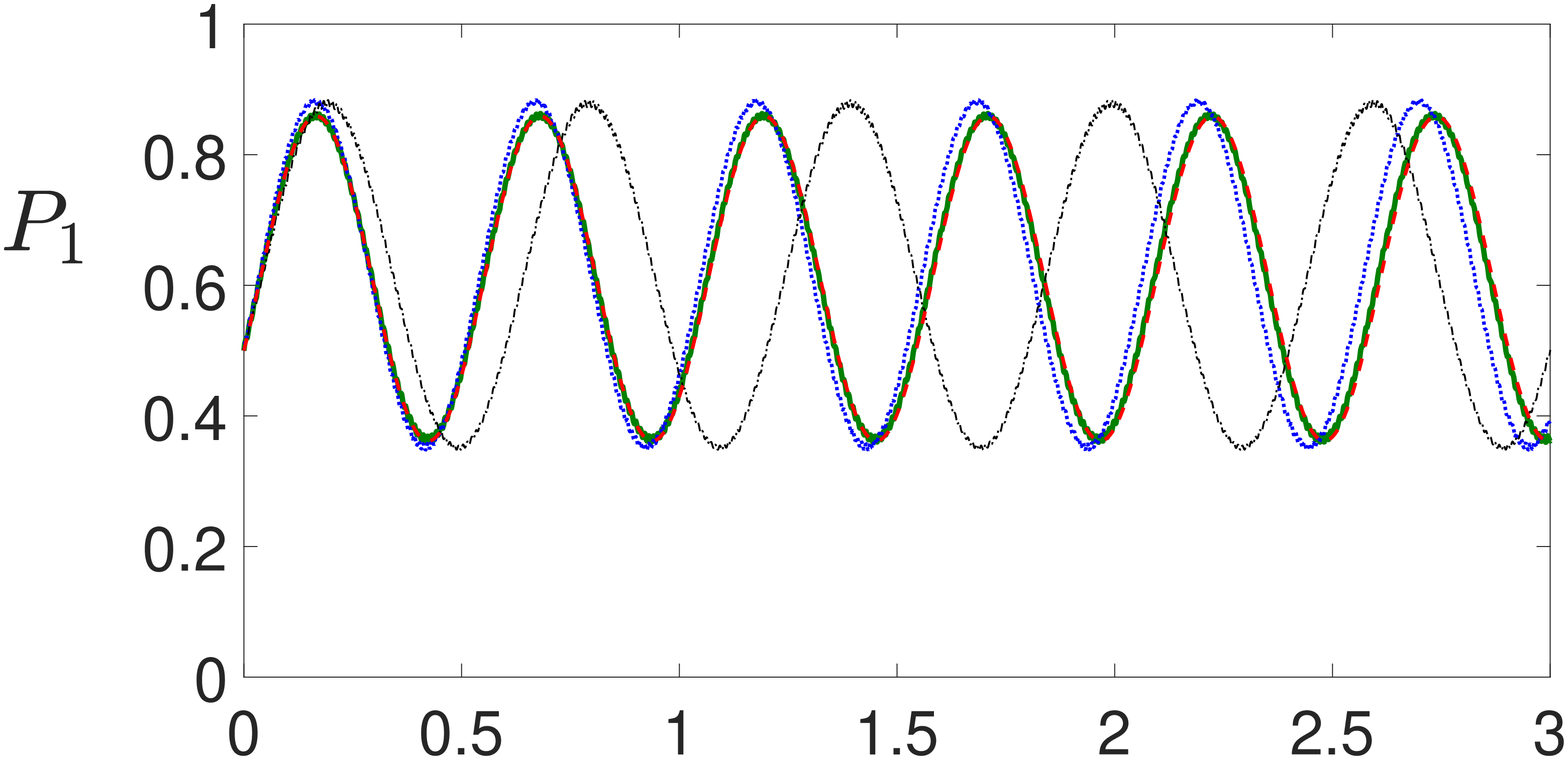}
	\includegraphics[width=0.45\textwidth]{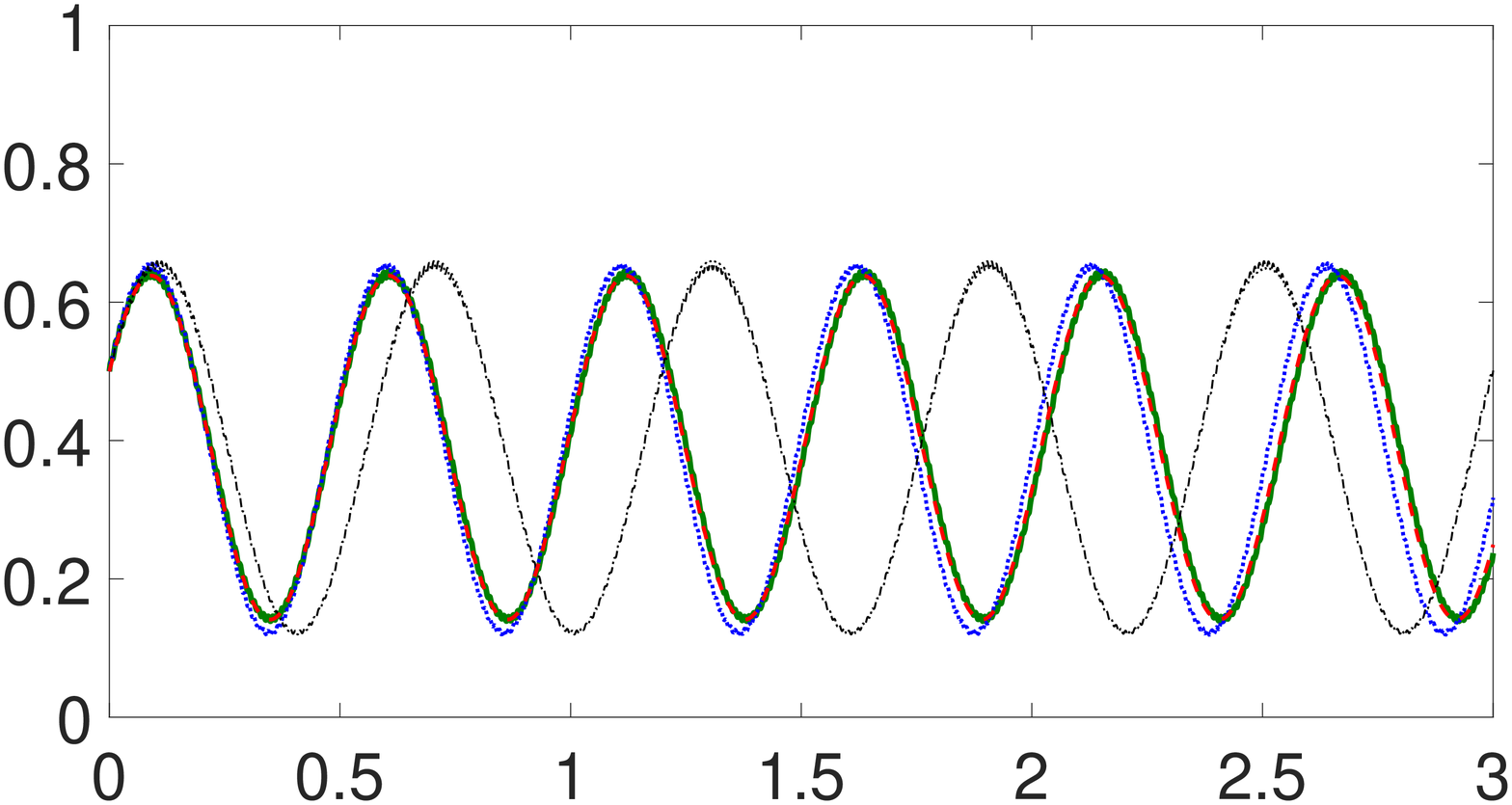}\\
	\vspace{-0.5cm}
	\includegraphics[width=0.45\textwidth]{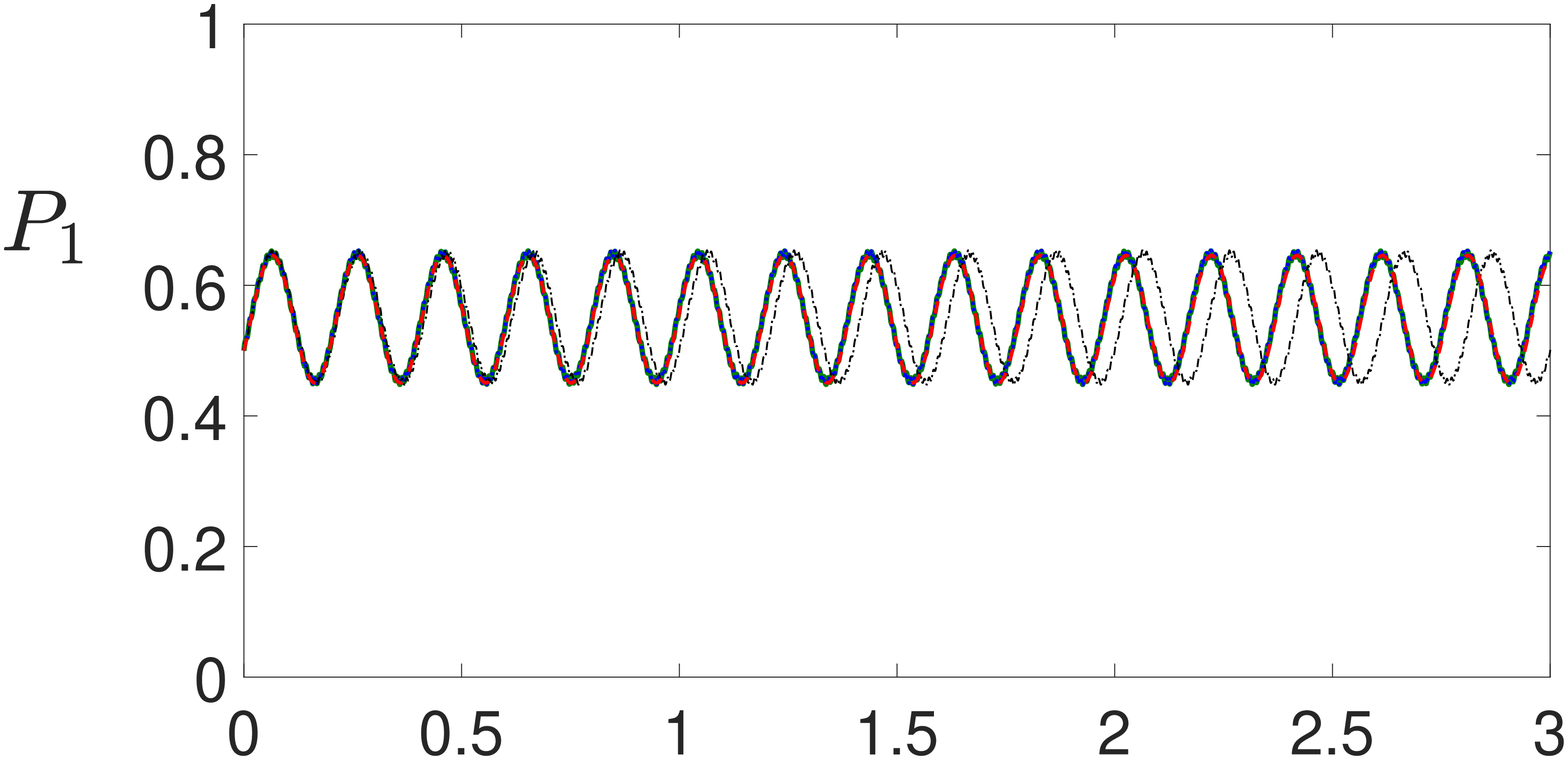}
	\includegraphics[width=0.45\textwidth]{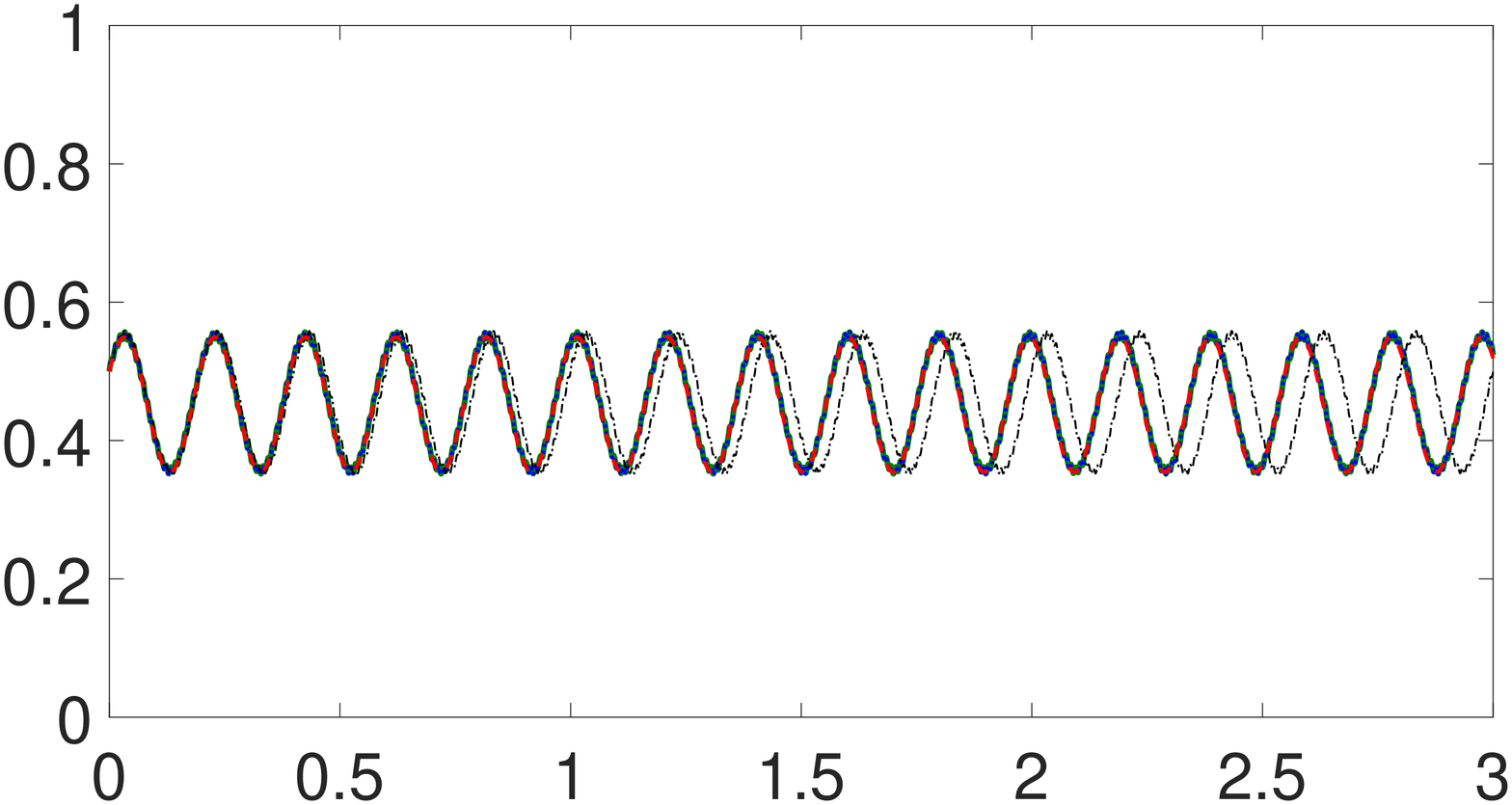}\\
	\vspace{-0.5cm}
	\includegraphics[width=0.45\textwidth]{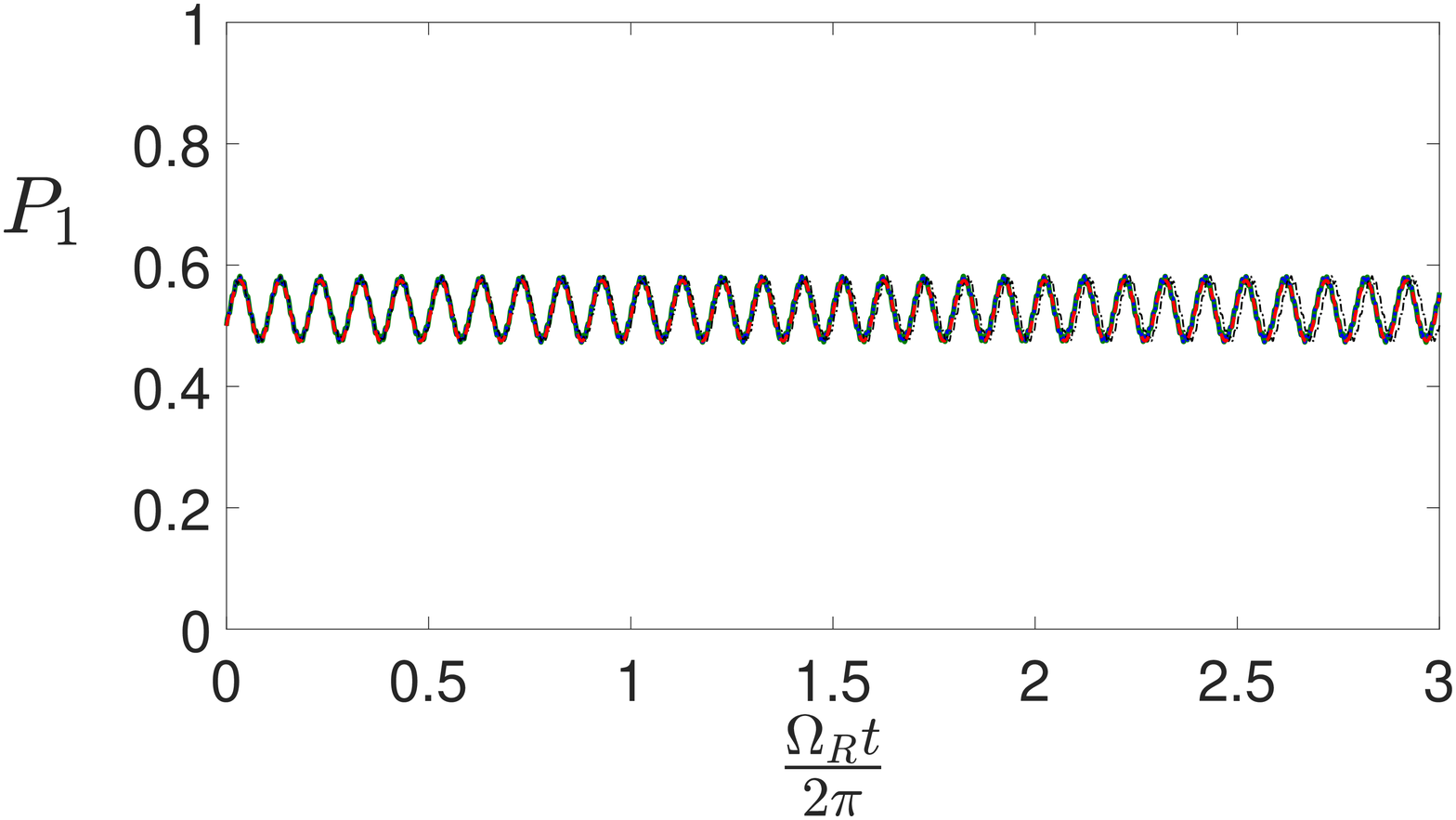}
	\includegraphics[width=0.45\textwidth]{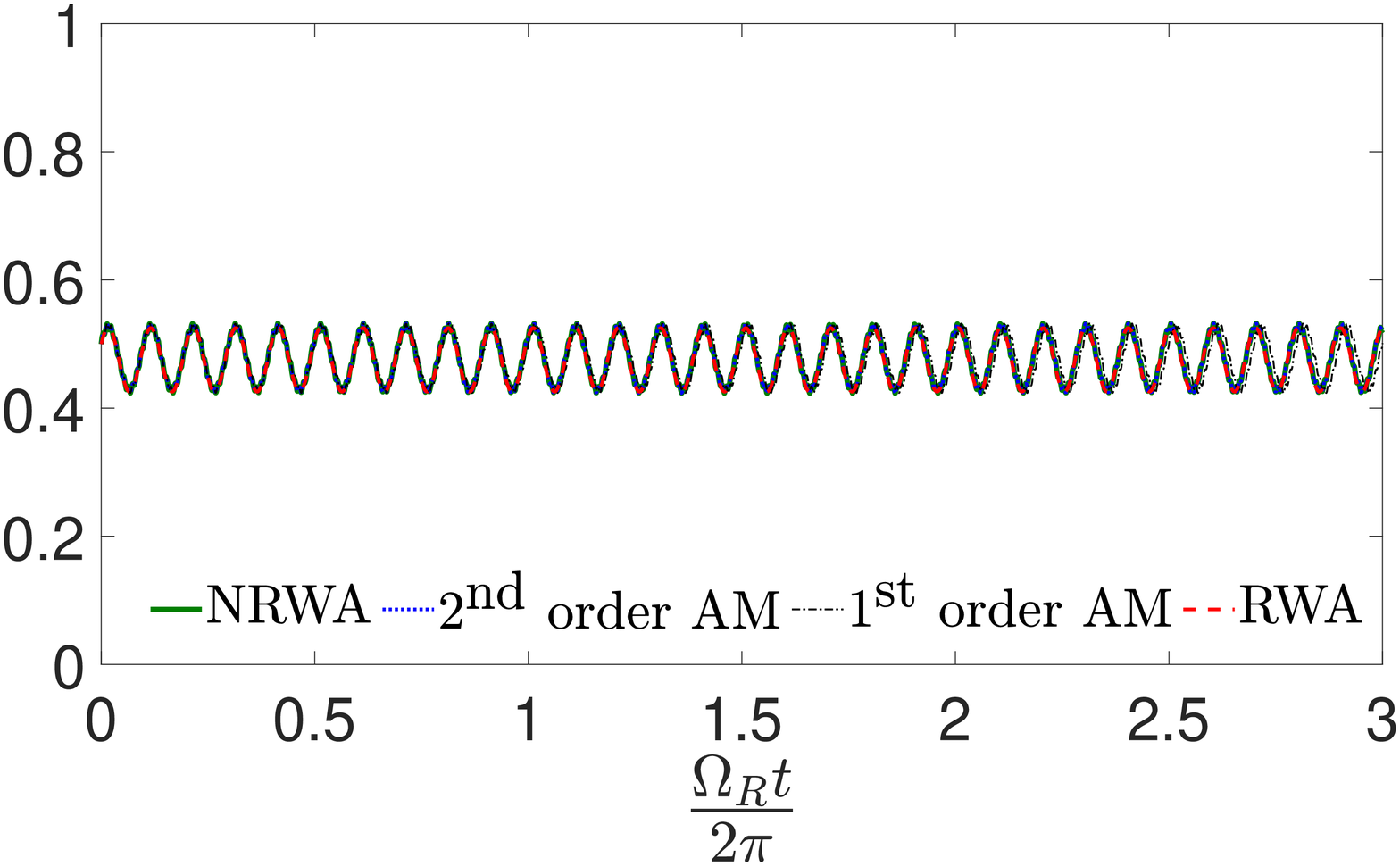}
	\vspace{-0.3cm}
	\caption{$\Delta \ne 0$. $P_1$ vs. $\frac{\Omega_{\textrm{R}} t}{2\pi}$, 
		for $\epsilon_2 = \frac{\Omega_{\textrm{R}}}{\Sigma} = 0.01$,
		varying $\epsilon_1 = \frac{\Omega_{\textrm{R}}}{\Delta}$ 
		with $\theta-\phi = \frac{\pi}{3}$. 
		(a) $\epsilon_1 = -0.9$. (b) $\epsilon_1 = 0.9$. 
		(c) $\epsilon_1 = -0.6$. (d) $\epsilon_1 = 0.6$. 
		(e) $\epsilon_1 = -0.2$. (f) $\epsilon_1 = 0.2$. 
		(g) $\epsilon_1 = -0.1$. (h) $\epsilon_1 = 0.1$.
		Lines correspond to 
		NRWA (continuous {\color{ForestGreen} ---}), 
		RWA (dashed {\color{red} $--$}), 
		second order AM (dotted {\color{blue} $\cdots$}), 
		first order AM (dash-dotted $\cdot -$).}
	\label{fig:epsilonDeltaphasi}
\end{figure*}

In Fig.~\ref{fig:epsilonDeltaphasi} we modify $\epsilon_1 = \frac{\Omega_{\textrm{R}}}{\Delta}$, keeping $\epsilon_2 = \frac{\Omega_{\textrm{R}}}{\Sigma} = 0.01$ 
with $\theta-\phi = \frac{\pi}{3}$. 
We observe strong oscillations, depending of course on the magnitude of $\epsilon_1$, although the initial probabilities are equal, a pure coherent phenomenon, due to the initial phase difference of the wave functions.
Decreasing $|\epsilon_1|$, second order AM approaches NRWA.

The discussion on the effect of the relative magnitude of $\epsilon_1, \epsilon_2, \epsilon_3$ (Subsubsec.~\ref{subsubsec:AM-RWA-NS:non-resonance10}), applies here, too.  

\begin{figure*}
	\centering
	\vspace{-0.2cm}
	\includegraphics[width=0.45\textwidth]{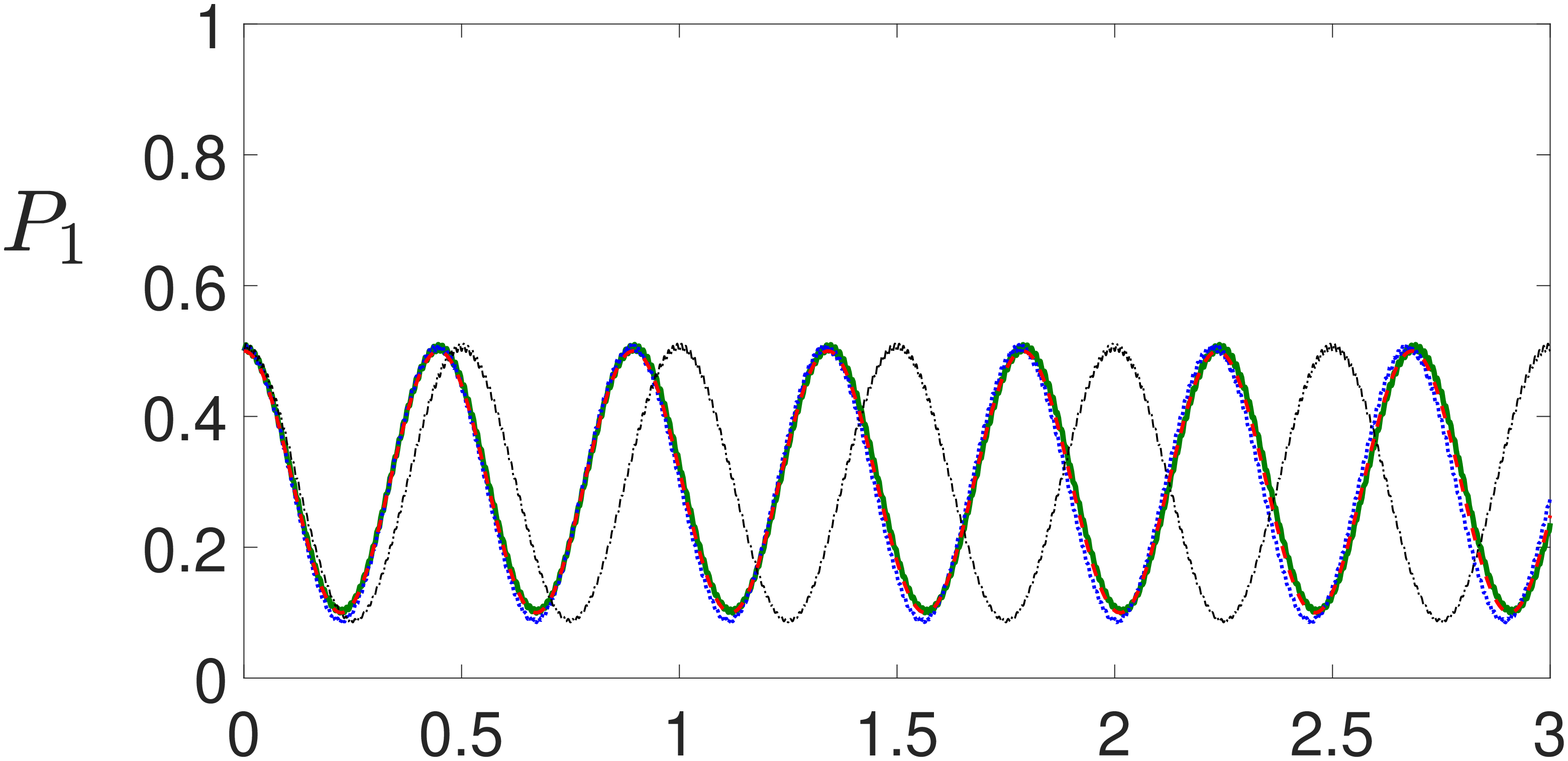}
	\includegraphics[width=0.45\textwidth]{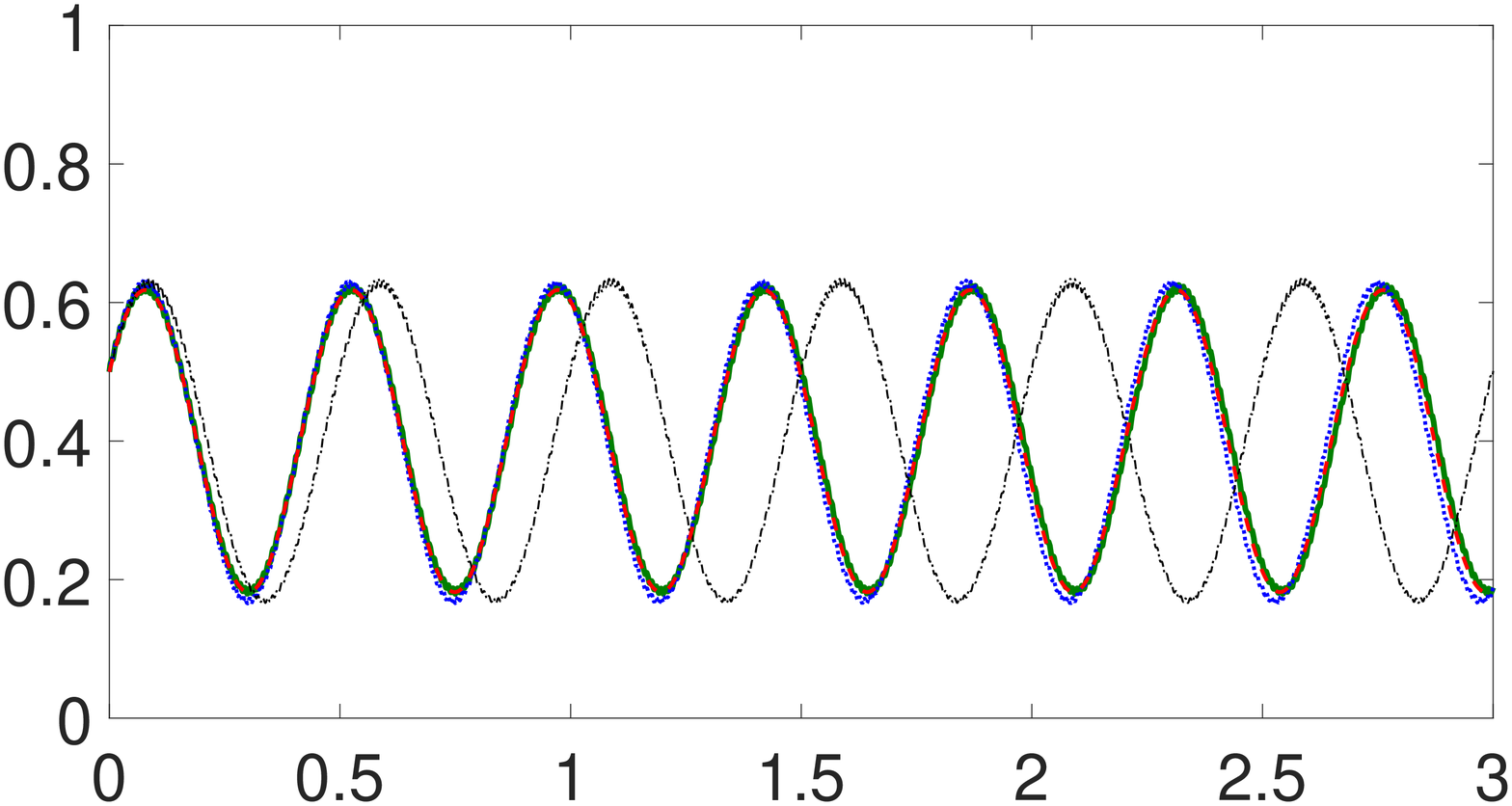}
	\vspace{-0.6cm}
	\includegraphics[width=0.45\textwidth]{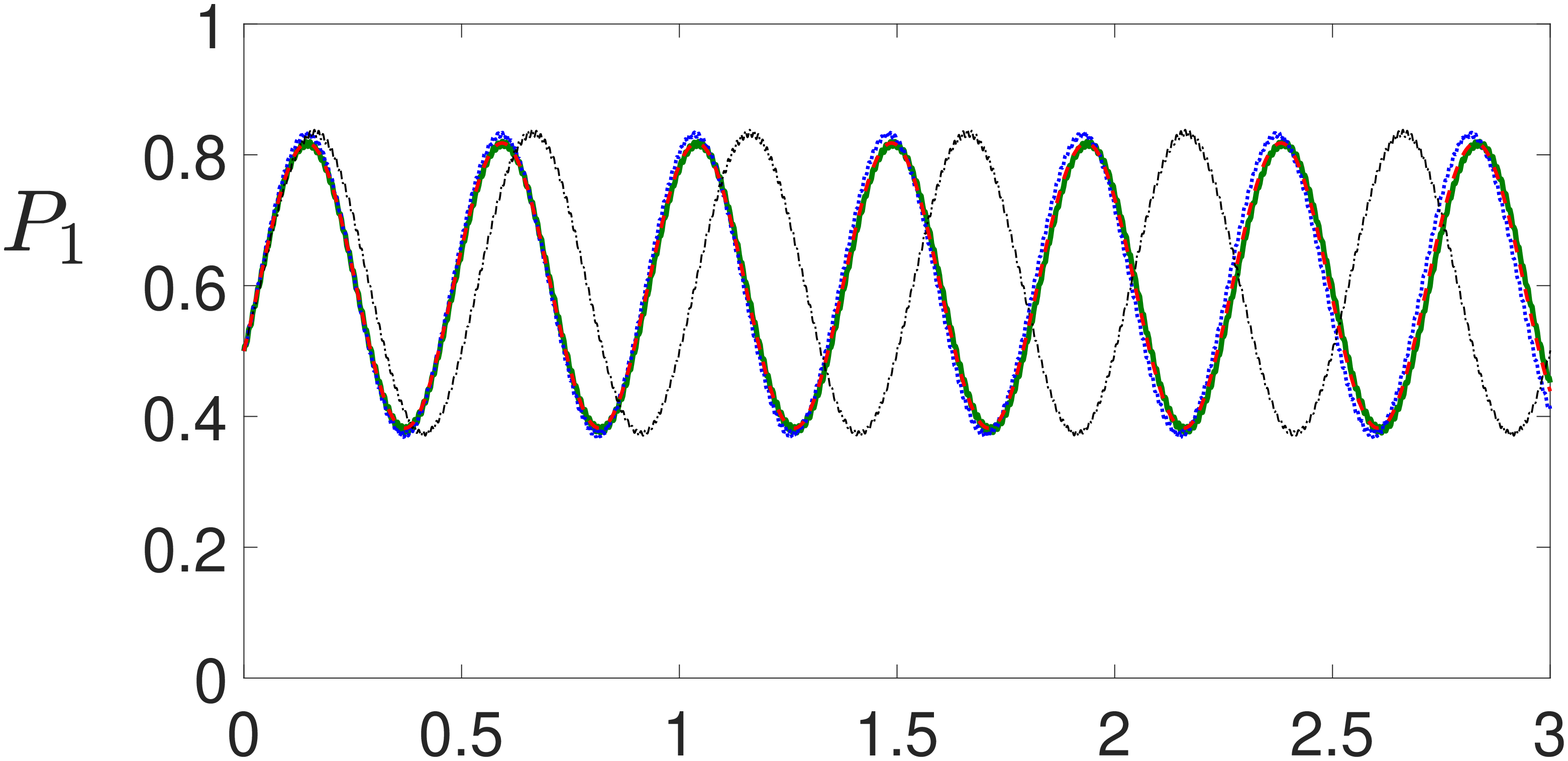}
	\includegraphics[width=0.45\textwidth]{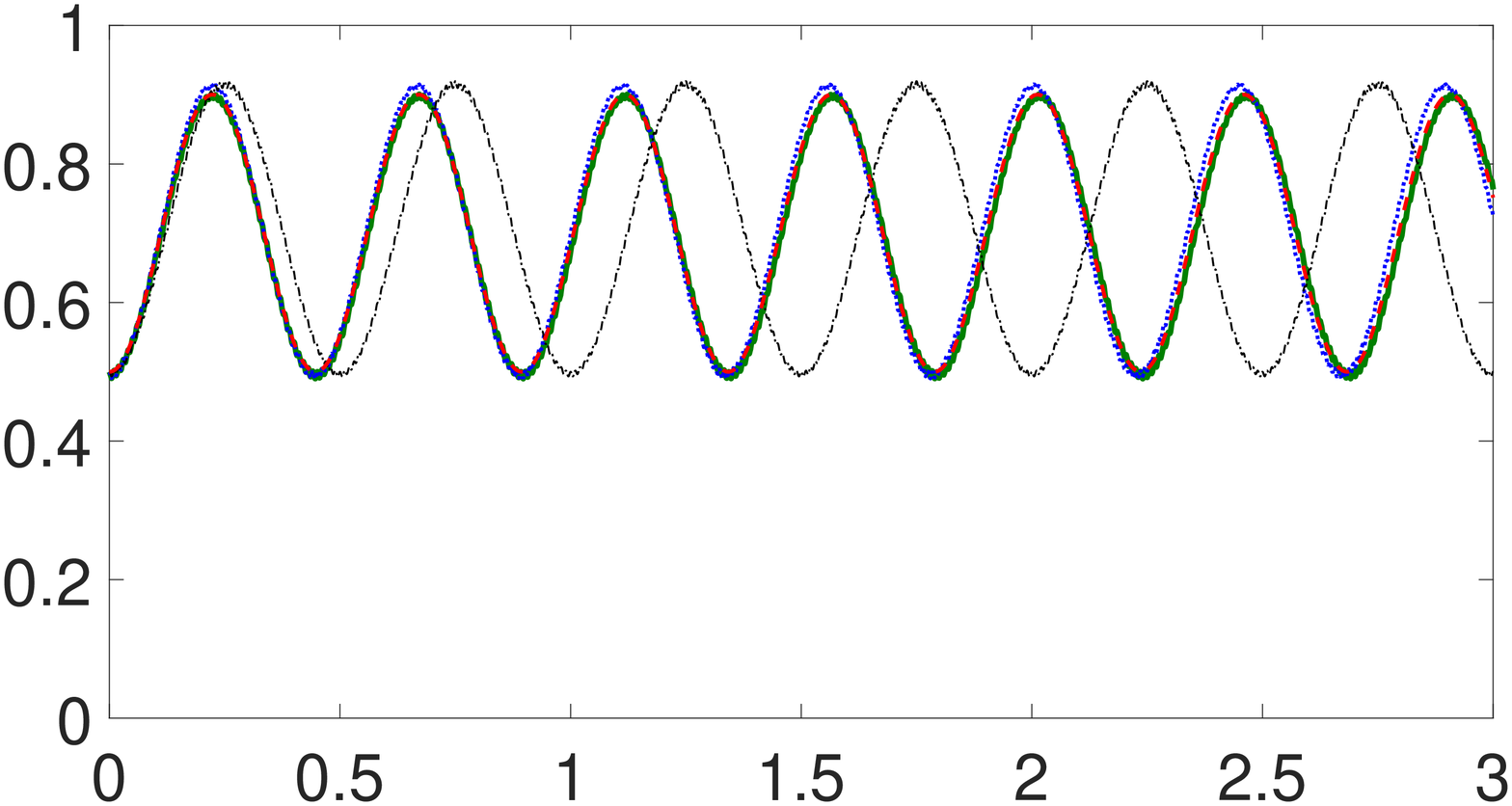}
	\vspace{-0.6cm}
	\includegraphics[width=0.45\textwidth]{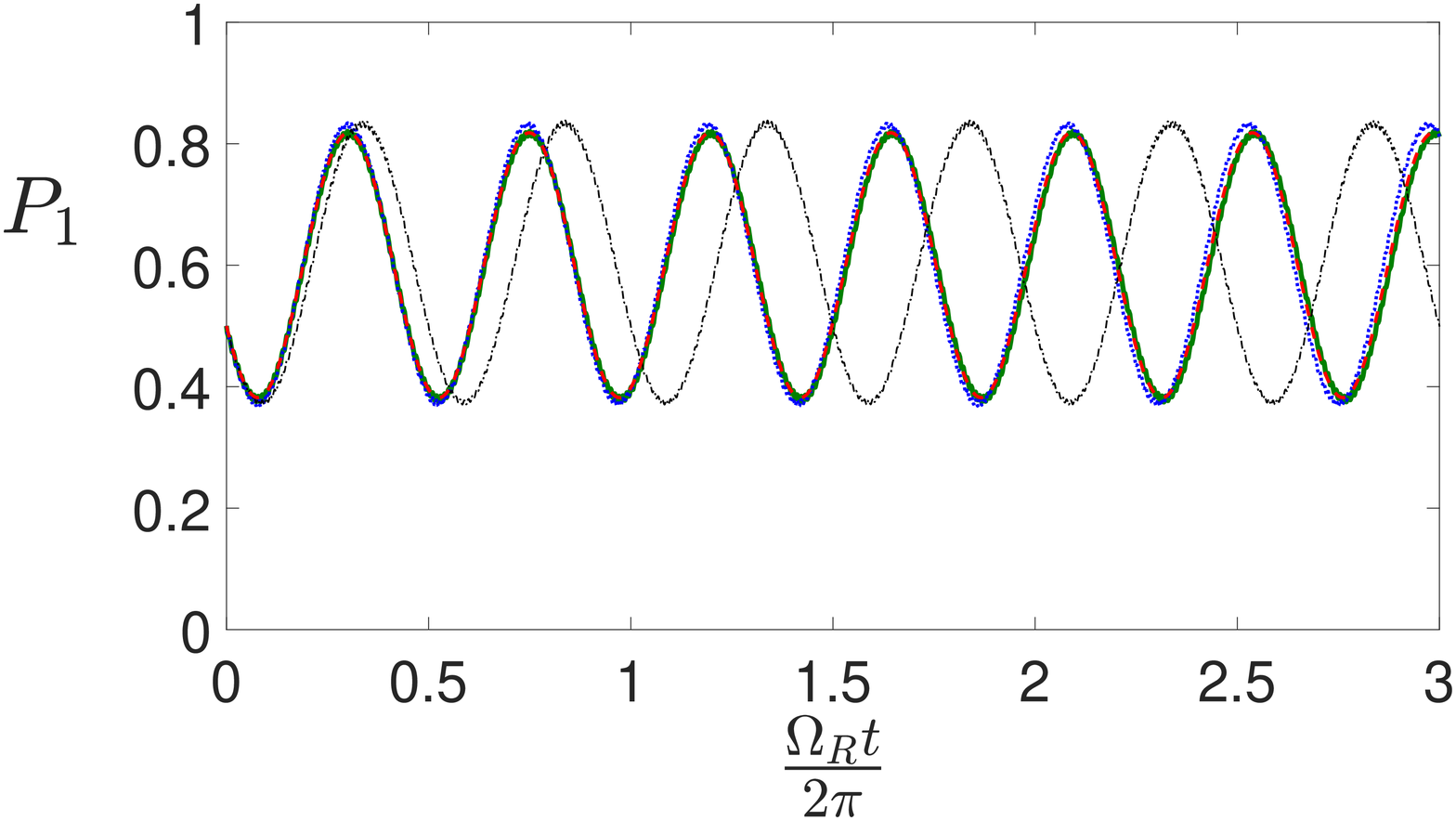}
	\includegraphics[width=0.45\textwidth]{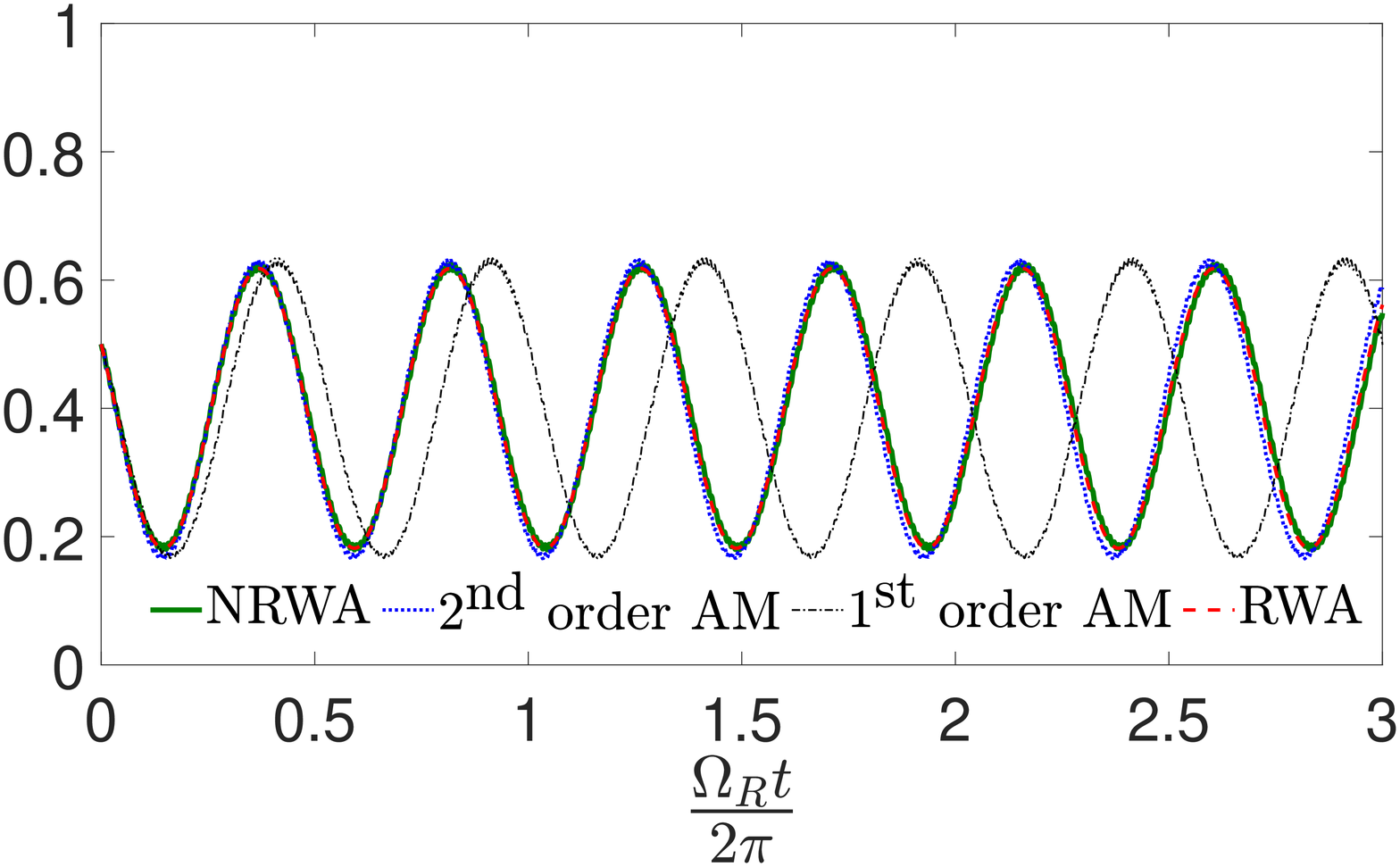}
	\caption{$\Delta \ne 0$. $P_1$ vs. $\frac{\Omega_{\textrm{R}} t}{2\pi}$  
		for $\epsilon_1 = \frac{\Omega_{\textrm{R}}}{\Delta} = 0.5$ and
		$\epsilon_2 = \frac{\Omega_{\textrm{R}}}{\Sigma} = 0.01$, varying $\theta - \phi$.
		(a) $\theta - \phi = 0$.
		(b) $\theta - \phi = \frac{\pi}{3}$.
		(c) $\theta - \phi = \frac{2\pi}{3}$. 
		(d) $\theta - \phi = \pi$. 
		(e) $\theta - \phi = \frac{4\pi}{3}$.
		(f) $\theta - \phi = \frac{5\pi}{3}$.
		Lines correspond to 
		NRWA (continuous {\color{ForestGreen} ---}), 
		RWA (dashed {\color{red} $--$}), 
		second order AM (dotted {\color{blue} $\cdots$}), 
		first order AM (dash-dotted $\cdot -$).}
	\label{fig:phasi}
\end{figure*}

In Fig.~\ref{fig:phasi} we keep $\epsilon_1 = 0.5$ and $\epsilon_1 = 0.01$, varying the initial phase difference of the wave functions, $\theta -\phi$. We observe another aspect of coherence, a vertical and horizontal displacement of the oscillations.

\subsubsection{Resonance} 
\label{subsubsec:AM-RWA-NS:resonancehh} 
\begin{figure*}[t!]
	\centering
	\vspace{0.2cm}
	\includegraphics[width=0.45\textwidth]{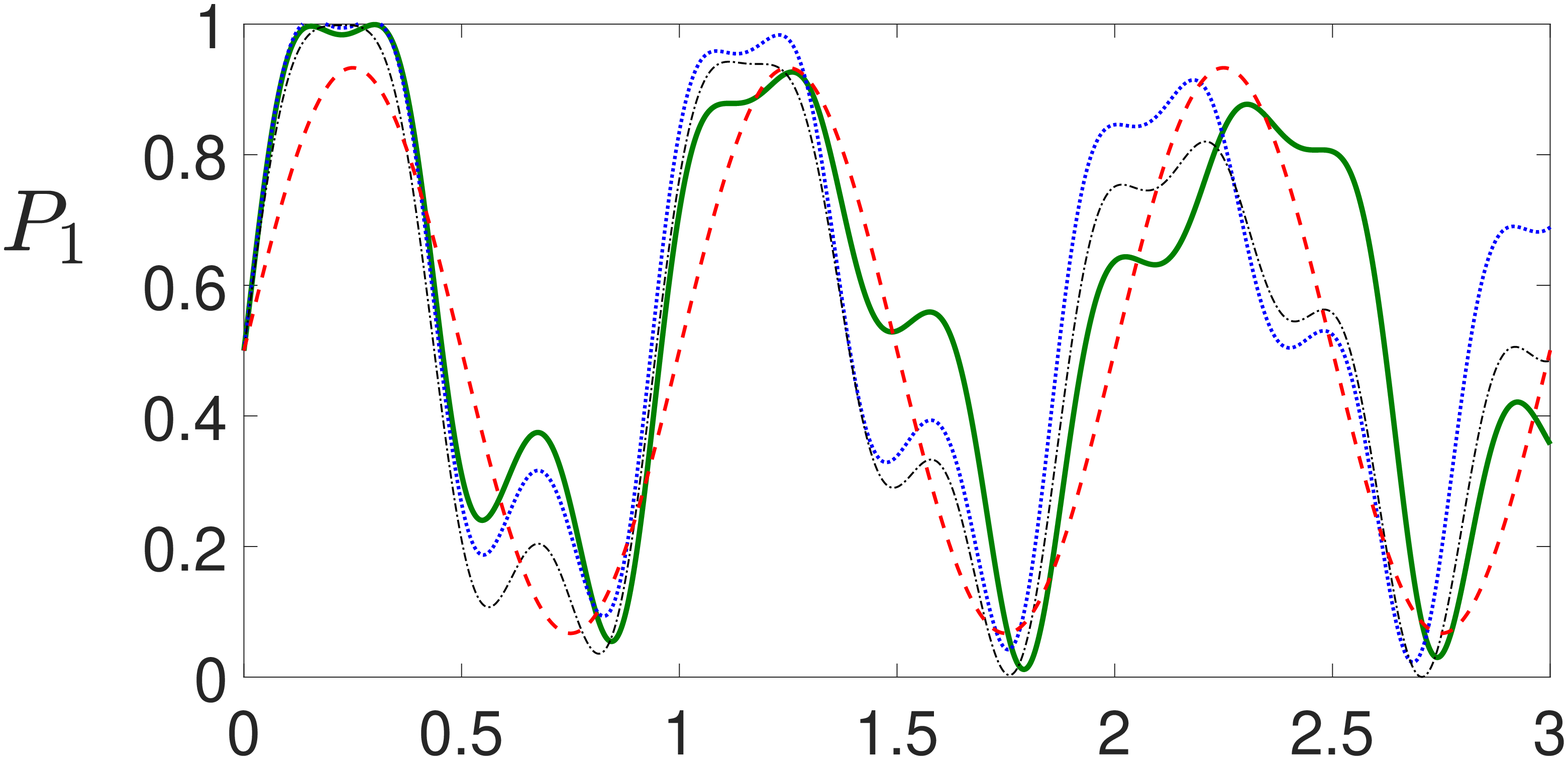}
	\includegraphics[width=0.45\textwidth]{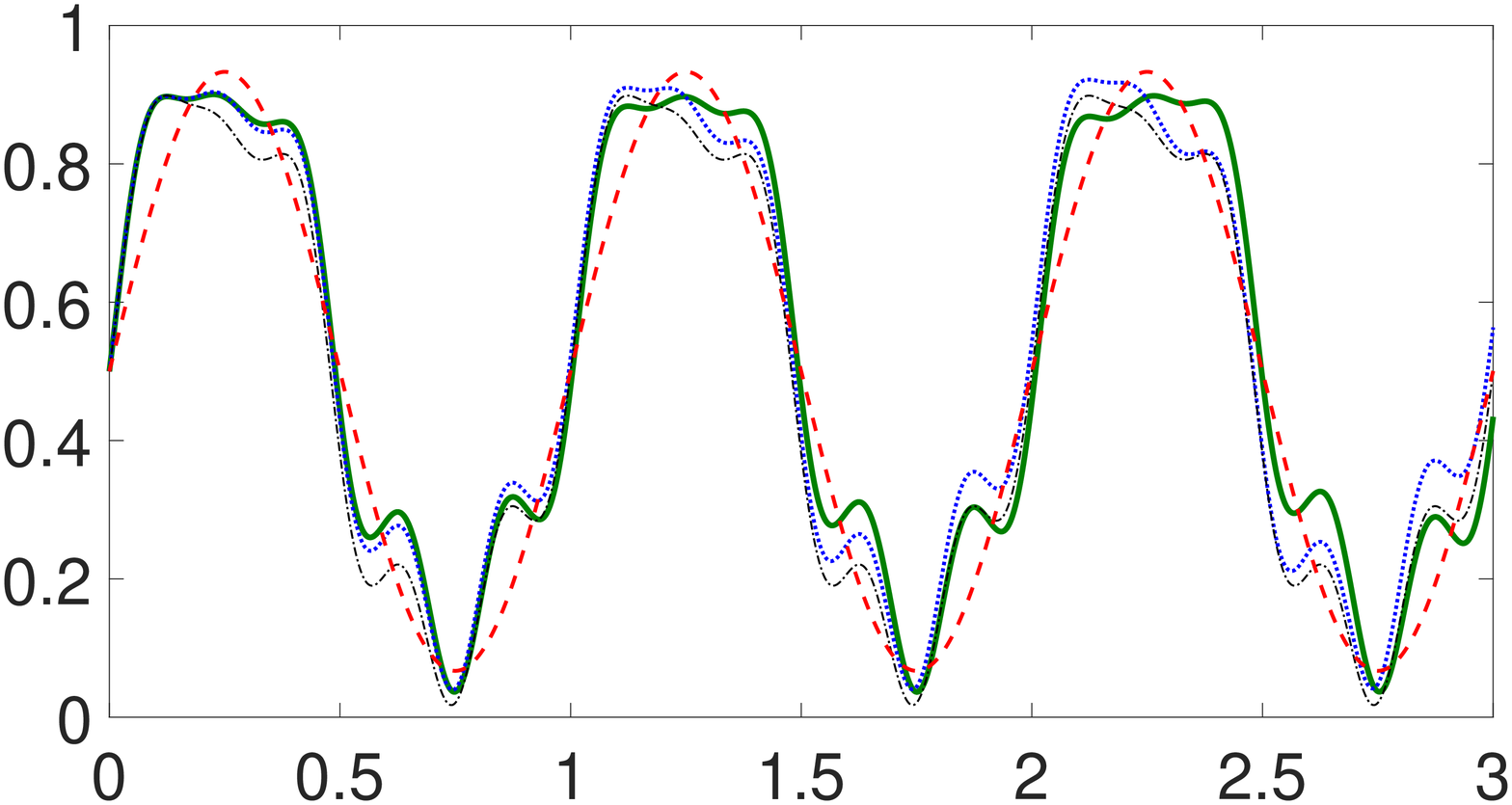}\\
	\vspace{-0.5cm}
	\includegraphics[width=0.45\textwidth]{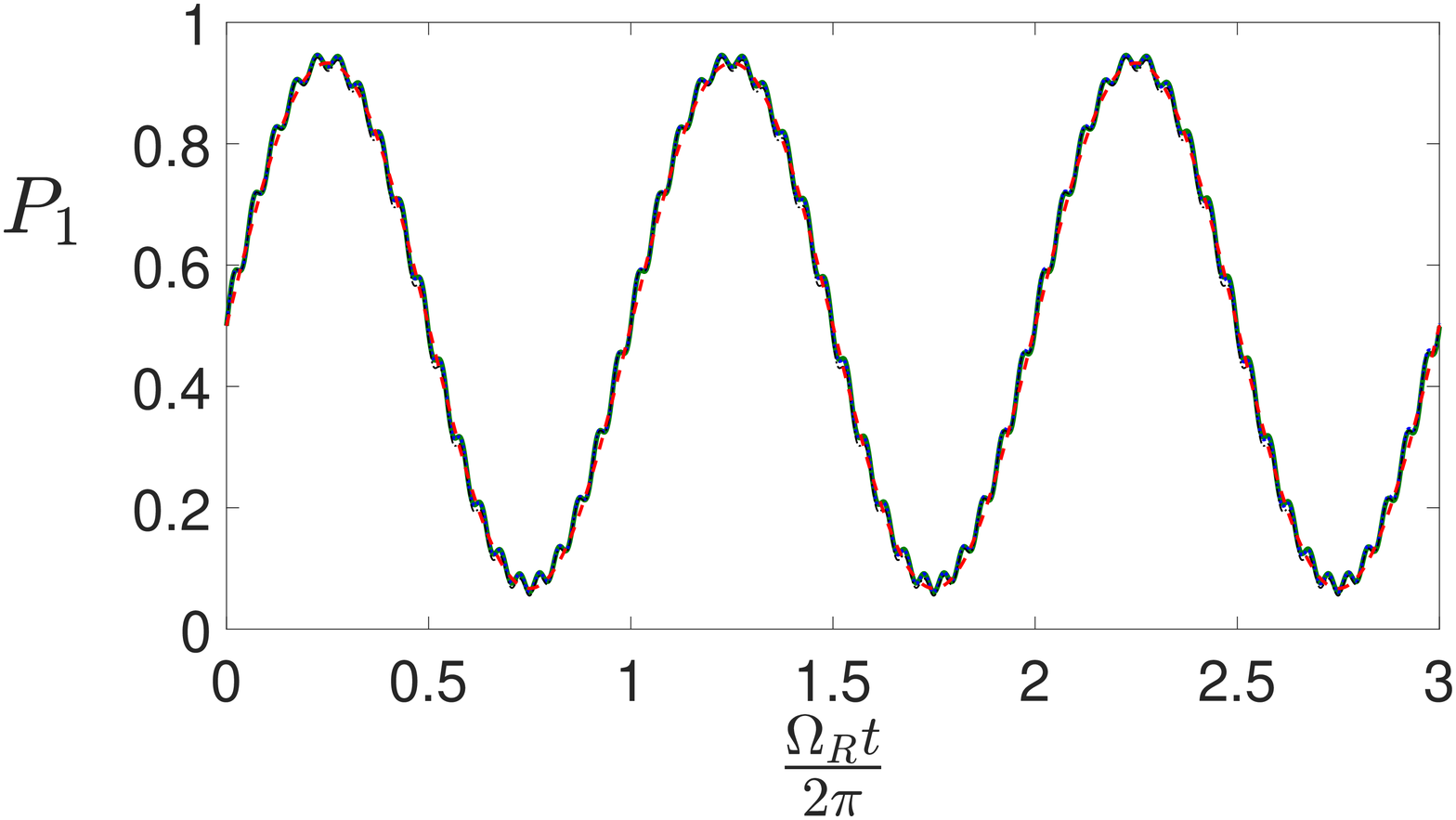}
	\includegraphics[width=0.45\textwidth]{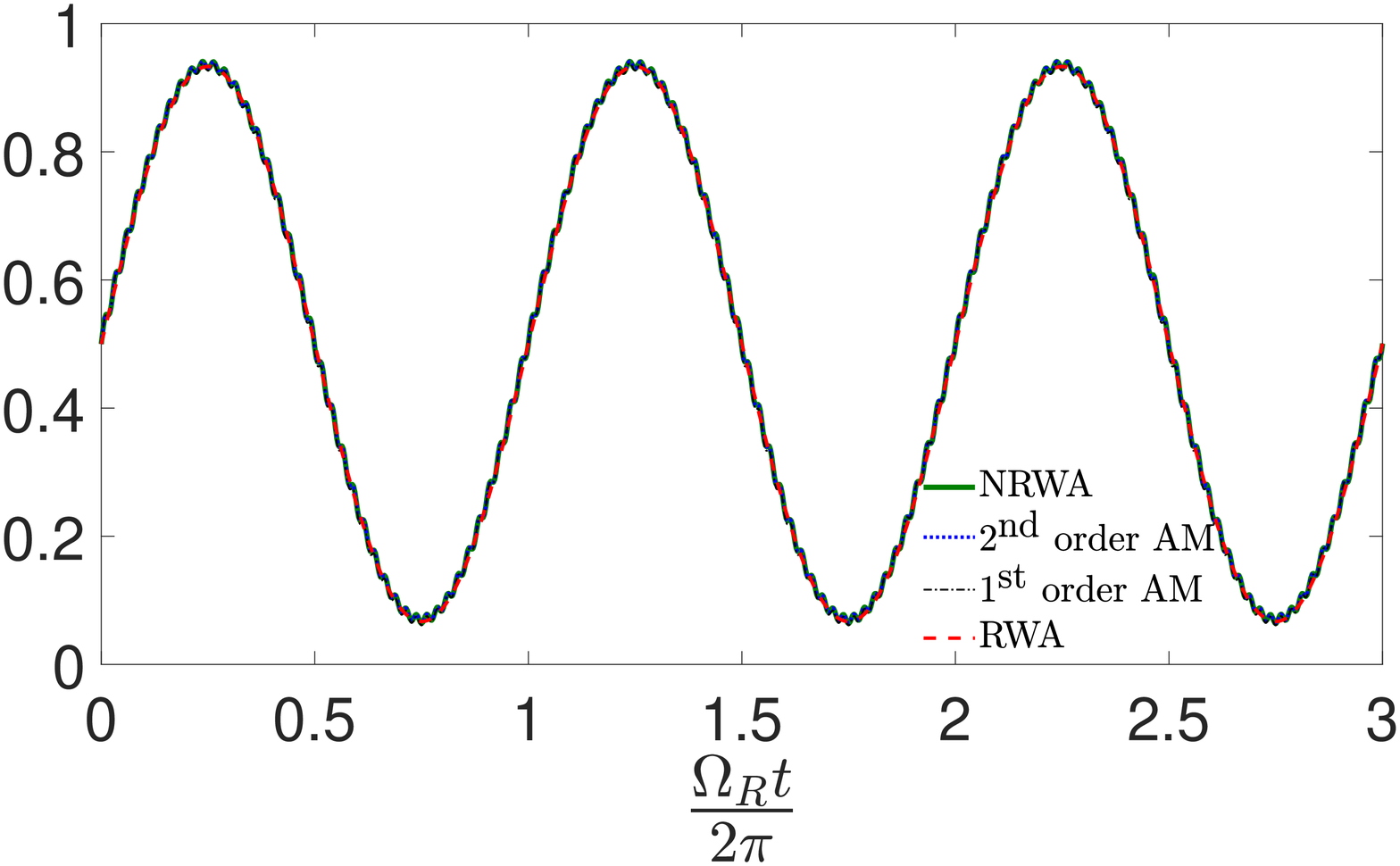}
	\vspace{-0.2cm}
	\caption{$\Delta = 0$. $P_1$ vs. $\frac{\Omega_{\textrm{R}} t}{2\pi}$, varying $\epsilon=\frac{\Omega_{\textrm{R}}}{\omega}$ 
		with $\theta-\phi = \frac{\pi}{3}$. 
		(a) $\epsilon=0.9$. (b) $\epsilon=0.5$. 
		(c) $\epsilon=0.1$. (d) $\epsilon=0.05$.
		Lines correspond to 
		NRWA (continuous {\color{ForestGreen} ---}), 
		RWA (dashed {\color{red} $--$}), 
		second order AM (dotted {\color{blue} $\cdots$}), 
		first order AM (dash-dotted $\cdot -$).}
	\label{fig:epsilonResonancephasi}
\end{figure*}

In Fig.~\ref{fig:epsilonResonancephasi} we modify  $\epsilon=\frac{\Omega_{\textrm{R}}}{\omega}$, for initial 
phase difference, $\theta-\phi = \frac{\pi}{3}$. Now, since we are in resonance, oscillations are particularly strong, of the order of one. As $\epsilon$ gets smaller, AM is identified with NRWA.

\begin{figure*}
	\centering
	\vspace{-0.5cm}
	\includegraphics[width=0.45\textwidth]{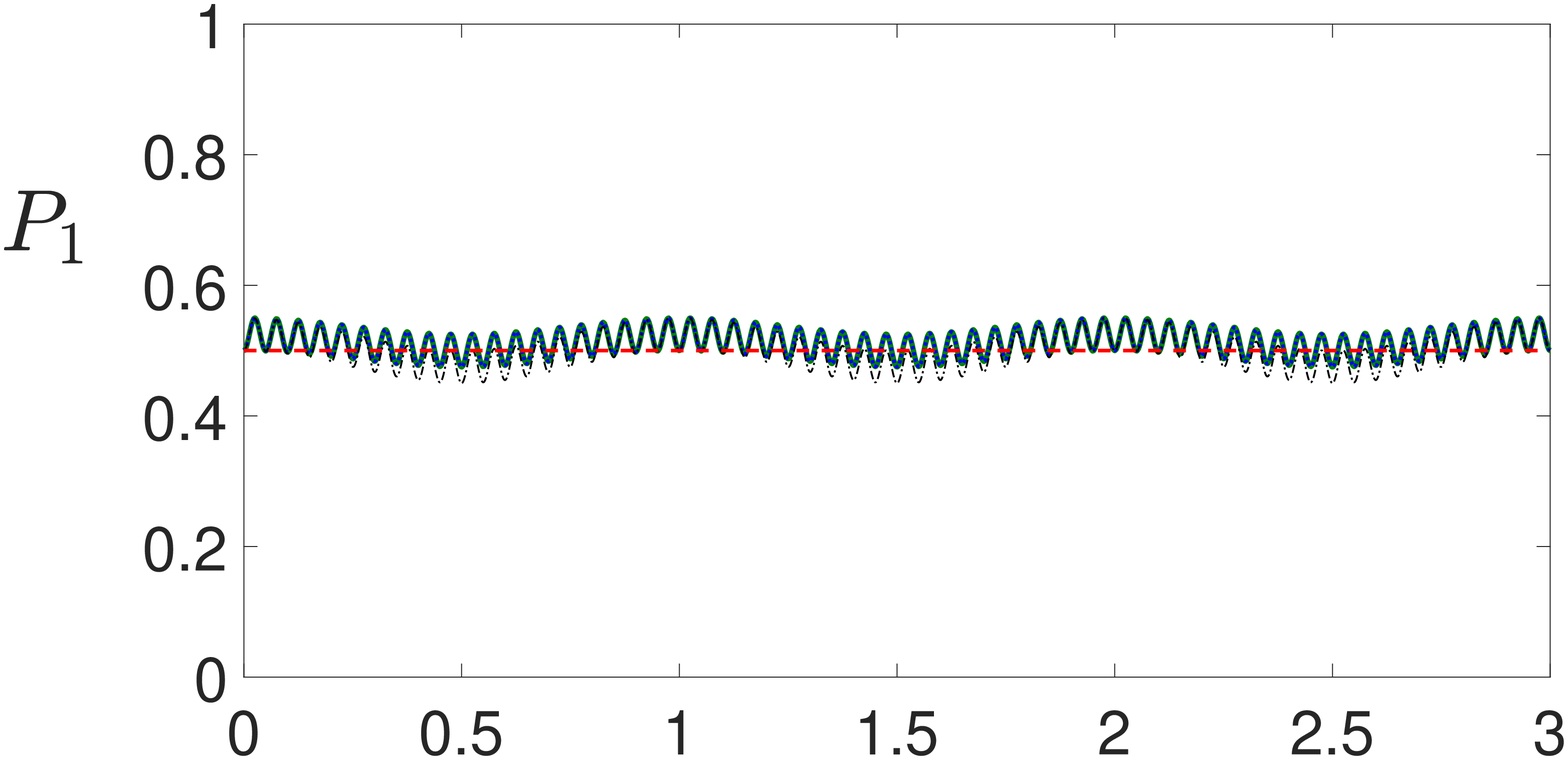}
	\includegraphics[width=0.45\textwidth]{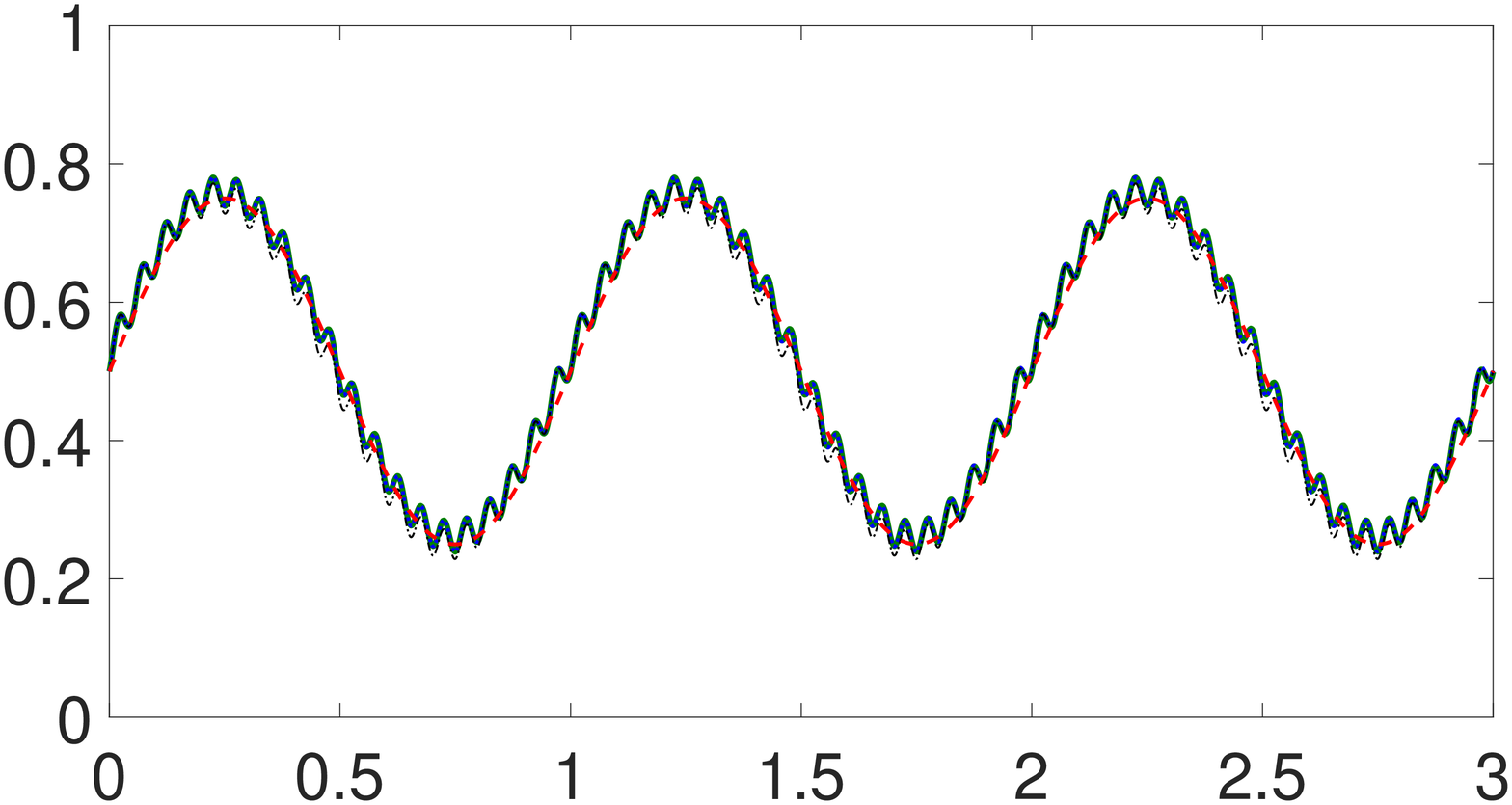}\\
	\vspace{-0.5cm}
	\includegraphics[width=0.45\textwidth]{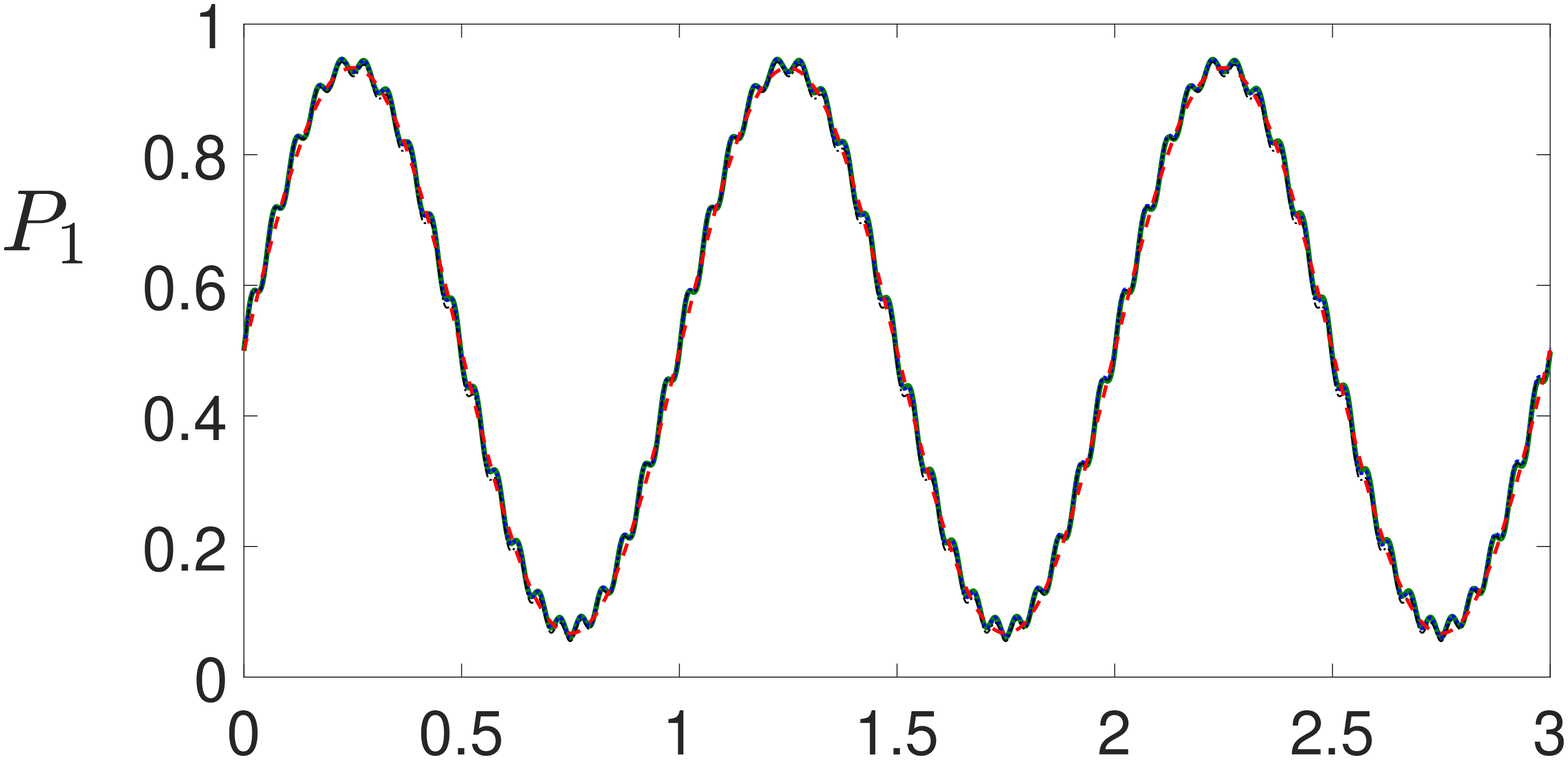}
	\includegraphics[width=0.45\textwidth]{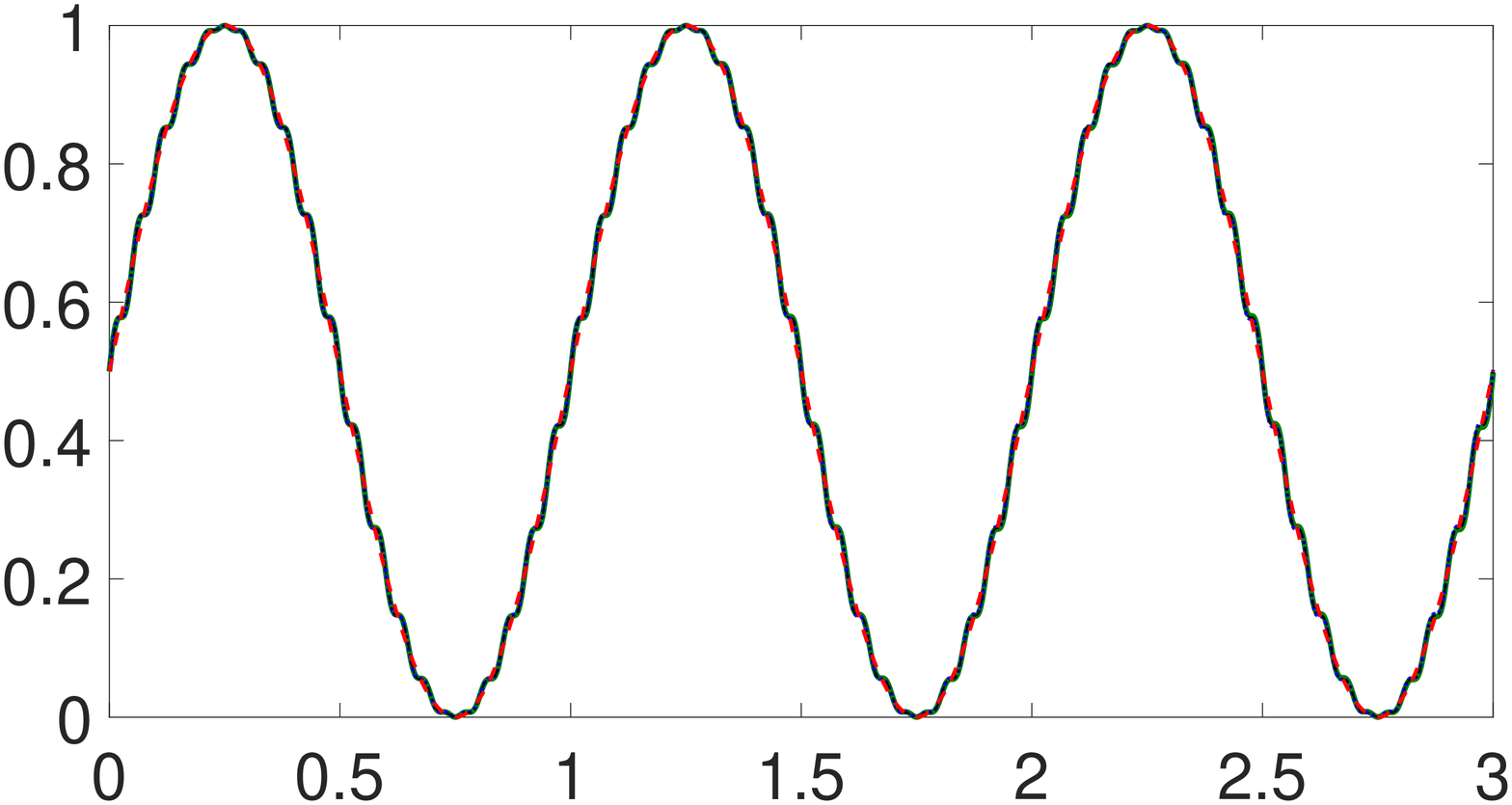}\\
	\vspace{-0.5cm}
	\includegraphics[width=0.45\textwidth]{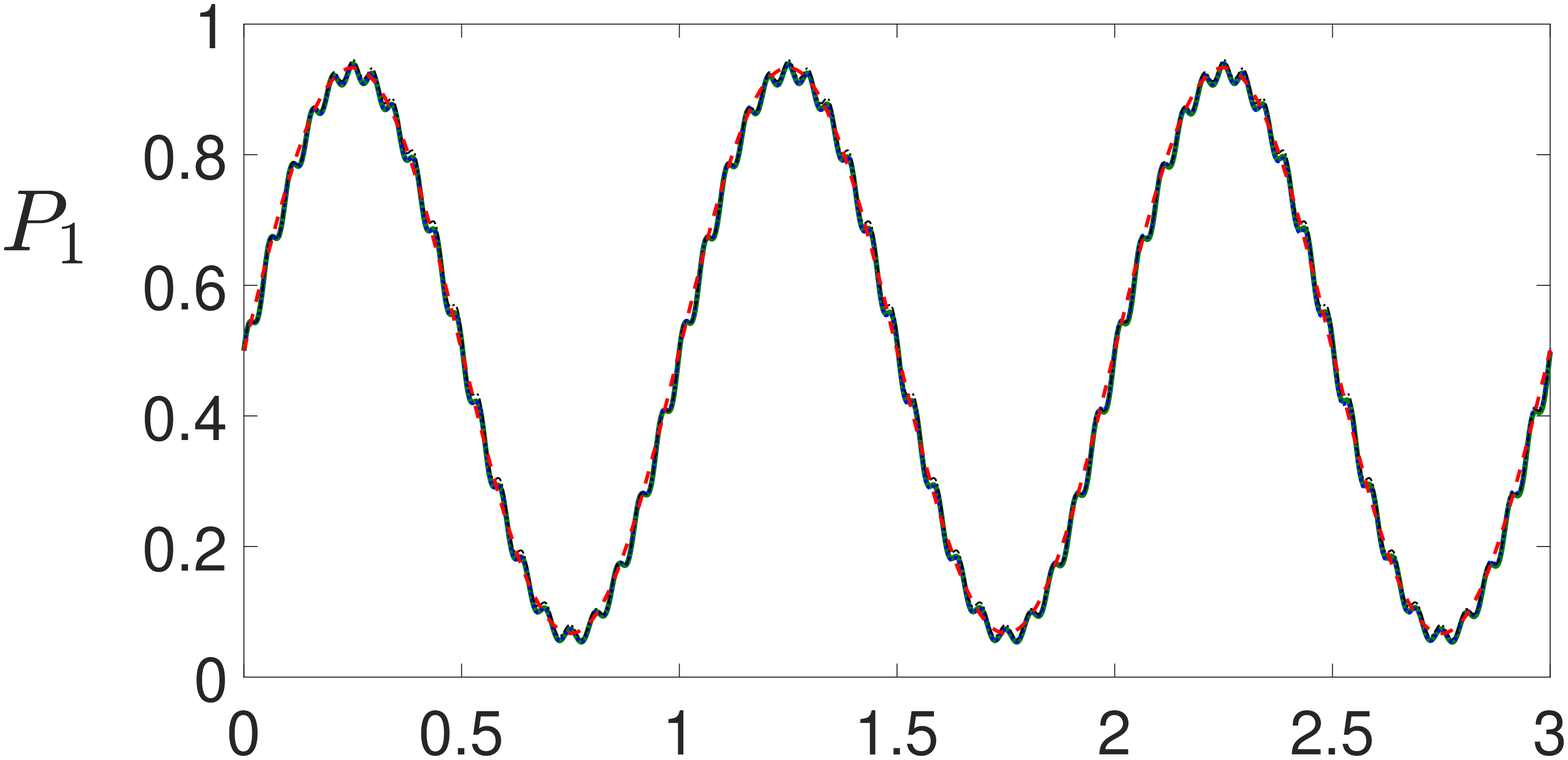}
	\includegraphics[width=0.45\textwidth]{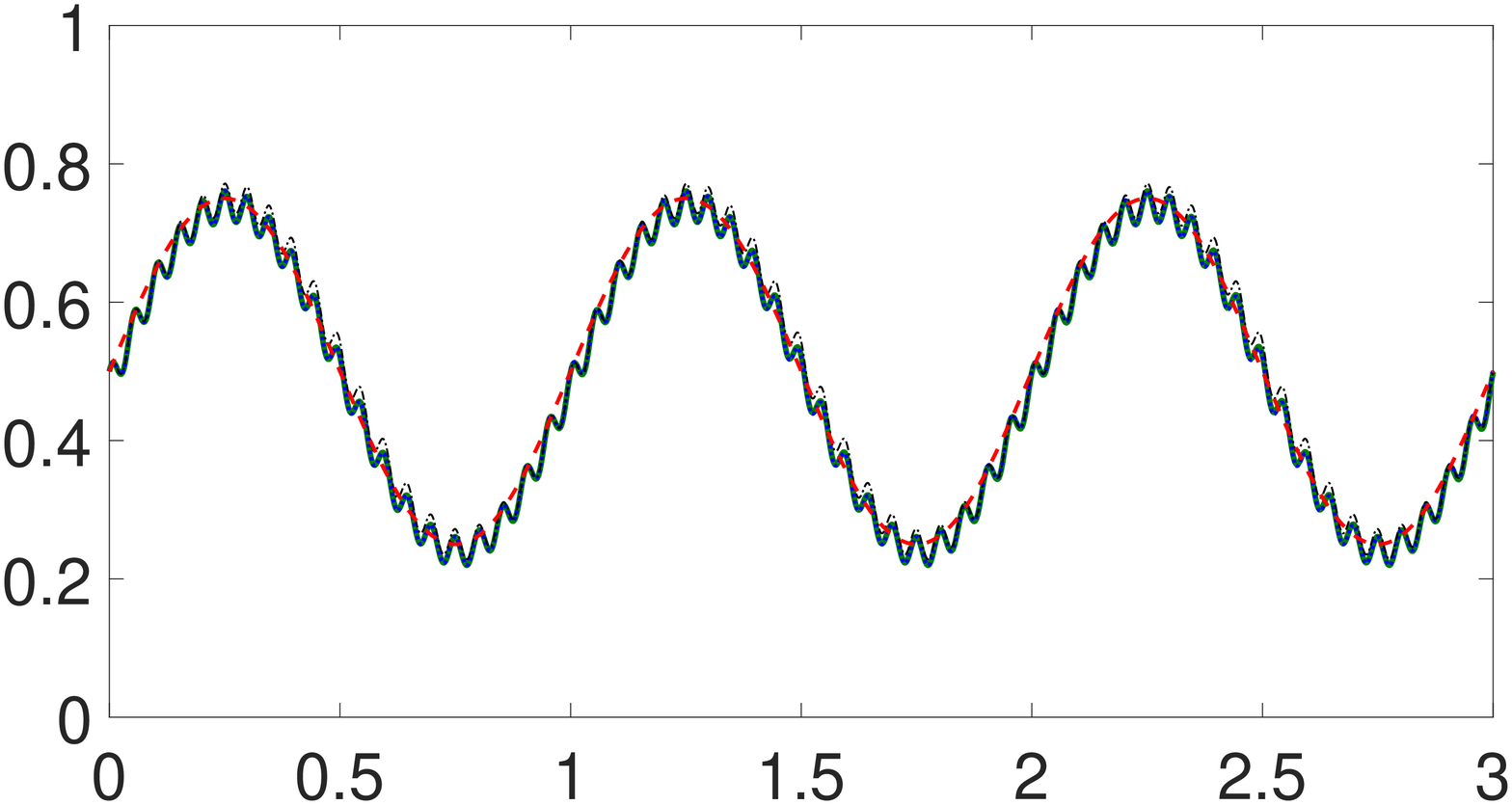}\\
	\vspace{-0.5cm}
	\includegraphics[width=0.45\textwidth]{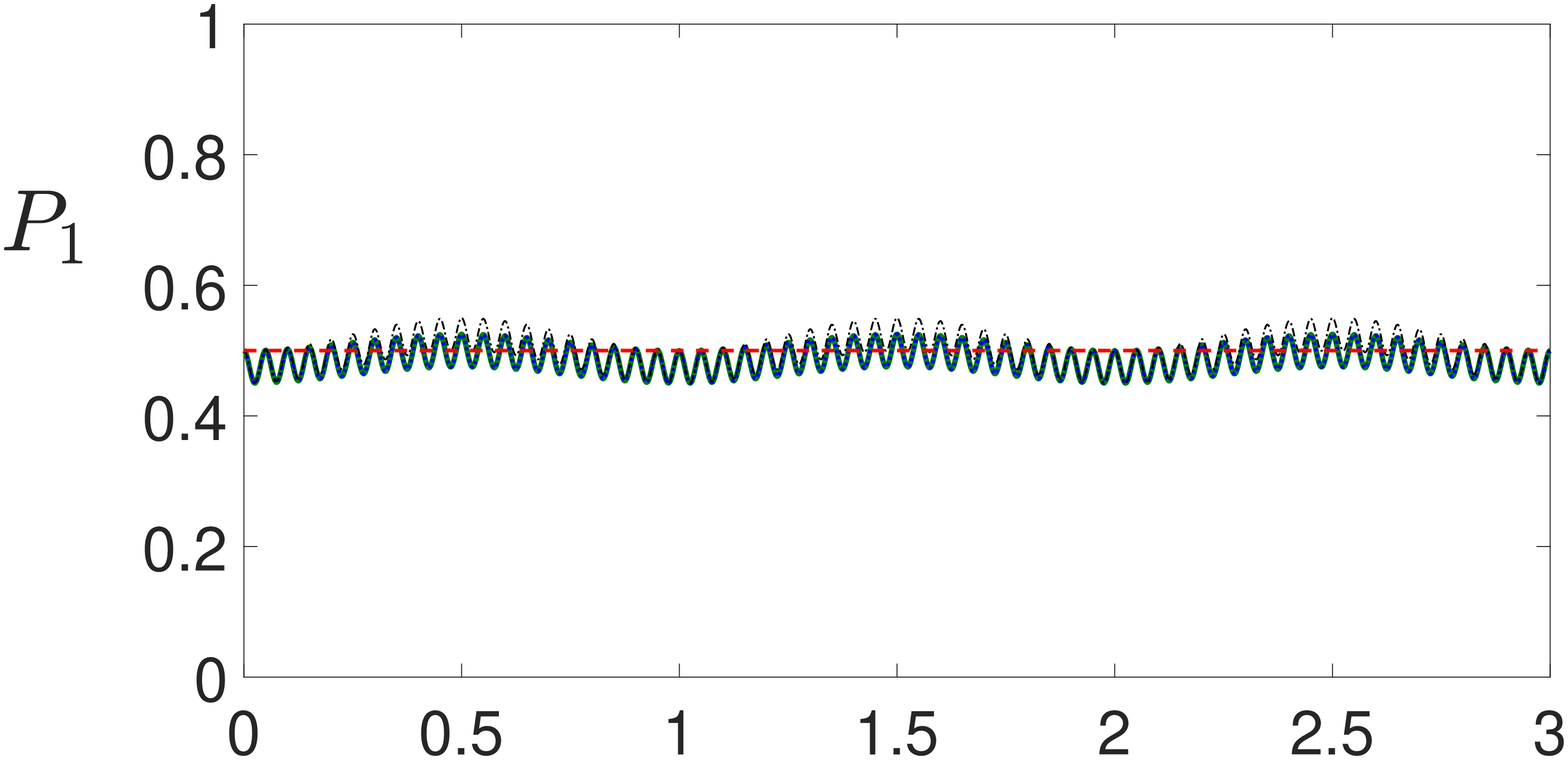}
	\includegraphics[width=0.45\textwidth]{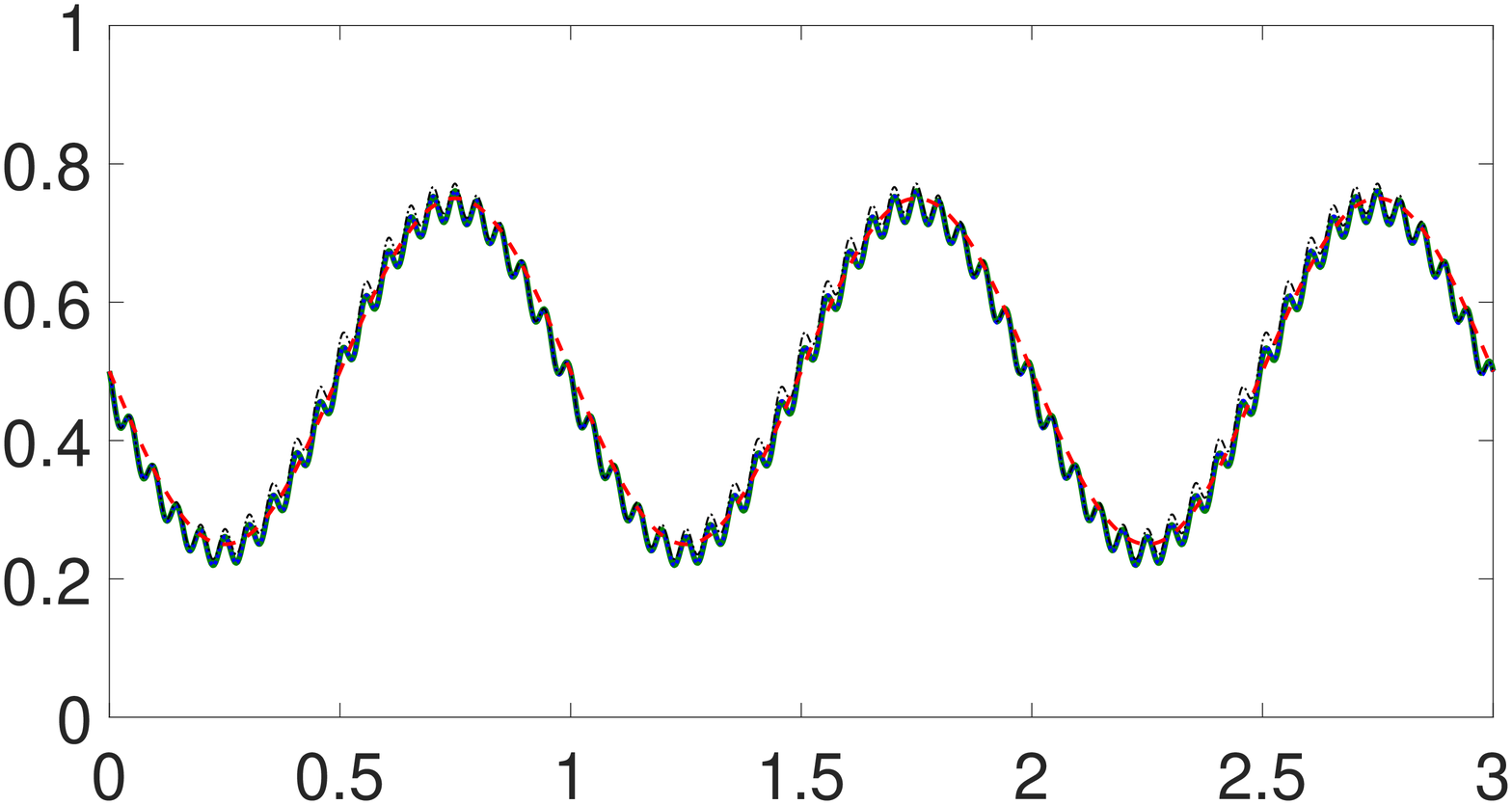}\\
	\vspace{-0.5cm}
	\includegraphics[width=0.45\textwidth]{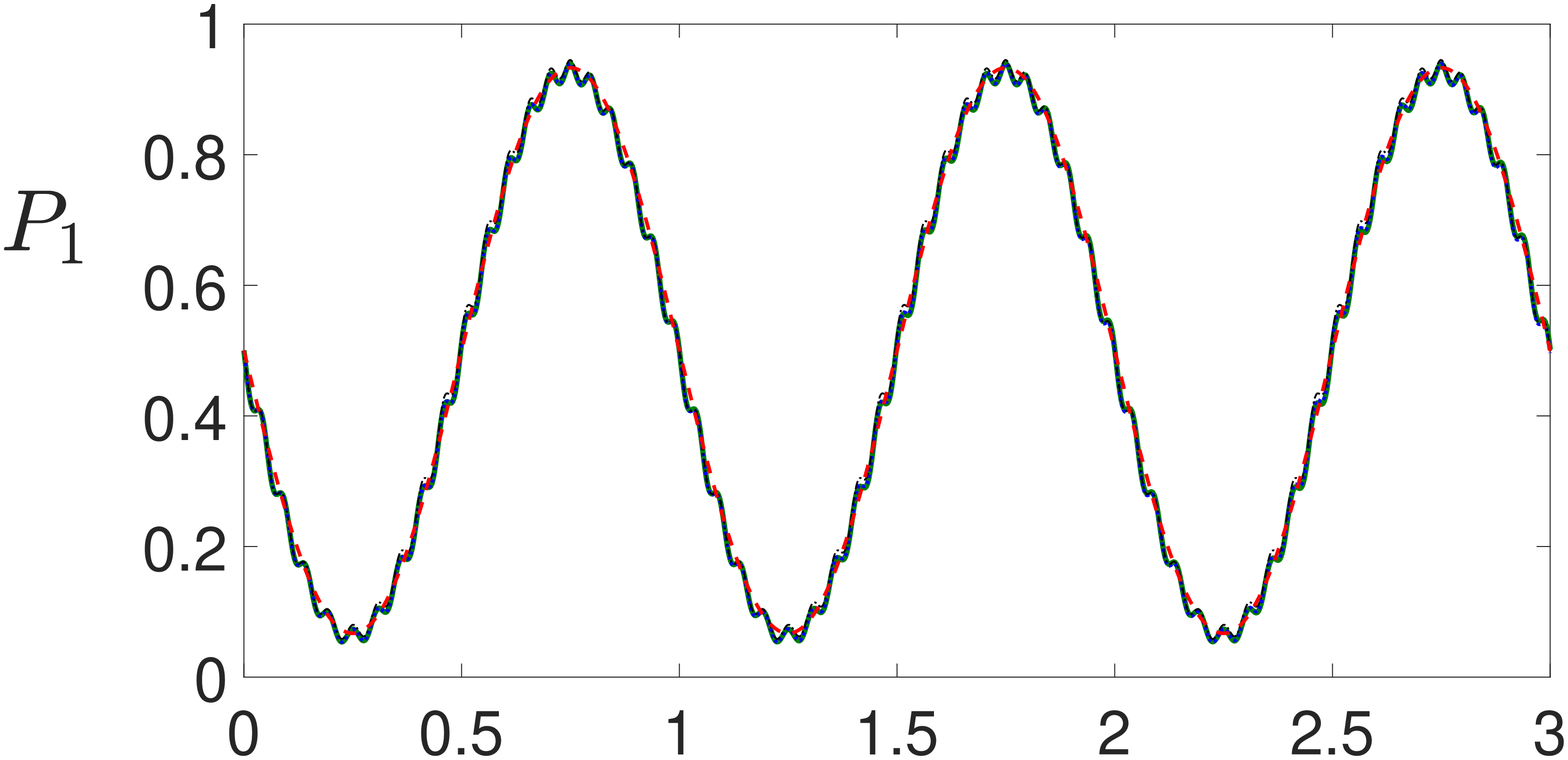}
	\includegraphics[width=0.45\textwidth]{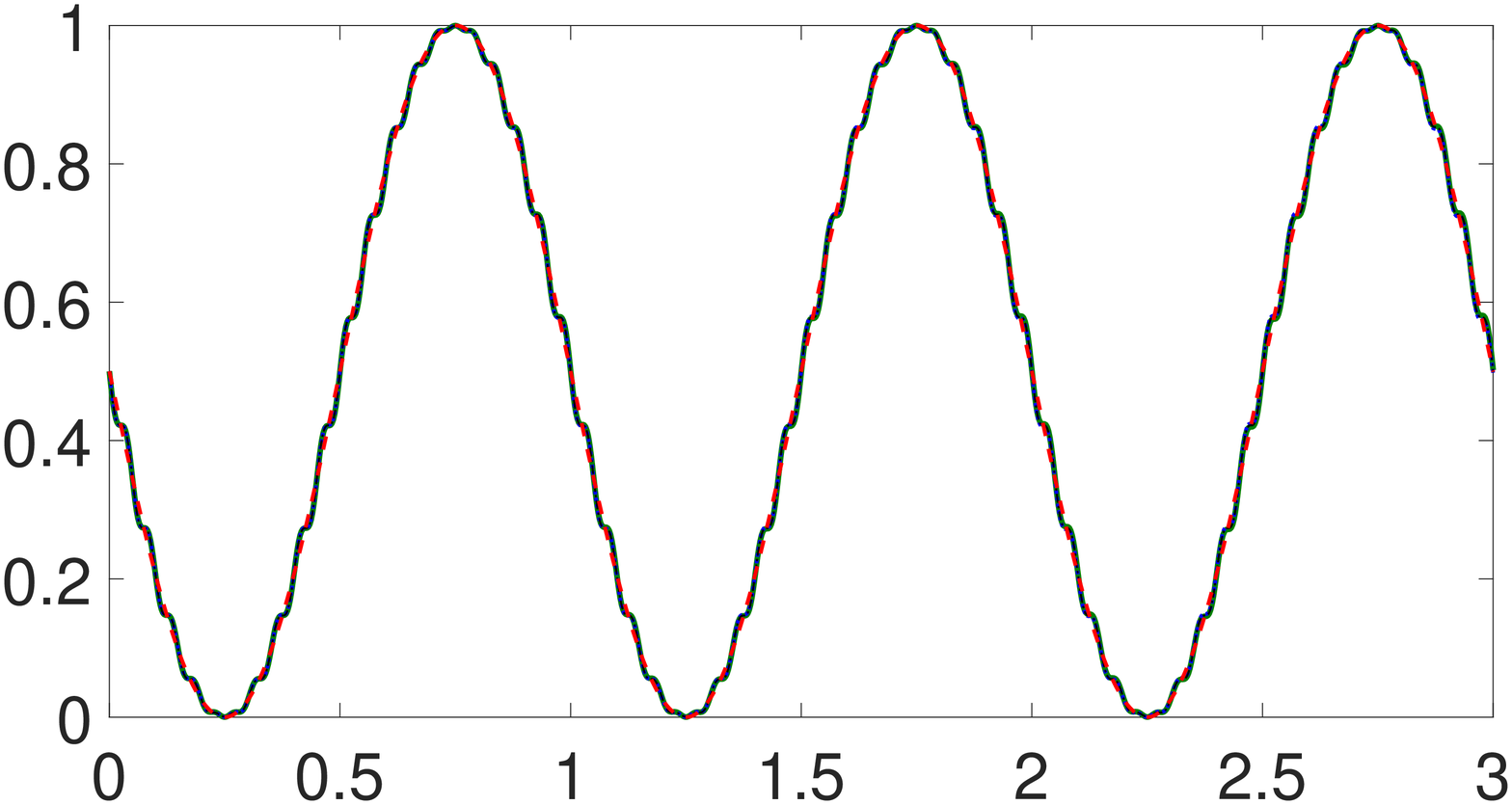}\\
	\vspace{-0.5cm}
	\includegraphics[width=0.45\textwidth]{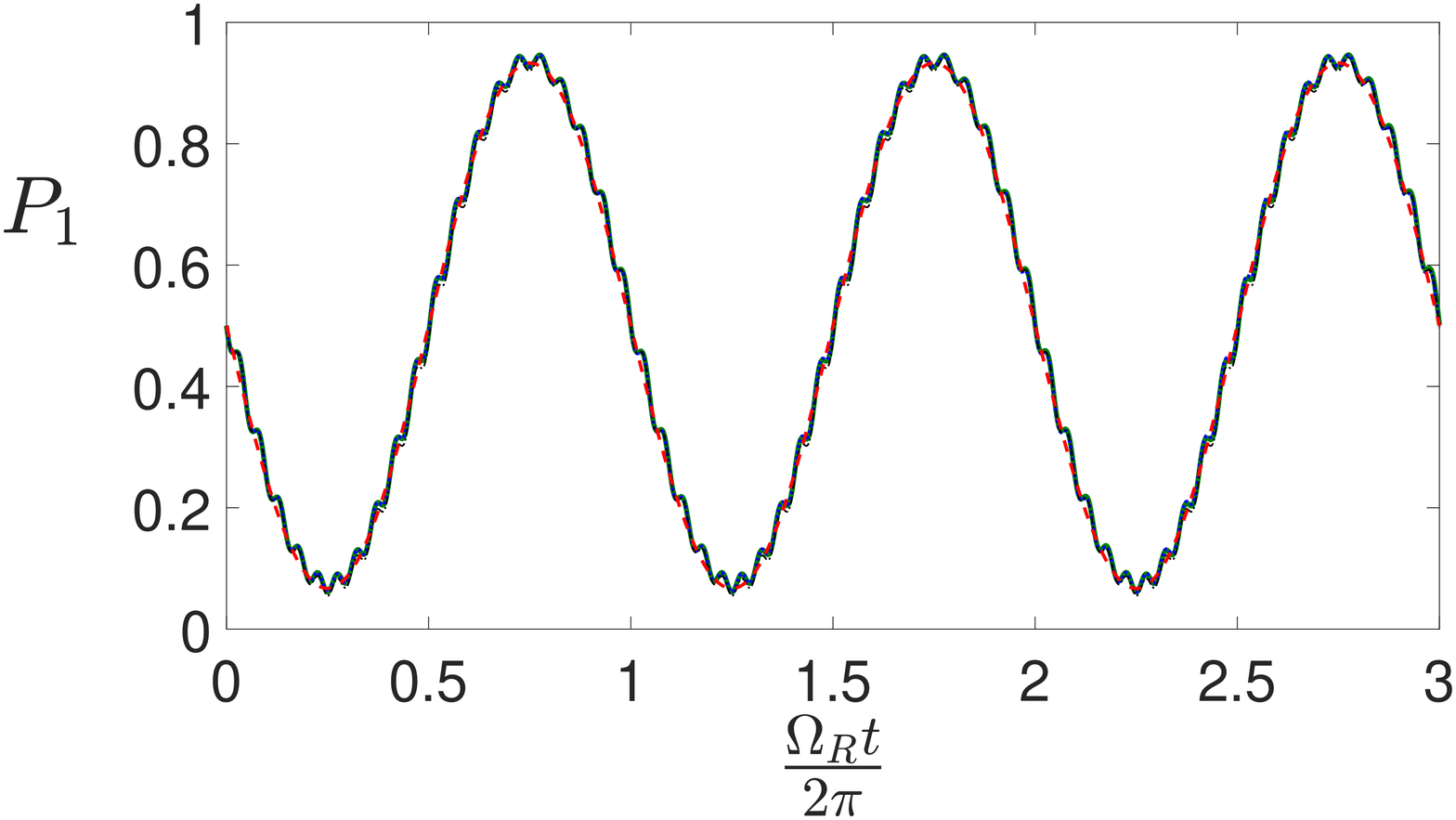}
	\includegraphics[width=0.45\textwidth]{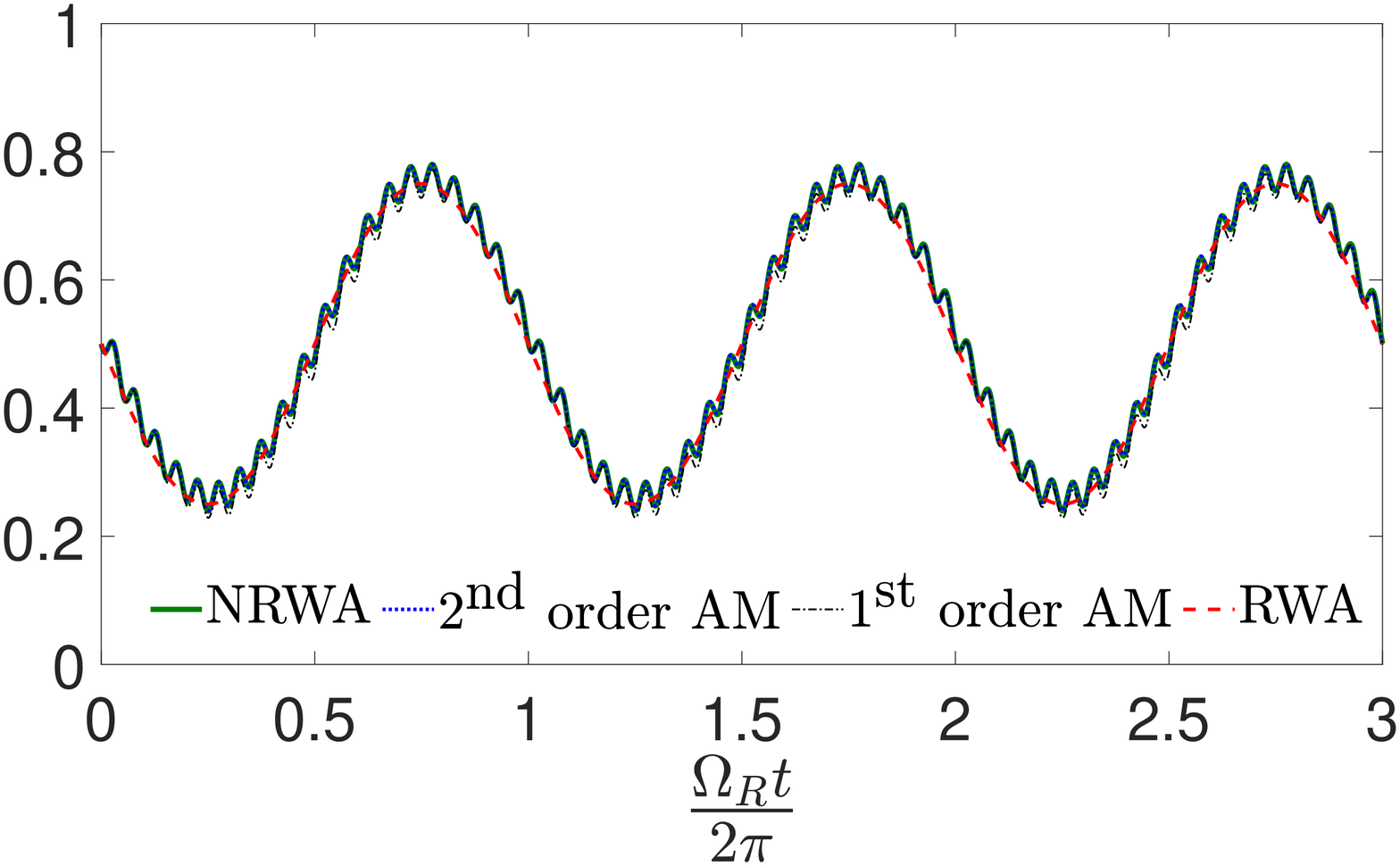}
	\vspace{-0.2cm}
	\caption{$\Delta = 0$. $P_1$ vs. $\frac{\Omega_{\textrm{R}} t}{2\pi}$, $\epsilon = \frac{\Omega_{\textrm{R}}}{\omega} = 0.1$, varying $\theta - \phi$.
		(a) $\theta - \phi = 0$.
		(b) $\theta - \phi = \frac{\pi}{6}$. 
		(c) $\theta - \phi = \frac{\pi}{3}$.
		(d) $\theta - \phi = \frac{\pi}{2}$. 
		(e) $\theta - \phi = \frac{2\pi}{3}$. 
		(f) $\theta - \phi = \frac{5\pi}{6}$. 
		(g) $\theta - \phi = \pi$. 
		(h) $\theta - \phi = \frac{7\pi}{6}$.
		(i) $\theta - \phi = \frac{4\pi}{3}$.
		(k) $\theta - \phi = \frac{3\pi}{2}$.
		(l) $\theta - \phi = \frac{5\pi}{3}$.
		(m) $\theta - \phi = \frac{11\pi}{6}$.
		Lines refer to
		NRWA (continuous {\color{ForestGreen} ---}), 
		RWA (dashed {\color{red} $--$}), 
		second order AM (dotted {\color{blue} $\cdots$}), 
		first order AM (dash-dotted $\cdot -$).} 
	\label{fig:phasiResonance}
\end{figure*}

In Fig.~\ref{fig:phasiResonance} we keep $\epsilon = 0.1$, varying the initial phase difference, $\theta - \phi$. We observe that the amplitude of the oscillations can be readily manipulated this way.


\subsection{Non-resonant AM vs. resonant AM} 
\label{subsec:comparison} 
The reader might wonder why we have introduced two different versions of the AM, one for non-resonance (Subsec.~\ref{subsec:NR}) and another for resonance (Subsec.~\ref{subsec:R}). 
In the discussion at the beginning of Sec.~\ref{sec:AM}, we have already explained the reason: When $\Delta$ becomes very small,  $\frac{\Omega_{\textrm{R}}}{\Delta}$ gets very large so that non-resonant AM is not successful anymore. Therefore, in resonance, the AM has to be manipulated in a different way. 

\begin{figure*}[t!]
	\centering
	\includegraphics[width=0.45\textwidth]{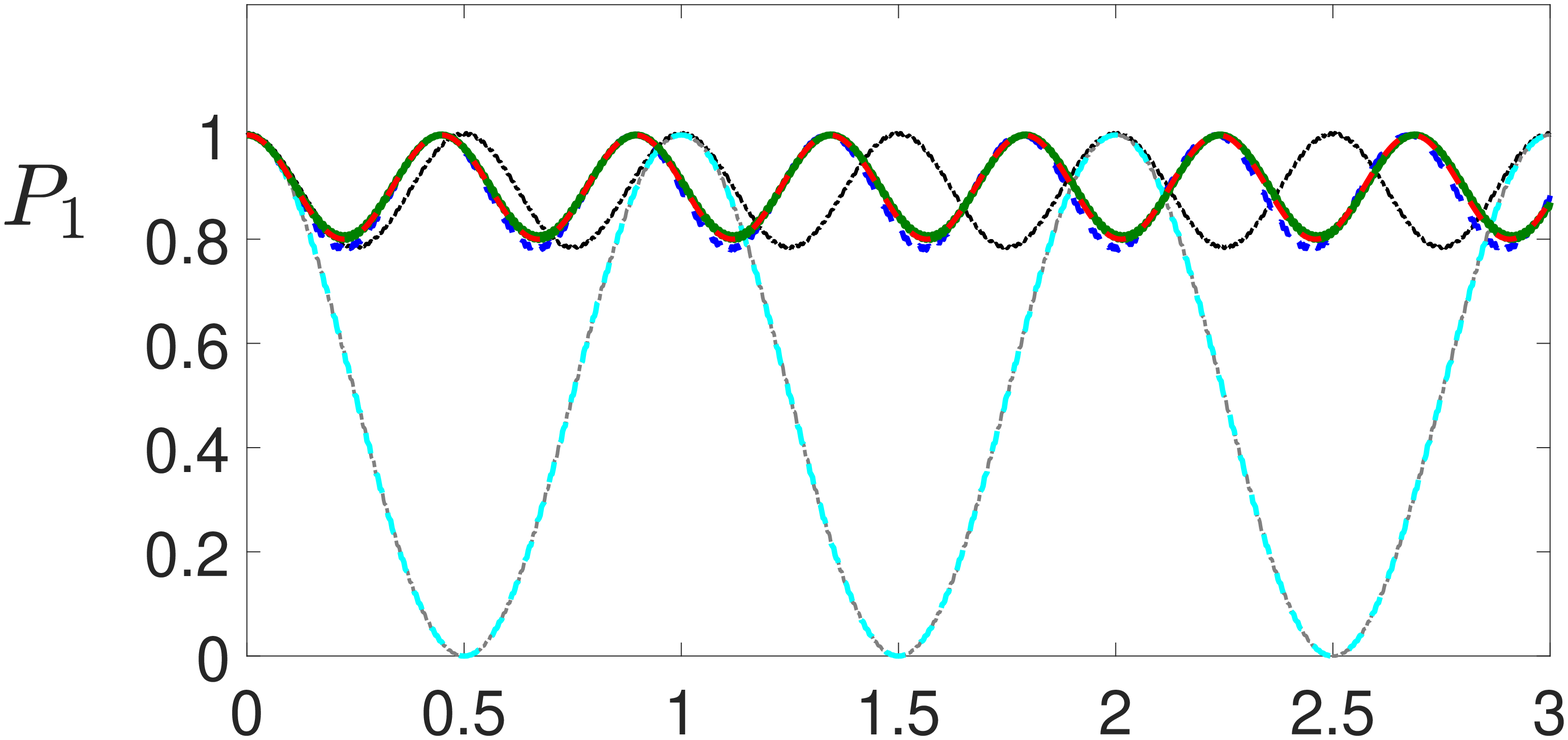}
	\includegraphics[width=0.45\textwidth]{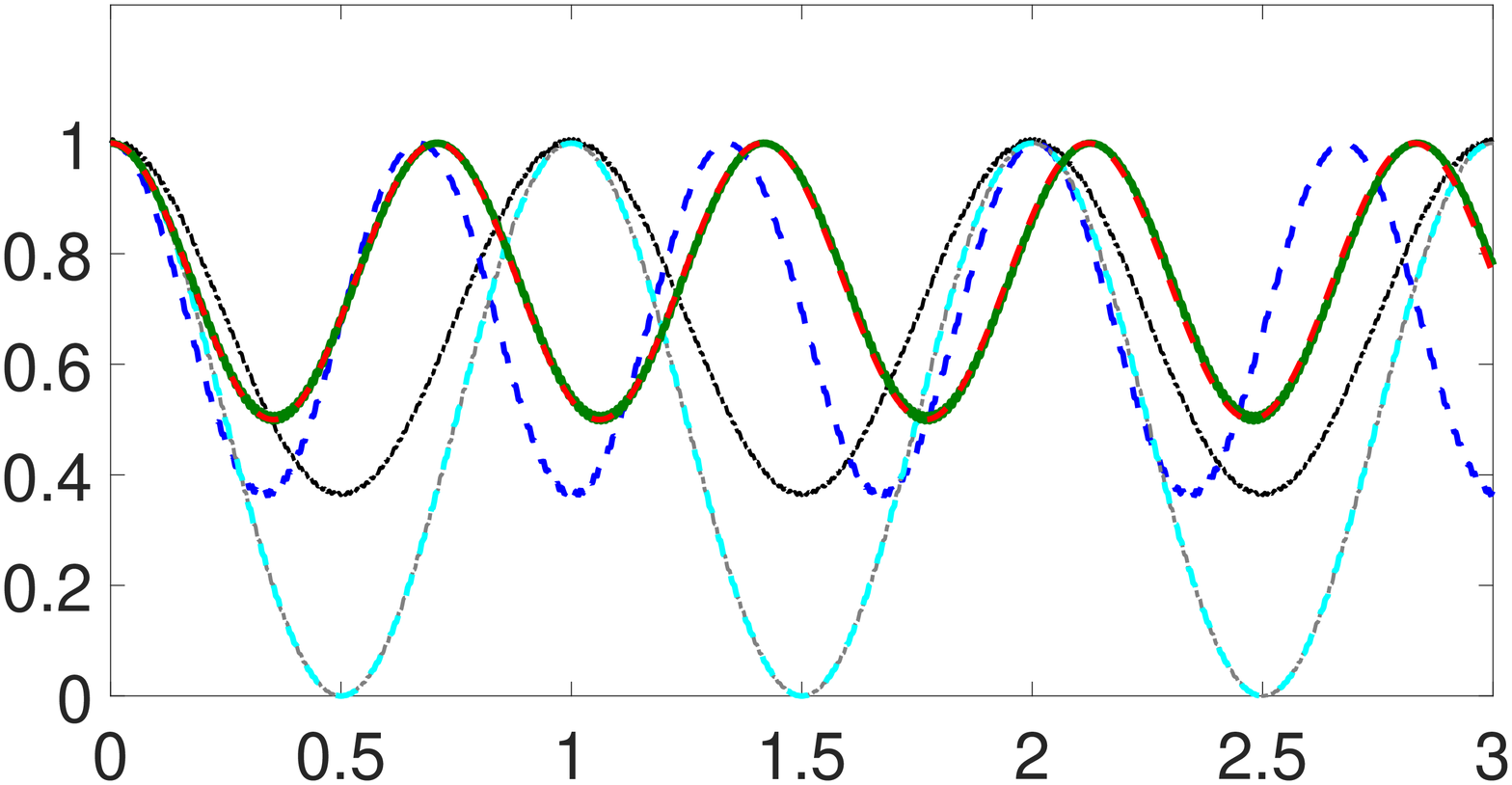}\\
	\vspace{-0.5cm}
	\includegraphics[width=0.45\textwidth]{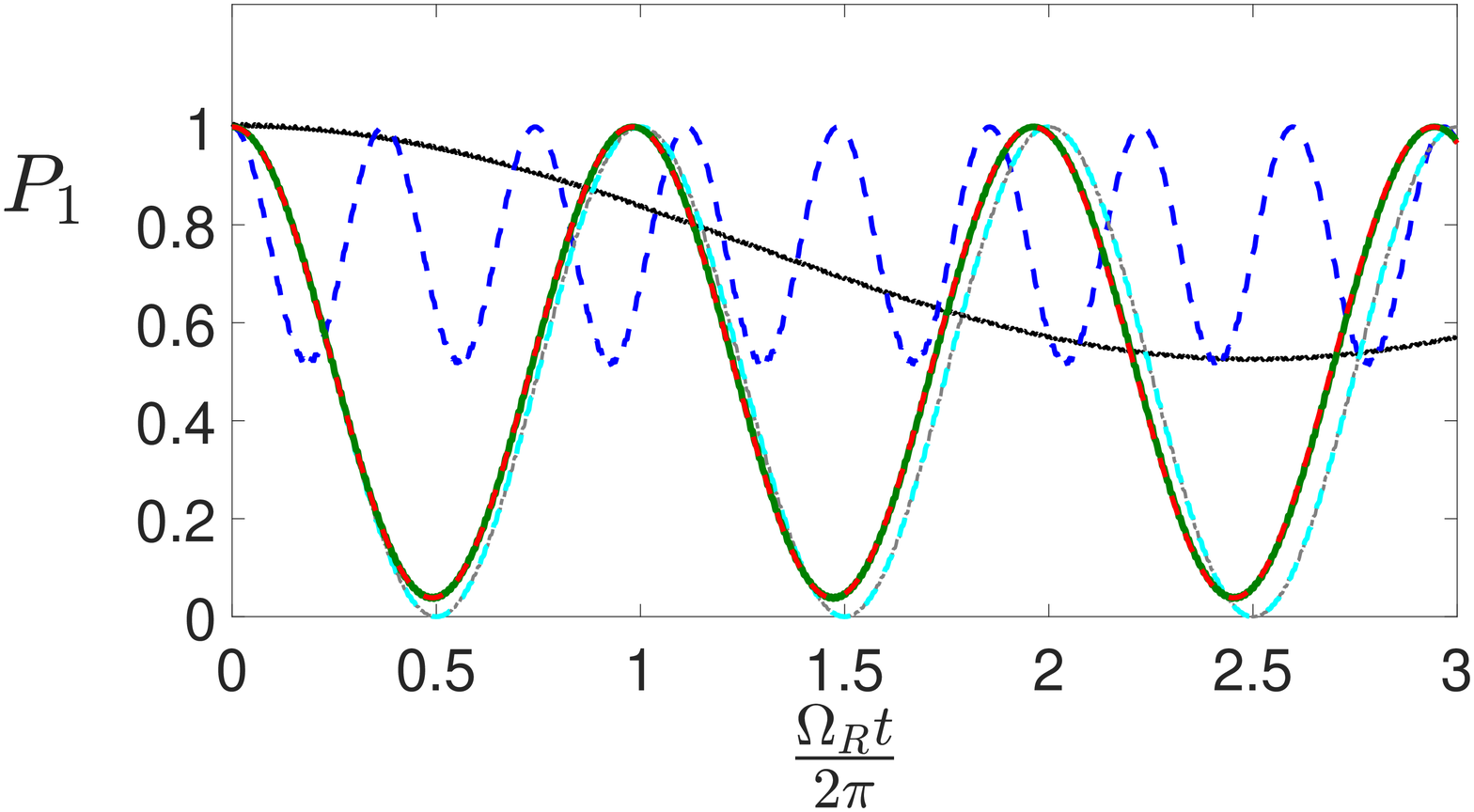}
	\includegraphics[width=0.45\textwidth]{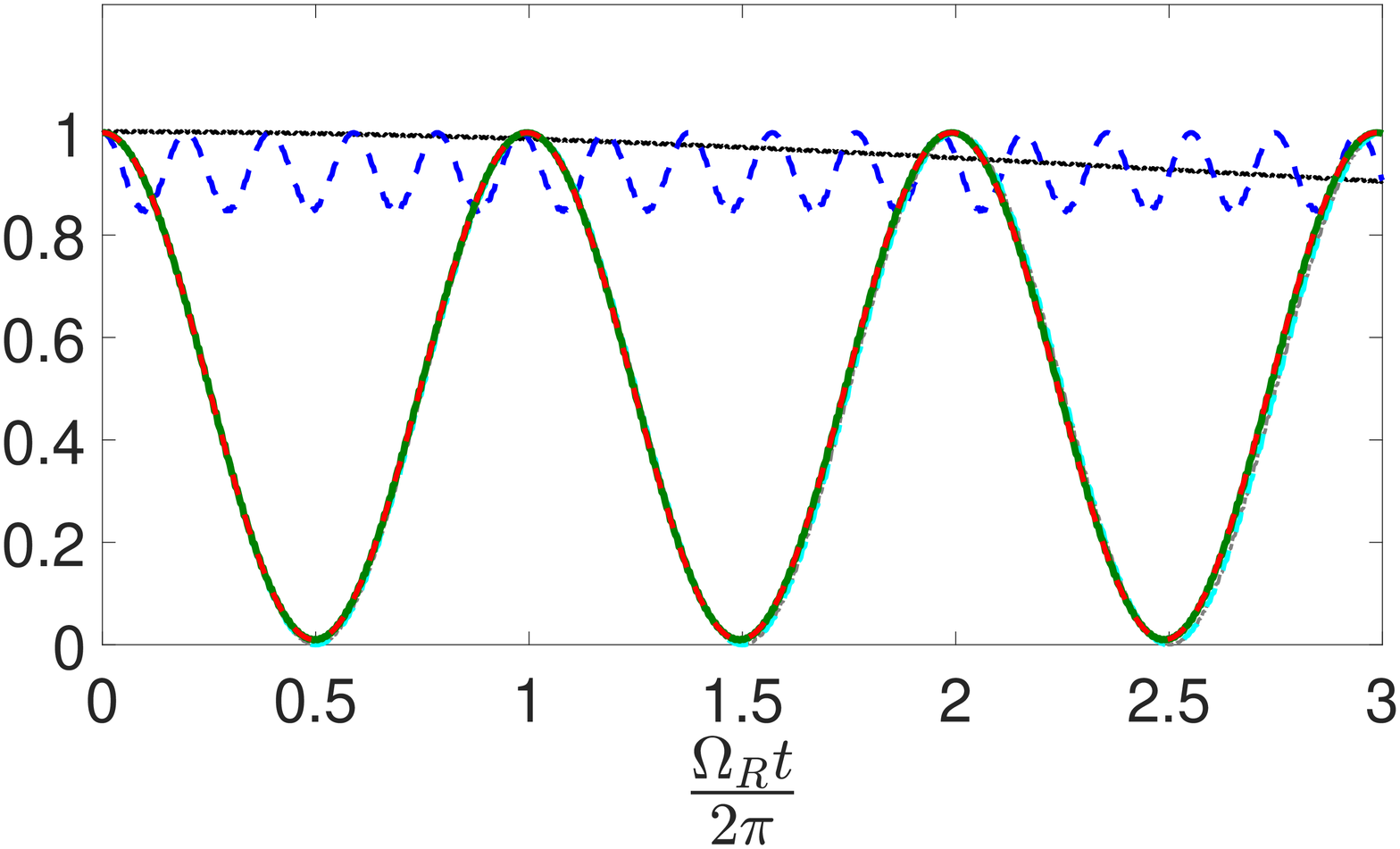}\\
	\vspace{-0.2cm}
	\caption{$\Delta \neq 0$. $P_1$ vs. $\frac{\Omega_{\textrm{R}} t}{2\pi}$, keeping  $\epsilon_2=\frac{\Omega_{\textrm{R}}}{\Sigma} = 0.01$ and
		varying $\epsilon_1=\frac{\Omega_{\textrm{R}}}{\Delta}$. 
		(a) $\epsilon_1 = 0.5$. 
		(b) $\epsilon_1 = 1$. 
		(c) $\epsilon = 5$. 
		(d) $\epsilon_1 = 10$.
		Lines correspond to 
		NRWA (continuous {\color{ForestGreen} ---}), 
		RWA (dashed {\color{red} $--$}), 
		non-resonance second order AM (dashed {\color{blue} $--$}), 
		non-resonance first order AM (dash-dotted $\cdot -$),
		resonance second order AM (dashed {\color{cyan} $--$}), 
		resonance first order AM (dash-dotted {\color{gray} $\cdot -$}).}
	\label{fig:NRandR}
\end{figure*}

Here we give a few examples. In Fig.~\ref{fig:NRandR}, we vary $\epsilon_1 = \frac{\Omega_{\textrm{R}}}{\Delta}$ and keep $\epsilon_2=\frac{\Omega_{\textrm{R}}}{\Sigma}=0.01$. We observe that for $\epsilon_1 < 1$, the second order AM for non-resonance is closer to the numerical solution (NRWA) than the second order AM for resonance. However, for $\epsilon_1 > 1$, $\Delta$ is so small that the second order AM for resonance comes closer to NRWA than the second order AM for non-resonance. We have already mentioned that the first order AM is usually far from NRWA, and we include it in the figures just to underline this fact.

\section{Conclusion} 
\label{sec:Conclusion} 
To explore the potential of the Averaging Method (AM) to handle coupled differential equations, we chose an archetypical system, i.e., Rabi oscillations in a two-level system (2LS), in the semiclasical approximation. 

We illustrated the need to manipulate in a different manner non-resonant AM from resonant AM. 

We introduced three dimensionless ratios $\epsilon$, namely, 
$\frac{\Omega_{\textrm{R}}}{\Delta}$,
$\frac{\Omega_{\textrm{R}}}{\Sigma}$ and
$\frac{\Omega_{\textrm{R}}}{\omega}$,  
to use properly the AM at non-resonance ($\Delta \ne 0$). 

In manipulating resonance ($\Delta = 0$), 
we had to consider that unavoidably, 
when $\Delta$ becomes smaller, at some point, $\frac{\Omega_{\textrm{R}}}{\Delta}$ gets so large that non-resonant AM is not successful anymore. 
Therefore, resonance had to be treated in a different way, 
using just one type of $\epsilon$, i.e.,
$\frac{\Omega_{\textrm{R}}}{\omega}$.

We compared first and second order AM with the full numerical solution (NRWA) as well as with Rotating Wave Approximation (RWA). 

First order AM is usually away from the numerical solution. 
However, second order AM can closely approach NRWA, provided the $\epsilon$ ratios are small enough. We explored the range where second order AM is successful. We also explored the range where RWA is successful.

Finally, we investigated various coherent phenomena, at resonance and at non resonance, putting a phase difference in the initial wave functions. Even with equal initial probabilities at the two levels, but with phase difference, strong oscillations can be generated and manipulated.


\newpage

\appendix{} 
\section{More Equations for first order AM} 
\label{Appendix1stAM} 
We proceed with detailed calculations.
\begin{equation}
d\mathbf{x}(t) = 
d\mathbf{y}(t) + \epsilon \; d\mathbf{w}[\mathbf{y}(t),t],
\end{equation}
\begin{equation}
d\mathbf{w}[\mathbf{y}(t),t] = 
\frac{\partial \mathbf{w}}{\partial \mathbf{y}} d\mathbf{y} + \frac{\partial \mathbf{w}}{\partial t}dt.
\end{equation}
By $\frac{\partial \mathbf{w}}{\partial t}$ we denote the derivative of $\mathbf{w}$ relative to $t$, keeping $\mathbf{y}(t)$ constant, hypothesizing that $\mathbf{y}(t)$ is a slowly varying function. Hence, 
\begin{equation}
\dot{\mathbf{x}} = \dot{\mathbf{y}} + 
\epsilon \frac{\partial \mathbf{w}}{\partial \mathbf{y}} \dot{\mathbf{y}} + 
\epsilon \frac{\partial \mathbf{w}}{\partial t}. 
\end{equation}
Therefore, Eq.~\eqref{standardform} becomes
\begin{equation}\label{AM2}
\dot{\mathbf{y}} + 
\epsilon \frac{\partial \mathbf{w}}{\partial \mathbf{y}} \dot{\mathbf{y}} + 
\epsilon \frac{\partial \mathbf{w}}{\partial t} = 
\epsilon f(\mathbf{y} + \epsilon \mathbf{w},t) + 
\epsilon^2 g(\mathbf{y} + \epsilon \mathbf{w},t).
\end{equation}
Using Eq.~\eqref{f}, we obtain
\begin{equation}\label{transformationfirst+f}
f(\mathbf{y} + \epsilon \mathbf{w},t) =
\overline{f}(\mathbf{y} + \epsilon \mathbf{w}) + 
\widetilde{f}(\mathbf{y} + \epsilon \mathbf{w}, t).
\end{equation}
Using a Taylor expansion vs. $\mathbf{y}$ we obtain
\begin{align}\label{taylor1}
&\overline{f}(\mathbf{y} + \epsilon \mathbf{w}) =
\overline{f}(\mathbf{y}) 
+ \epsilon \mathbf{w} \frac{\partial \overline{f}(\mathbf{y})}{\partial \mathbf{y}} + 
\mathcal{O}(\epsilon^2 \mathbf{w}^2),  \\ \label{taylor2}
&\widetilde{f}(\mathbf{y} + \epsilon \mathbf{w},t) = 
\widetilde{f}(\mathbf{y},t) + \epsilon \mathbf{w} \frac{\partial \widetilde{f}(\mathbf{y},t)}{\partial \mathbf{y}} + 
\mathcal{O}(\epsilon^2 \mathbf{w}^2), \\ \label{taylor3}
&g(\mathbf{y} + \epsilon \mathbf{w},t) = 
g(\mathbf{y},t) + \epsilon \mathbf{w} \frac{\partial g(\mathbf{y},t)}{\partial \mathbf{y}} + 
\mathcal{O}(\epsilon^2 \mathbf{w}^2).
\end{align}
Eq.~\eqref{AM2}, using Eqs.~\eqref{f}, \eqref{taylor1}, \eqref{taylor2}, \eqref{taylor3}, becomes 
\begin{align}\label{becomes}
& \dot{\mathbf{y}} + 
\epsilon \frac{\partial \mathbf{w}}{\partial \mathbf{y}} \dot{\mathbf{y}} + 
\epsilon \frac{\partial \mathbf{w}}{\partial t} = 
\epsilon \left(\overline{f}(\mathbf{y}) + 
\widetilde{f}(\mathbf{y},t)\right) + \\
& \epsilon^2 \left(\frac{\partial \overline{f}(\mathbf{y})}{\partial \mathbf{y}} \mathbf{w} + 
\frac{\partial \widetilde{f}(\mathbf{y},t)}{\partial \mathbf{y}} \mathbf{w} + g(\mathbf{y},t)\right) + \mathcal{O}(\epsilon^3)  \nonumber
\end{align}
Rearranging,
\begin{align}\label{Rearranging}
& (I+ \epsilon \frac{\partial \mathbf{w}}{\partial \mathbf{y}})\dot{\mathbf{y}} = 
\epsilon \left(\overline{f}(\mathbf{y}) + 
\widetilde{f}(\mathbf{y},t) - \frac{\partial \mathbf{w}}{\partial t} \right) + \\
& \epsilon^2 \left(\frac{\partial \overline{f}(\mathbf{y})}{\partial \mathbf{y}} \mathbf{w} + 
\frac{\partial \widetilde{f}(\mathbf{y},t)}{\partial \mathbf{y}} \mathbf{w} + g(\mathbf{y},t) \right) + \mathcal{O}(\epsilon^3) \nonumber
\end{align}
$I$ is the unit relevant to the nature of $\mathbf{y}$. 
If $\mathbf{y}$ is a simple function of $t$, $I=1$. 
If $\mathbf{y}$ is a column matrix, as in our case, 
$I = 
\begin{bmatrix}
1 & 0 \\
0 & 1
\end{bmatrix}$.
Now, the use of Eq.~\eqref{w} to simplify Eq.~\eqref{Rearranging} is obvious.
\begin{align}
& \dot{\mathbf{y}} = 
\left(I + \epsilon \frac{\partial \mathbf{w}}{\partial \mathbf{y}}\right)^{-1} \\ \nonumber
& \left(\epsilon \overline{f}(\mathbf{y}) + \epsilon^2 \left(\frac{\partial \overline{f}(\mathbf{y})}{\partial \mathbf{y}} \mathbf{w} + 
\frac{\partial \widetilde{f}(\mathbf{y},t)}{\partial \mathbf{y}} \mathbf{w} + g(\mathbf{y},t) \right) + \mathcal{O}(\epsilon^3)\right),
\end{align}
\begin{equation}
\left(I + \epsilon \frac{\partial \mathbf{w}}{\partial \mathbf{y}}\right)^{-1} = 
I - \epsilon\frac{\partial \mathbf{w}}{\partial \mathbf{y}} +  \mathcal{O}(\epsilon^2) .
\end{equation}

\section{More Equations for second order AM} 
\label{Appendix2ndAM} 
We proceed with detailed calculations.
\begin{equation}
d\mathbf{y}(t) = 
d\mathbf{z}(t) + \epsilon^2 \; d\mathbf{u}[\mathbf{z}(t),t],
\end{equation}
\begin{equation}
d\mathbf{u}[\mathbf{z}(t),t] = 
\frac{\partial \mathbf{u}}{\partial \mathbf{z}} d\mathbf{z} + \frac{\partial \mathbf{u}}{\partial t}dt.
\end{equation}
By $\frac{\partial \mathbf{u}}{\partial t}$ we denote the derivative of $\mathbf{u}$ relative to $t$, keeping $\mathbf{z}(t)$ constant, hypothesizing that $\mathbf{z}(t)$ is a slowly varying function. Hence,
\begin{equation}
\dot{\mathbf{y}} = \dot{\mathbf{z}} + 
\epsilon^2 \frac{\partial \mathbf{u}}{\partial \mathbf{z}} \dot{\mathbf{z}} + 
\epsilon^2 \frac{\partial \mathbf{u}}{\partial t}. 
\end{equation}
Therefore, Eq. \eqref{AMsecondorderstart} becomes
\begin{align} \label{AMsecondorderstart2}
&\dot{\mathbf{z}} + 
\epsilon^2 \frac{\partial \mathbf{u}}{\partial \mathbf{z}} \dot{\mathbf{z}} + 
\epsilon^2 \frac{\partial \mathbf{u}}{\partial t} = \\ 
& \epsilon \overline{f}(\mathbf{z}+\epsilon^2 \mathbf{u}) + \epsilon^2\left( \overline{h}(\mathbf{z}+\epsilon^2\mathbf{u}) + \widetilde{h}(\mathbf{z}+\epsilon^2\mathbf{u},t)\right)+\mathcal{O}(\epsilon^3) \nonumber
\end{align}
Using a Taylor expansion vs. $\mathbf{z}$ we obtain
\begin{align}\label{taylor4}
&\overline{f}(\mathbf{z} + \epsilon^2 \mathbf{u}) = \overline{f}(\mathbf{z}) + \epsilon^2 \mathbf{u} \frac{\partial \overline{f}(\mathbf{z})}{\partial \mathbf{z}} + \mathcal{O}(\epsilon^4 \mathbf{u}^2), \\ \label{taylor5}
&\overline{h}(\mathbf{z} + \epsilon^2 \mathbf{u}) = \overline{h}(\mathbf{z}) + \epsilon^2 \mathbf{u} \frac{\partial \overline{h}(\mathbf{z})}{\partial \mathbf{z}} + \mathcal{O}(\epsilon^4 \mathbf{u}^2), \\ \label{taylor6}
&\widetilde{h}(\mathbf{z} + \epsilon^2 \mathbf{u},t) = \widetilde{h}(\mathbf{z},t) + \epsilon^2 \mathbf{u} \frac{\partial \widetilde{h}(\mathbf{z},t)}{\partial \mathbf{z}} + \mathcal{O}(\epsilon^4 \mathbf{u}^2)
\end{align}

Eq.~\eqref{AMsecondorderstart2} using Eqs~\eqref{h}, \eqref{taylor4}, \eqref{taylor5}, \eqref{taylor6}, becomes
\begin{align}
&\dot{\mathbf{z}} + 
\epsilon^2 \frac{\partial \mathbf{u}}{\partial \mathbf{z}} \dot{\mathbf{z}} + 
\epsilon^2 \frac{\partial \mathbf{u}}{\partial t} = \\ 
& \epsilon \overline{f}(\mathbf{z}) + \epsilon^2\left(\overline{h}(\mathbf{z}) + \widetilde{h}(\mathbf{z},t)\right) + \mathcal{O}(\epsilon^3) \nonumber
\end{align}
Rearranging
\begin{align} \label{Rearranging2}
& (I + \epsilon^2 \frac{\partial \mathbf{u}}{\partial \mathbf{z}})\dot{\mathbf{z}} = \epsilon \overline{f}(\mathbf{z}) + \\
& \epsilon^2 \left(\overline{h}(\mathbf{z}) + \widetilde{h}(\mathbf{z},t)-\frac{\partial \mathbf{u}}{\partial t}\right) + \mathcal{O}(\epsilon^3) \nonumber
\end{align}
$I$ is the unit relevant to the nature of $\mathbf{z}$. 
If $\mathbf{z}$ is a simple function of $t$, $I=1$. 
If $\mathbf{z}$ is a column matrix, as in our case, 
$I = 
\begin{bmatrix}
1 & 0 \\
0 & 1
\end{bmatrix}$.
Now, the use of Eq.~\eqref{u} to simplify Eq.~\eqref{Rearranging2} is obvious.
\begin{align}
\dot{\mathbf{z}} = (I + \epsilon^2 \frac{\partial \mathbf{u}}{\partial \mathbf{z}})^{-1} 
\left(\epsilon \overline{f}(\mathbf{z}) + \epsilon^2 \overline{h}(\mathbf{z}) + \mathcal{O}(\epsilon^3) \right)
\end{align}
\begin{equation}
\left(I + \epsilon^2 \frac{\partial \mathbf{u}}{\partial \mathbf{z}}\right)^{-1} = 
I - \epsilon^2 \frac{\partial \mathbf{u}}{\partial \mathbf{z}} +  \mathcal{O}(\epsilon^4) .
\end{equation}

\bibliography{ChS}

\end{document}